\documentclass[%
 aip,
 amsmath,amssymb,
 reprint,%
]{revtex4-1}

\usepackage{graphicx}% Include figure files
\usepackage{dcolumn}% Align table columns on decimal point
\usepackage{bm}% bold math
\usepackage[utf8]{inputenc}
\usepackage[T1]{fontenc}
\usepackage{mathptmx}
\usepackage{siunitx}

\usepackage[utf8]{inputenc}
\usepackage[T1]{fontenc}
\usepackage{mathptmx}
\usepackage{siunitx}
\usepackage{tikz} 
\usetikzlibrary{positioning}
\usetikzlibrary{shapes,arrows,positioning,automata,backgrounds,calc,er,patterns}
\usepackage{tikz-feynman}
\tikzfeynmanset{compat=1.0.0}
\usepackage{dirtytalk}
\usepackage{hyperref}

\newcommand{\beq}{\begin{equation}}
\newcommand{\eeq}{\end{equation}}
\newcommand{\barr}{\begin{eqnarray}}
\newcommand{\earr}{\end{eqnarray}}
\newcommand{\bseq}{\begin{subequations}}
\newcommand{\eseq}{\end{subequations}}
\newcommand{\ket}[1]{|#1\rangle}
\newcommand{\braket}[2]{\langle #1|#2\rangle}

\newcommand{\expectation}[3]{\langle #1|#2|#3\rangle}

\newcommand{\vett}[1]{\textbf{#1}}
\newcommand{\uvett}[1]{\hat{\textbf{#1}}}

\begin{document}

\preprint{AIP/123-QED}

\title{Path Integrals: From Quantum Mechanics to Photonics}
% Force line breaks with \\

\author{Charles W. Robson}
 \affiliation{Photonics Laboratory, Tampere University, P.O. Box 692, FI-33014, Tampere, Finland}
\author{Yaraslau Tamashevich}%
 \affiliation{Photonics Laboratory, Tampere University, P.O. Box 692, FI-33014, Tampere, Finland}
\author{Tapio T. Rantala}
\affiliation{Department of Physics, Tampere University, P.O. Box 692, FI-33014, Tampere, Finland}
\author{Marco Ornigotti}
 \email{marco.ornigotti@tuni.fi}
 \affiliation{Photonics Laboratory, Tampere University, P.O. Box 692, FI-33014, Tampere, Finland}

\date{\today}% It is always \today, today,
             %  but any date may be explicitly specified

\begin{abstract}
The path integral formulation of quantum mechanics, \emph{\emph{i.e.}}, the idea that the evolution of a quantum system is determined as a sum over all the possible trajectories that would take the system from the initial to its final state of its dynamical evolution, is perhaps the most elegant and universal framework developed in theoretical physics, second only to the Standard Model of particle physics. In this tutorial, we retrace the steps that led to the creation of such a remarkable framework, discuss its foundations, and present some of the classical examples of problems that can be solved using the path integral formalism, as a way to introduce the readers to the topic, and help them get familiar with the formalism. Then, we focus our attention on the use of path integrals in optics and photonics, and discuss in detail how they have been used in the past to approach several problems, ranging from the propagation of light in inhomogeneous media, to parametric amplification, and quantum nonlinear optics in arbitrary media. To complement this, we also briefly present the Path Integral Monte Carlo (PIMC) method, as a valuable computational resource for condensed matter physics, and discuss its potential applications and advantages if used in photonics.
\end{abstract}

\maketitle

\section{Introduction}

The path integral formulation of quantum mechanics, developed in the mid-twentieth century, is not only a remarkable synthesis of several of the core ideas of theoretical physics, but it is also a powerful computational technique for the analysis of a huge variety of physical systems in very different contexts, such as quantum mechanics \cite{feynman48,hibbs,zinn,feyn_thes,kleinert, schulman}, quantum field theory \cite{fai,das,rivers,maggiore}, gauge field theory \cite{fadeev, frampton}, black hole physics \cite{schulman}, quantum gravity \cite{hawking1978,Birr82}, string theory \cite{polchinski}, topology \cite{schulman,topoPath}condensed matter physics \cite{condMatt1,condMatt2,condMatt2bis,condMatt3,condMatt4,condMatt5,condMatt6}, statistical mechanics \cite{feynmanStatisticalMechanics,bhagwat,hibbs,schulman}, polymer physics \cite{kleinert, schulman}, financial markets \cite{kleinert}, optical communications \cite{pathcomm}, atomic physics \cite{atomic1,atomic2}, spectroscopy \cite{spectroscopy}, light propagation in turbid media \cite{turbid} and classical \cite{gomez-reino1987,claudioQG,classicalOptics1,classicalOptics2,classicalOptics3,classicalOptics4} and quantum \cite{hillery,hillery2,ornigotti2019} optics, amongst others.

The underlying concepts of the path integral approach are sometimes considered difficult to grasp. Indeed, from a philosophical standpoint, its underpinnings are extraordinary – in describing a system, all possible paths between its initial and final state must be taken into account mathematically. This idea of summing over all paths has previously been characterised ontologically as \emph{everything that can happen does happen}  \cite{cox}.

As befits the subject, the history of path integrals is rather circuitous and does not follow a straight line from Feynman’s seminal work on the subject, published  in 1948 \cite{feynman48}, to the present day. Already in the 1920s, the mathematician Norbert Wiener had developed a method to treat Brownian motion and diffusion using a technique of integrating over paths, with formal similarities to Feynman’s eventual construction but in a purely classical context \cite{wiener1,brush,demichev}. During the early part of the following decade, the general ideas of de Broglie and Schr\"odinger, that waves can be associated with particle dynamics \cite{crossroad}, motivated  Dirac to publish a paper in the \emph{Physikalische Zeitschrift der Sowjetunion} (Physical Journal of the Soviet Union) setting the stage for future developments by proposing the Lagrangian as a more natural basis for a theory of quantum mechanics, as opposed to a Hamiltonian-based method which he argued was less fundamental due to its non-relativistic form \cite{dirac1}. This led him to propose the key idea that a quantum mechanical transition amplitude for a particle is given by a phase factor controlled by the action along that particle’s path \cite{dirac2}. This also led him to state, that the classical path of a quantum system can be interpreted as resulting from the constructive interference of all such paths. In other terms, the action $S$ of a given system counts, \emph{de-facto}, the number of waves of the path in units of Planck's constant $h$, and, therefore, for each path, the phase of the related wave  \footnote{It is not difficult, using an electromagnetic analogy, to see that in this context, Planck's constant $h$ plays the role of wavelength of the path. A plane wave propagating, say, along the $x$-direction, in fact, is associated with a phase factor of the form $\exp{(i\,2\pi\,x/\lambda)}$. The phase factor carried by each path i, instead, goven by $\exp{(i\,2\pi\,S/h)}$. A simple comparison allows the reader to identify $\lambda\rightarrow h$.} is given by $\exp{(i\,2\pi\,S/h)}=\exp{(iS/\hbar)}$. This allows the evaluation of the interference pattern of the particle dynamics, and essentially represents what is nowadays commonly understood as path integral \cite{hibbs, kleinert}.  

The pivotal step in the development of the theory occurred in the 1940s, when Feynman formulated his version of quantum mechanics, a `third way' following the earlier, well-known Schr{\" o}dinger and Heisenberg alternatives \cite{feyn_thes}. Based on integration over all paths between initial and final physical states, with each path contributing an action-dependent phase as Dirac had proposed, Feynman's formulation culminated in his important 1948 paper entitled \emph{Space-time approach to non-relativistic quantum mechanics} \cite{feynman48}.

From a conceptual point of view, one of the most interesting consequences of path integrals is that they can provide a deep understanding of the relation between quantum and classical mechanics, as the limit $h\rightarrow 0$ emerges naturally from the formalism as the classical limit of the theory \cite{zinn}.

Although the concepts underlying the path integral formalism may at first appear quite alien, they are deeply profound. Even if its utility were limited, the beauty of the idea would still merit a wide audience. The fact is, however, that the path integral has immense value as a practical computational technique in the physical sciences, and it can be applied to solve problems in many diverse areas, even beyond physics. Two examples showing the remarkable breadth of its applicability are its use in quantum gravity, where the ``sum over histories” is interpreted as a sum over all different spacetime configurations interpolating between an initial and final state of the universe \cite{hawking1978}, and in financial market modelling, where the formalism has proven useful, since the time dependence of asset prices can be represented by  fluctuating paths \cite{kleinert}. 

The universality of the technique has allowed scientists to tackle many problems, and gain tremendous physical insight into them. In statistical physics, for example, path integrals conveyed the basic framework for the first formulation of the renormalisation group transformation, and they are largely employed to study systems with random distribution of impurities \cite{schulman}. In particle physics, they allow one to understand and properly account for the presence of instantons \cite{weinberg}. In quantum field theory they provide the natural framework to quantise gauge fields \cite{fadeev, frampton, ammon}. In chemical, atomic, and nuclear physics, on the other hand, they have been applied to various semiclassical schemes for scattering theory \cite{schulman}. Through path integrals, topological and geometrical features of classical and quantum fields can be readily investigated and be used to create novel forms of perturbative and nonperturbative analysis of fundamental processes of Nature \cite{baez, polchinski, green}.  In addition to that, path integrals allow one to re-interpret established results, such as, for example, the BCS theory of superconductivity \cite{fai, BCS1}, from a novel, more insightful, perspective.

Explicit, analytical solutions to problems formulated in terms of path integrals, however, are scarce, and only available for very simple systems, such as a free particle, or the ubiquitous harmonic oscillator \cite{hibbs}. The complexity of the path integral formalism, in fact, increases very rapidly to overwhelming levels of difficulties for many simple problems. As an example of that, the simplest quantum system, \emph{i.e.}, a single hydrogen atom, required nearly 40 years to be fully solved in terms of path integrals \cite{kleinertHatom}.

On the other hand, it is amidst complex and computationally challenging problems that path integrals show their true potential, providing a simple, insightful and intuitive perspective on the physical principles regulating such processes. To do that, numerical techniques, such as Monte Carlo methods \cite{monte1,monte2,creutz,gattringer,usersguide,condMatt2bis,condMatt6}, and the computational power of modern supercomputers are crucial to their successful implementation.  

Most of the practical applications to numerical simulations of quantum particles with path integral approaches are systems with finite temperature in thermodynamical equilibrium with one of the statistical ensembles \cite{feynmanStatisticalMechanics}. Finite temperature equilibrium involves dynamics of constituent particles and exchange of energy with environment or heat bath.  These are the central factors in condensed matter physics with phase transitions, conductivities and other processes related to interactions between constituent particles.  Typically, one assumes canonical ensembles, where both the number of particles and the volume they occupy are kept constant at a given temperature $T$, but other ensembles can be chosen where needed.

For a many-particle system in finite temperature, there is no wave function, but the mixed state can be described with a density matrix, and it turns out that it can be written in terms of path integrals in imaginary time \cite{hibbs,feynmanStatisticalMechanics}.  The expectation values of observables are then evaluated by using the trace of the density matrix, which in space basis means finite closed loop paths.  Thus, in terms of path integrals each of the particles propagate from a position in space in imaginary time back to the same position, the time period being inversely proportional to the temperature.  Then, with Metropolis Monte Carlo it is possible to sample particle paths with correct weight in predefined temperature and collect data enough for convergence of expectation values of relevant operators.

This approach based on imaginary time propagation is called Path Integral Monte Carlo method (PIMC). David Ceperley and coworkers have carried out seminal development work and a sizable number of PIMC simulations of various many-particle systems ranging from superfluid He \cite{condMatt1} and neutron matter \cite{condMatt7} to electrons and hydrogen in extreme conditions \cite{condMatt8}, including both bosonic and fermonic particles. 

In recent years, one of the authors of this tutorial (TTR) has significantly contributed to taking PIMC simulations to new though simple quantum particle systems, such as small atoms \cite{tapio1,tapio2}, molecules \cite{tapio3,tapio4,tapio5,tapio6}, a chemical reaction \cite{tapio7} and quantum dots \cite{tapio8,tapio9,tapio10}, with the ultimate goal of providing a more accurate description of their electronic structure and related properties including many-body effects, and how they change with temperature \cite{tapio2,tapio11,tapio12}. In this context, PIMC has proven to be a very reliable and excellent method to calculate the electric polarisabilities of atomic and molecular systems, leading therefore to an accurate estimation of the optical properties of both individual small quantum systems and collections of them, in the form of dilute gases \cite{tapio6,tapio13,tapio14}. 

A different approach, based on a real, rather than imaginary, time path integral (RTPI) has been recently proposed as a way to describe the full quantum dynamics of a quantum system at zero-Kelvin, and to also characterise the evolution of its eigenstates \cite{tapio15,tapio9,tapio16,tapio17,tapio13,tapio18}. A combination of PIMC and RTPI therefore gives the possibility to have a comprehensive tool to study the properties of complex systems and their classical and quantum evolution. This feature in particular might prove to be very useful not only in chemistry and condensed matter physics, where this technique fluorished in the past decades, but also as a viable mean to understand and design the properties of materials of interest for photonics. 

A fully integrable simulation platform, that allows control of both electronic and photonic properties of matter \emph{exactly}, without the necessity to revert to approximations or effective theories, in fact, would constitute a tremendous resource towards the optimisation of integrated photonic systems. 

It is interesting to notice, that throughout the last 30 years, path integrals have been used to describe several problems in classical and quantum optics, such as the propagation of light in gradient index media \cite{gomez-reino1987}, the estimation of the channel capacity of classical and quantum fiber-based communication networks \cite{pathcomm}, parametric amplification \cite{hillery,hillery2}, light-matter interaction beyond the rotating wave approximation \cite{atomic1}, decoherence and dephasing in nonlinear spectroscopy \cite{spectroscopy}, and the effect of retardation in radiative damping \cite{atomic2}. Path integrals have also been employed to link the nonparaxial propagation of light with different models for quantum gravity \cite{claudioQG}. All these works share the common thread of employing nonrelativistic path integrals to calculate the propagator (\emph{i.e.}, the Green's function) of the electromagnetic field in different contexts, and use this information to solve the problem at hand. A different approach, based on path integrals in quantum field theory and Feynman diagrams, has been recently introduced as a viable way to handle classical \cite{bechler} and quantum \cite{ornigotti2019} optical phenomena in arbitrary media. 

However, the benefits of path integrals in photonics, namely their ability to calculate both the properties of matter and its interaction with light in an exact way, without the need of any approximation both on the matter and light side, and the new physical insight that this could bring to photonics, remains to date uncharted territory. 

In this tutorial, we aim at introducing the concept and methods of path integrals to the reader unfamiliar with the field, and to provide researchers in optics and photonics with a reference point for both analytical and numerical methods involving path integrals, with the hope that this will provide a powerful and practical toolkit that could be used in the future to tackle challenging problems in photonics. An accurate description of the diverse interactions between light and matter, emerging from the interplay of fundamental particles and fields, calls naturally for the use of quantum physics, and its degree of complexity grows very quickly. Path integrals are a natural way to study these interactions, and they actively take advantage of the complexity of the problem. Using them in photonics might then lead to novel methods to exploit complicated light-matter dynamics in photonic systems.

The tutorial is split into two main parts: Part 1, comprising Sections \ref{sect1} through \ref{sect4bis}, covers the basics of the path integral approach in physics, and presents examples on how they can be used to solve problems in classical and quantum optics, as well as how to employ PIMC to determine optical properties of materials. In particular, Sect. \ref{sect1} covers the fundamentals of the path integral approach in physics, treating core concepts such as the principle of least action, classical and quantum probabilities, and a brief description of the mathematics of integration over infinite number of paths. Basic examples, comprising the dynamics of a free quantum particle, the quantum harmonic oscillator, and diffraction from a double slit are covered in Sect. \ref{sect2}. To conclude Part I, two examples of the use of path integrals in classical and quantum optics are presented, namely the propagation of light in an inhomogeneous medium and how this could be related to the physics of a harmonic oscillator with a time-dependent frequency, in Sect. \ref{sect3}, and the investigation of degenerate parametric down conversion presented in Sect. \ref{sect4} (based on Ref. \citenum{hillery}), respectively. Finally, Sect. \ref{sect4bis} briefly discusses how PIMC can be used to predict the optical properties of matter, and presents some perspectives on the use of this computational resource for photonics.

Part 2, on the other hand, including Sections \ref{sect5} through \ref{sect9}, deals with the basics of path integrals in quantum field theory (QFT), and presents an application of such framework to the case of the dynamics of the electromagnetic field in arbitrary media. In particular, Sect. \ref{sect5} briefly introduces the concept of path integral for quantum fields, and makes use of the simple case of a scalar field as an example to calculate the relevant quantities and establish the formalism. After having done that, Sect. \ref{sect6} discussed how to include nonlinear interactions in the formalism, and introduces Feynman diagrams. The results from these two sections are then intuitively and qualitatively generalised for the case of a vector field in Sect. \ref{sect7}, as a reference point for Sect. \ref{sect8}, where these results are applied to the particular case of an electromagnetic field propagating in a dispersive medium of arbitrary shape. The last section of Part II, namely Sect. \ref{sect9} then presents two explicit examples, of how path integrals can be used in quantum optics. The first example presents how to describe the onset of spontaneous parametric down conversion (SPDC) in lossy media through path integrals, while the second example deals with the calculation of the rate of spontaneous emission of a quantum emitter surrounded by a dispersive medium.

In the spirit of the educational purpose of a tutorial, and given the mathematical complexity of path integrals especially concerning the concepts introduced in Part 2, we also provide, in Appendix \ref{appendixA}, a step-by-step guide on how to deal with path integral calculations for the explicit case of the electromagnetic field in arbitrary media. We hope this would serve as a good reference and guide to better understand the techniques and methods presented below.

Finally conclusions and future perspectives are then given in Sect. \ref{sect10}.

\section{Fundamentals of Path Integrals}\label{sect1}

\subsection{Probability Amplitudes: Classical vs. Quantum}\label{sectCLvsQU}
\begin{figure*}[!t]
\begin{center}
\includegraphics[width=\textwidth]{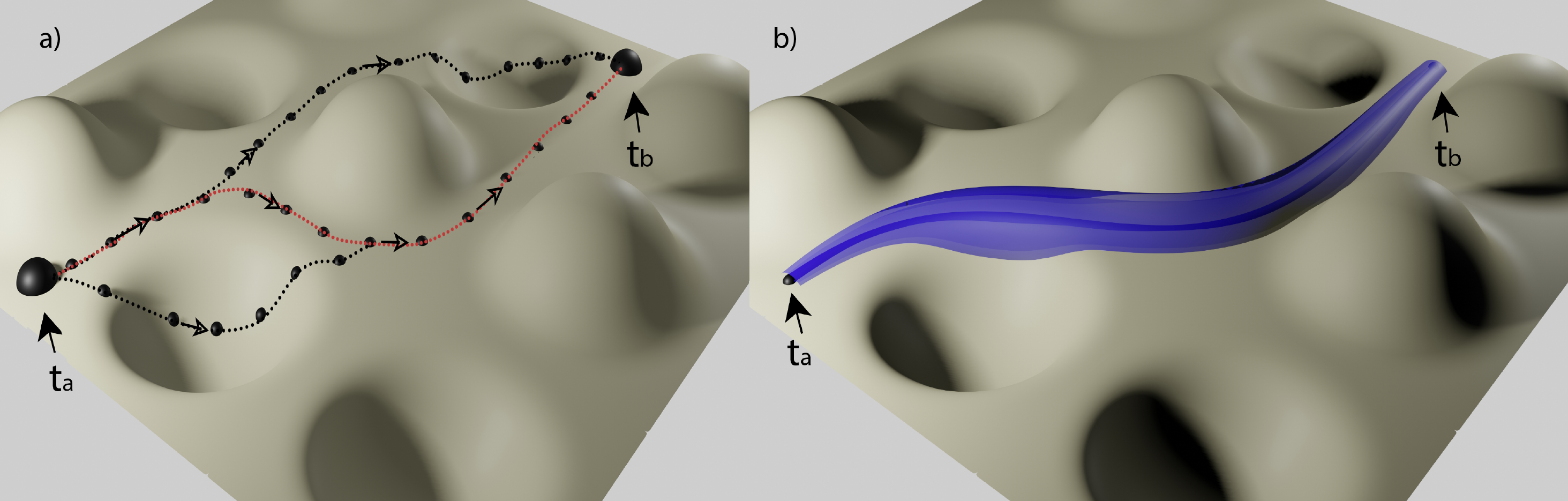}
\caption{Pictorial representation of the evolution of a classical (left) and quantum (right) system from an initial state at time $t_a$, to a final state at time $t_b$ over a certain potential landscape. (a) While many different paths link the initial and the final state of the classical system, once the initial conditions have been fixed, only the path with minimal action (red line in the figure) is the actual one undertaken by the system. (b) For a quantum system, on the other hand, the classical path (blue tube) minimising the action is interpreted as the path along which the interference of all the contributing paths linking $t_a$ and $t_b$ is maximum. }
\label{figure1action}
\end{center}
\end{figure*}
Quantum physics is an abstract theory, whose specific features beyond classical physics are typically only spectroscopically observable.  A good starting point to find the underlying differences between the two seemingly different worlds of classical and quantum physics, is represented by their different interpretation of the concept of probability.  This, in fact, turns out to be a direct manifestation of the wave nature of quantum particles, and thus, the fundamental issue that we need to incorporate in the study of the dynamics of quantum systems.

The necessity of a change in viewpoint concerning probability, and the consequent definition, for quantum physics, of a complex-valued probability amplitude, emerges very clearly within the context of the least action principle. Let us consider the situation depicted in Fig. \ref{figure1action}, where a classical [panel (a)] and a quantum [panel (b)] particle are evolving from an initial time $t_a$ to a final time $t_b$. 

For the classical, deterministic, system, although many different paths joining $t_a$ with $t_b$ are available, only the stationary paths with least action [red line in Fig. \ref{figure1action} (a)]  gives a significant contribution to its dynamics, and determine, ultimately, its equations of motion.

For a quantum system, on the other hand, its intrinsic wave nature (and, ultimately the uncertainty principle) prevent it for following one single path, and the classical path [blue tube in Fig. \ref{figure1action} (b)] must be interpreted as the one with maximum constructive interference coming from all possible paths.
%
%For such quantum probability or probability density we need to define a complex valued probability amplitude.  This allows emergence of probability distributions from all interfering paths, whereas with the classical deterministic dynamics, only the stationary paths with least action contribute.  From the wave perspective, the classical path is the one with maximum constructive interference.

To understand this better, let us consider a classical particle first propagating from a point $x_a$ to a point  $x_c$, and subsequently to a third point $x_b$.  If we denote with $P(c,a)\equiv P(x_a\rightarrow x_c)$ the probability for the particle to propagate from $x_a$ to $x_c$, and, similarly, with $P(b,c)\equiv P(x_c\rightarrow x_b)$ the probability for the particle to propagate from $x_c$ to $x_b$, the (classical) conditional probability for the particle to propagate from $x_a$ to $x_b$ by going through $x_c$ reads
%Classically, in case of different possible alternatives for the intermediate $c$, the conditional probability for propagation from $a$ to $b$ is given as
\beq
	\label{eq:CProba1}
       P(b,a) = \sum_c P(b,c) P(c,a),
\eeq
where the summation over $c$ takes into account all possible alternatives for the intermediate state $c$. Note, that the definition of conditional probability given above, \emph{i.e.}, $P(b,a)$, differs slightly from the usual one, which reads $P(a|b)$. This notation, however, is fully equivalent with the traditional one, and will turn out to be of more practical use for the purpose of this work.
%we do not use conventional notations for the conditional probabilities, here.  The chosen notations turn out to be more practical for our purpose.

The above definition can be readily generalised to the case of continuous variables, \emph{i.e.}, to probability densities, by promoting $P(a)$ and $P(b)$ to probability density functions, and to interpret $a$ and $b$ as two sets of coordinates $\{x_a\}$ and $\{x_b\}$ for the particle to occupy at a given time $t_a$ and $t_b$, respectively \footnote{we implicitly assumed that the two set of coordinates span a one dimensional space, for simplicity. Extension to higher dimensions can be, in fact, easily carried out.}. In this case, then, the summation over all possible alternatives $c$ in Eq. \eqref{eq:CProba1} becomes an integral over the set of coordinates $\{x_a\}$, \emph{i.e.},

%Now, let us assume one dimensional space for simplicity.  Then, we can generalize the above probabilities to probability densities $P(a)$ and $P(b)$, where both $a$ and $b$ are two sets of coordinates \{$x_a$\} and \{$x_b$\} for the particle to occupy at given times $t_a$ and $t_b$.   Thus, we can write an integral equation
\beq
	\label{eq:CProba2}
       P(b) = \int_a  P(b,a) P(a) \, {\rm d}x_a,
\eeq
where the subscript $a$ on the integral indicates that the integration has to take into account all the possible values of the integrating coordinate $x_a$.
%For the classical dynamics of a particle the initial condition is typically given as a single coordinate $a_0$ at time $t_a$.  Then, the probability distribution $P(a_0)$ is Dirac delta function, deterministically implying another delta function specifying the end point of the path, the coordinate at time $t_b$, subject to all interactions $c$ on the path between these two positions..

Now, let us extend the concepts introduced above to the case of a quantum particle with wave nature. To do that, let us first rewrite Eq. \eqref{eq:CProba1} for the probability amplitude $\psi$ associated to the quantum particle as
\beq
	\label{eq:QProba1}
       \psi(b,a)\equiv K(b,a) = \sum_c \psi(b,c) \psi(c,a) .
\eeq
The probability amplitude $\psi(b,a)$ defined above constitutes the basic quantity from which the dynamics of a quantum particle can be derived, and it is usually referred to in literature as the kernel (or propagator, or Green's function) of the quantum system at-hand. The conventional symbol for that in path integral language is $K(b,a)$, and we then adopt this notation for the rest of this manuscript. 

The kernel defined above has a simple physical interpretation. In fact, it can be thought as the impulse response of the system at-hand \cite{hibbs, byronFuller}. Moreover, if we know the probability amplitude of the system at a given initial state $\psi(a)$, we can immediately evolve it to a final state characterised by a probability amplitude $\psi(b)$ using the following relation
\beq
	\label{eq:QProba2}
        \psi(b) = \int_a K(b,a) \psi(a) \, {\rm d}x_a.
\eeq
Finally, the experimentally observable classical probability distributions are found as squares of the absolute values of the probability amplitudes, $P(a) = | \psi(a)|^2$ and $P(b) = | \psi(b)|^2$ at times $t_a$ (initial state) and $t_b$ (final state), respectively. With this definition, the probability amplitudes appearing above can be readily interpreted as the wave function of the quantum system. Equation \eqref{eq:QProba1}, in particular, hints at the interpretation of the wave function of a quantum system as the sum (or, better said, interference) of all possible paths linking the initial and final states of the considered evolution. 
%
%With this interpretation we know, what the mysterious wave functions really are. 
%
\subsection{Lagrangian, Action and Path Integral}

Contrary to the canonical formulation of quantum mechanics, which bases its premises on the Hamiltonian function $H$ of the system, and therefore on the concept of total energy \cite{sakurai}, the path integral formalism starts from the Lagrangian function, generally defined as the difference between kinetic and potential energy of the system  \cite{arnold}, \emph{i.e.}, $L=T-V$. For this reason, the path integral formalism is often referred to as the \emph{third formulation} of quantum mechanics, with the first being the matrix mechanics developed by Heisenberg, Born and Jordan in 1925 \cite{heisenberg1,born1,matrixMechanics}, while the second one being the familiar Hamiltonian formulation developed by Schr\"odinger in 1926 \cite{schroedinger1}. 

Although the Hamiltonian and Lagrangian formulation of physical problems are essentially, from the point of view of physical meaning, equivalent, the latter is more elegant, and, thanks also to its natural appearance in the path integral formalism, has been adopted as the natural framework for more complicated theories, such as QFT \cite{brown,das}, particle physics \cite{kaku}, and string theory \cite{green}.

For simplicity of notations, we proceed in using the one dimensional space with the coordinate $x$, but the generalisation to three dimensions is trivial.  Then, for a particle with mass $m$ in motion on the path $x(t)$ with velocity $\dot{x}(t)$ in a potential $V(x,t)$ the classical Lagrangian is
\beq
	\label{eq:Lagrangian}
        L(\dot{x},x,t) = {1 \over 2} m \dot{x}^2 - V(x,t).
\eeq
On a way to both finding the classical equation of motion and concurrently incorporating the wave nature of dynamics of the particle, on the path $x(t)$ from $t_a$ to $t_b$ we define the action
\beq
	\label{eq:Action}
	S(b,a) = \int_{t_a}^{t_b}\,dt\, L(\dot{x},x,t).
\eeq
In Lagrangian mechanics the action is a parameter related to the path length, whose optimisation will lead to the equations of motion. This procedure, \emph{i.e.}, optimising the path $x(t)$ such that $\delta S=0$ is called ``the principle of least action", and leads to the Euler–Lagrange equation \cite{landau, arnold}
\beq
	\label{eq:Euler}
	 \frac{\rm d}{{\rm d} t} \left( \frac{\partial L}{\partial \dot{x}} \right) - \frac{\partial L}{\partial x} = 0.
\eeq
If we now substitute Eq. \eqref{eq:Lagrangian} into the equation above, we  find the following differential equation 
\beq
m  \ddot{x}  = - \frac{\partial V(x)}{\partial x},
\eeq
which we immediately recognise as Newton's classical equation of motion. This constitutes the fundamental ingredient for defining the classical and quantum probabilities as in Eqs. \eqref{eq:CProba1} and \eqref{eq:QProba1}. However, for the dynamics of a classical system, the integral approach above is redundant, and the traditional approach based on Newton's equation of motion is favourable one in most cases.
%
%Thus, we have found the ingredients for formulation of the probabilities in Eqs.~(1–2), though it is obvious, that for the classical dynamics the integral equation approach is clumsy compared to the Newton's differential equation, in most cases.
%
Similarly, in quantum mechanics Schr\"odinger equation is the best approach for simple problems. However, there are several sophisticated cases where Eqs. \eqref{eq:QProba1} and \eqref{eq:QProba2} are more practical and easy to handle. For these more complicated scenarios, therefore, we need to find the kernel $K(b,a)$, and the right way to do that is to incorporate in the ``classical" path-based approach above, the information of the wave nature of quantum particles, \emph{i.e.}, to introduce interference between paths.
%
%Therefore, we search for the kernel $K(b,a)$, also called propagator or Green's function, by incorporation of the wave nature of particle dynamics, {\it \emph{i.e.}} the interference of paths.
%

Following the ideas of Dirac \cite{dirac1} and Feynman \cite{feynman48, feyn_thes}, we consider the action $S$ as a (classical) measure of the path length, and Planck's constant $h$ as the wave length.  Then, we can assign a wave $\exp{\left[i 2 \pi S(x(t),a)/h\right]}$ to the path $x(t)$ and follow the phase of the waves to find the interference effects.  In particular, we want to take into account the contributions of all possible waves of the form $\exp{\left[i S(b,a)/\hbar\right]}$ to the probability amplitude $K(b,a)$.

The sum (or integral) over the contributions of all possible paths is called the path integral, \emph{i.e.}, 
\beq\label{eq:pathIntegral}
K(b,a) = \int_a^b\, Dx(t)\,\exp{\left[\frac{i}{\hbar}S(b,a)\right]},
\eeq
where the notation ${ Dx(t)}$ indicates integration over all paths from $a = (x_a,t_a)$ to $b = (x_b,t_b)$, which, following Ref. \citenum{hibbs}, can be defined as
\beq
\int_a^b\,Dx(t) = \lim_{\varepsilon\rightarrow 0}\frac{1}{A(\varepsilon)}\int\,\frac{dx_1}{A(\varepsilon)}\,\int\,\frac{dx_2}{A(\varepsilon)}\,\cdots\,\int\,\frac{dx_{N-1}}{A(\varepsilon)},
\eeq
where $A(\varepsilon)$ is a suitable normalisation factor that ensures the limit to properly converge (an example of it is given in the next section, but the explicit for of $A(\varepsilon)$ might change depending on the problem at-hand), and $\varepsilon$ comes from the discretisation of the time interval in the action $S(b,a)$ in $N$ finite points, i.e., $t_b-t_a=N\varepsilon$, with $t_0=t_a$, and $t_N=t_b$. This, on the other hand, implies a discretisation on the paths $x(t)$, which now are defined as $x_k=x(t_k)=x(t_0+k\varepsilon)$, with $x_0=x_a=x(t_a)$ and $x_N=x_b=x(t_b)$. This discretisation procedure allows us to consider each of the $N-1$ integrals above as standard Riemann/Lebesgue integrals in the variable $dx_k$. Then, when all $N-1$ integrals have been computed, the limit $\varepsilon\rightarrow 0$, together with the correct definition of the normalisation factor $A(\varepsilon)$, ensures convergence of the path integral $\int_a^b\,Dx(t)$, and justifies its definition as integration over all possible paths. Notice, moreover, that the same line of reasoning will allow us, in Sect. \ref{sect5} to define the path integral for fields.
	
The explicit form of the path integral above is usually defined by the form of the potential term $V(x)$ appearing in the Lagrangian \eqref{eq:Lagrangian}. For some potential functions there are analytical, closed-form, exact solutions to the Eq. \eqref{eq:pathIntegral}, but one needs to prepare for numerical methods with possible approximations in more general cases, such as multi-dimensional or many-body problems.  In what follows, we present some of the exact propagators.

The above result, obtained for a simple one dimensional system, can be readily extended to three dimensional and many particle systems. While the former is straightforward to work out, in the latter case, the quantum statistics of fermions and bosons needs to be taken into account explicitly, which makes the problem of finding the correct generalisation of the path integral to the many body case less trivial \cite{fetter}. 

In addition to that, the path integral approach also allows for an easy way to simulate the evolution of the density matrix for finite temperature equilibrium systems. An example of this will be discussed below in Sect. \ref{sect4}.
%
%The former is straightforward but in the latter case, the quantum statistics of fermions and bosons needs to be respected, which may be less trivial. Also, the density matrix of finite temperature equilibrium systems can be simulated, which will also be discussed below.
%
\section{Basic Examples for Quantum Particles}\label{sect2}

For the evaluation of the wave function from the integral equation \eqref{eq:QProba2} we need to calculate explicitly the kernel from the path integral, \emph{i.e.}, Eq. \eqref{eq:pathIntegral}. In this section we therefore consider the simplest kernels, with examples following the book of Feynman and Hibbs \cite{hibbs}.

In case the integrand is an exponential of a quadratic function (Gaussian integral) the kernel can always be evaluated using recursively the basic Gaussian integral formula \cite{byronFuller}
\beq\label{gaussianIntegral}
\int\,dx\,e^{-ax^2+bx}=\sqrt{\frac{\pi}{a}}\,e^{b^2/4a}.
\eeq
Another useful result to keep in mind is the fact, that  for given $x_a$ and $x_b$ endpoints,  the contributions from other than the classical path interfere destructively and vanish.  Thus, only the classical action $S_{cl}(x_b,x_a)$ contributes, leading to the major simplification for the kernel, \emph{i.e.},
 \beq\label{eq11}
 K(x_b,x_a) = F(t_{b} - t_{a}) \exp{\left[\frac{i}{\hbar}S_{cl}(x_b,x_a)\right]}.
  \eeq
As long as the action involves path variables only up to the second order, therefore, the exact propagators in the form above can be factored out from the path integral, leaving at most to calculate a prefactor of the form $F(t_b-t_a)$. 

To illustrate how one arrives at the result above, let us consider a general quadratic Lagrangian in $x$ and $\dot{x}$ of the form
\beq\label{lag1}
L(x,\dot{x},t) = a \dot{x}^{2} + b \dot{x} x + c x^{2} + d \dot{x} + e x + f,
\eeq
where $\{a,b,c,d,e,f\}$ are (arbitrary, but well-behaved) time-dependent coefficients.
The action corresponding to this Lagrangian is then the integral of Eq. \eqref{lag1} with respect to time between two fixed end points $t_a$ and $t_b$, as given by Eq. \eqref{eq:Action}.

Let us now assume, that $x_{cl}(t)$ is the classical path between the specified end points, i.e., the path for which $\delta S=0$ holds. We can then represent $x(t)$ in terms of deviations from the classical path $x_{cl}(t)$ by introducing the function $y(t)$ as
\beq\label{change}
x(t) = x_{cl}(t)+ y(t).
\eeq
This substitution means, that instead of defining a points on the path by its distance from and arbitrary coordinate axis, we measure instead the deviation $y(t)$ from the classical path. Moreover, at each time $t\in[t_a,t_b]$, the variables $x$ and $y$ differ only by the constant $x_{cl}(t)$, and therefore $dx_{i}$ = $dy_{i}$ for each point $t_{i}$. In general, then, it follows that $Dx(t) =  Dy(t)$. Notice, that as a consequence of Eq. \eqref{change}, $y(t_a)=0=y(t_b)$, as at the endpoints the path $x(t)$ coincides with the classical path $x_{cl}(t)$.

If we then use the change of variables defined in Eq. \eqref{change}, the Lagrangian \eqref{lag1} can be written as the sum of three terms as follows
\beq
L(x,\dot{x},t)=L_{cl}+L_y+L_{mix},
\eeq
where $L_{cl}$ ($L_y$) is just Eq. \eqref{lag1} with $\{x,\dot{x}\}\rightarrow\{x_{cl},\dot{x}_{cl}\}$ ($\{x,\dot{x}\}\rightarrow\{y,\dot{y}\}$), and
\beq
L_{mix}=\left(2a\dot{x}_{cl}+bx_{cl}\right)\dot{y}+\left(b\dot{x}_{cl}+2c x_{cl}\right)y.
\eeq
Similarly, the action can be then written as the sum of three terms, namely the classical action $S_{cl}(b,a)$, the action relative to the deviation $y(t)$, i.e., $S_y(b,a)$, and the mixed action $S_{mix}(b,a)$. Because of the fact that $y(t_a)=0=y(t_b)$, however, all the terms which contain linear terms in $y$ result in a vanishing integral. Thus, only the second-order terms in $y$ give rise to a nonzero contribution to the total action, which can now be written as
\beq
S(b,a)= S_{cl}(b,a) + \int_{t_{a}}^{t_{b}} \left( a\, \dot{y}^{2} + b\, \dot{y} y + c\, y^{2} \right) dt.
\eeq
Notice how $S_{cl}(b,a)$ does not depend on the deviation $y(t)$, and therefore the corresponding exponential can be treated as a constant, with respect to the path integration $Dy(t)$. The Kernel can then be written in the following form
\barr
K(b,a) &= &\int_{0}^{0}\,Dy(t)\,\exp{\left[ \left\{ \frac{i}{\hbar} \int_{t_{a}}^{t_{b}} (a\, \dot{y}^{2} + b\, \dot{y} y + c\, y^{2} ) dt \right\} \,\right]} \nonumber\\
 & \times & \exp{\left[ \frac{i}{\hbar} S_{cl}(b,a)\right]},
\earr
where the notation $\int_0^0\,Dy(t)$ is reminiscent of the fact that all the paths $y(t)$ obey the boundary condition $y(t_a)=0=y(t_b)$. The path integral above, then, can be written as a function of the time interval $(t_b-t_a)$ solely, i.e.,
\beq
F(t_b-t_a)=\int_{0}^{0}\,Dy(t)\,\exp{\left[ \left\{ \frac{i}{\hbar} \int_{t_{a}}^{t_{b}} (a\, \dot{y}^{2} + b\, \dot{y} y + c\, y^{2} ) dt \right\} \,\right]} .
\eeq
This, ultimately, allows us to write the kernel in the following simplified form
\beq
K(b,a) = F(t_{b} - t_{a}) \exp{\left[\frac{i}{\hbar}S_{cl}(x_b,x_a)\right]},
\eeq
which is equivalent to that of Eq. \eqref{eq11}.
\subsection{Path Integrals for a Free Quantum Particle}\label{freeParticle}
\begin{figure}[!t]
\begin{center}
\includegraphics[width=0.48\textwidth]{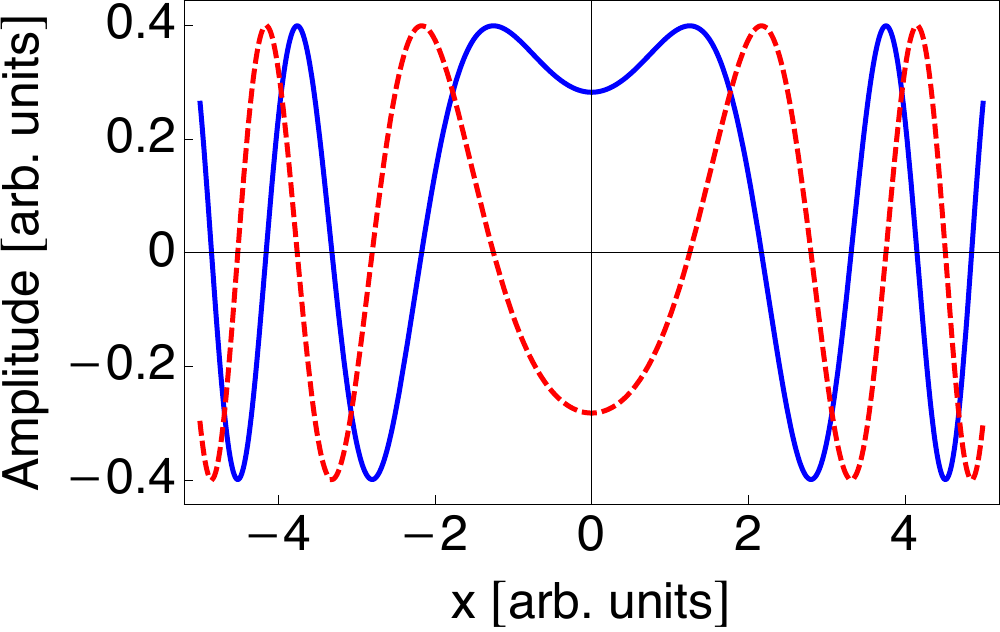}
\caption{Real (blue, solid line) and imaginary (red, dashed line) parts of the kernel $K(x_b,x_a)$ of a free quantum particle, as given by Eq. \eqref{eq:Path_integral_Free_Particle}, as a function of $x=x_b-x_a$, for a fixed time interval $t_b-t_a=T$. For this plot, we assumed $m=1=\hbar$, which corresponds to measure time in square meters, rather than seconds. Moreover, $T=1$ has also been used.}
\label{figKernelFree}
\end{center}
\end{figure}

The first example concerns the simplest quantum system, \emph{i.e.}, a quantum particle of mass $m$, freely propagating without experiencing any interaction. Following the assumptions made in the previous section, we discuss the case of a one dimensional free particle. The generalisation to an arbitrary number of dimensions can be readily done, since the dynamics of a free quantum particle in $D$ dimensions can be seen as the product of the independent evolution of $D$ one dimensional particles \cite{dirac2}.

The Lagrangian of a free particle of mass $m$ is given by
\beq
\label{eq:free_particle_lagrangian}
	L(x,\dot{x}) = \frac{m\, \dot{x}^2}{2},
\eeq
and the equation of motion deriving from the Euler-Lagrange equations \eqref{eq:Euler} is simply $m\,\ddot{x}=0$.The correspondent action $S = \int_{t_a}^{t_b} L(\dot{x},x,t) d t$ can be readily calculated explicitly by means of part integration, and has the following form
\beq\label{eq:Action_Free_Particle}
	S =\frac{ m(x_b - x_a)^2}{2(t_b - t_a)},
\eeq
where $x_{a,b}=x(t_{a,b})$. To calculate the path integral for a free particle, we need to consider all the possible paths the particle takes from the initial state $(t_a,x_a)$ to the final state $(t_b,x_b)$. To do that, we simply divide the time interval $T=t_b-t_a$ into $N$ smaller intervals of length $\varepsilon=t_{i+1}-t_i$ (such that $T=N\varepsilon$), calculate the action $S_i$ corresponding to the particle evolution within each single interval, such that
\beq
S=\sum_{i=1}^NS_i=\sum_{i=1}^N\frac{m}{2\varepsilon}(x_i-x_{i-1})^2,
\eeq
with $x_0\equiv x_a$ and $x_N\equiv x_b$. We then substitute this result into Eq. \eqref{eq:pathIntegral} and evaluate the path integral over the set of $N$ distinct trajectories, i.e,
\beq\label{equation12}
\int_a^b\,Dx(t)\hspace{0.5cm}\rightarrow\hspace{0.5cm}\frac{1}{A(\varepsilon)^N}\int\,dx_1\int\,dx_2\,\cdots\,\int\,dx_{N-1},
\eeq
where $A(\varepsilon)=\sqrt{(2\pi i \hbar \varepsilon)/m}$ is a factor included to ensure the integral to converge.  This factor, however, is not merely a normalisation factor, since it is complex, and therefore it contributes to the overall phase of the path integral. This is discussed in great detail in Ref. \citenum{hibbs}. Finally, we take the limit $N \rightarrow\infty$, to arrive at the following expression for the propagator of a free particle
\beq\label{eq:Path_integral_free_particle_full}
K(x_2,x_1) =\lim_{N \rightarrow\infty} \frac{1}{A(\varepsilon)^N}\int\,dx_1 \cdots \int\,dx_N \exp{\left(\frac{i}{\hbar}\sum_{i=1}^NS_i\right)}
\eeq
The integrals appearing above are Gaussian in the variables $x_i$, and can be then readily calculated one after another. To see how, let us first calculate explicitly the integral with respect to $x_1$. Once we have this result, we can perform the other integrations in cascade in the same manner. 

First, notice that the relevant term in the action, depending explicitly on $x_1$ (\emph{i.e.}, those obtained by setting $i=1$ and $i=2$ in the expression above) give rise to the following term
\barr
\frac{i}{\hbar}S_1&=&\frac{im}{2\hbar\varepsilon}\left[(x_2-x_1)^2+(x_1-x_0)^2\right]=\frac{im}{2\hbar\varepsilon}(x_2^2+x_0^2)\nonumber\\
&+&\frac{im}{\hbar\varepsilon}\left[x_1^2-(x_2+x_0)x_1\right],
\earr
and that, in particular, the first term does not depend on the integration variable $x_1$. Integrating the above quantity with respect to $x_1$ then gives
\barr\label{eq:free_particle_path_integral_calculation_first_factor}
&&\frac{m}{2\pi i\hbar\varepsilon}\int\,dx_1\,\exp{\left(\frac{i}{\hbar}S_1\right)}=\exp{\left[\frac{im}{2\hbar\varepsilon}(x_2^2+x_0^2)\right]}\nonumber\\
&\times&\frac{m}{2\pi i\hbar\varepsilon}\int\,dx_1\,\exp{\left\{\frac{im}{\hbar\varepsilon}\left[x_1^2-(x_2+x_0)x_1\right]\right\}}\nonumber\\
&=&\sqrt{\frac{m}{2\pi i\hbar (2\varepsilon)}}\exp{\left[\frac{im}{2\hbar(2\varepsilon)}(x_2-x_0)^2\right]},
\earr
where to pass from the second to the third line we have employed the change of variables $X=ix_1$ and used Eq. \eqref{gaussianIntegral}.

Next, we take into account the terms in the action depending explicitly on $x_2$, \emph{i.e.}, $S_2=S_1+(x_3-x_2)^2$, and we integrate with respect to $x_2$. We can do so, by simply taking the result of the integral of $S_1$ given in Eq. \eqref{eq:free_particle_path_integral_calculation_first_factor} and multiply it by
\beq\label{multiTerm}
\sqrt{\frac{m}{2 \pi i \hbar \varepsilon}}\exp \left[ \frac{i m}{2 \hbar \varepsilon} (x_3 - x_2)^2\right],
\eeq
and integrate again, this time over $x_2$. The result is similar to that of Eq. \eqref{eq:free_particle_path_integral_calculation_first_factor}, except that $ (x_2 - x_0)^2 $ becomes $ (x_3 - x_0)^2 $, and $2 \varepsilon\rightarrow 3 \varepsilon$. It is now clear, that we can solve the set of $N$ integrals in Eq. \eqref{eq:Path_integral_free_particle_full} by recursively applying terms of the form \eqref{multiTerm} and then performing Gaussian integration with respect to the variable $x_i$. After $N-1$ steps, we are left with the following result
\beq
\sqrt{\frac{m}{2 \pi i \hbar (N \varepsilon)}} \exp \left[ \frac{i m}{2 \hbar (N \varepsilon)} (x_N - x_0)^2\right].
\eeq
If we now notice, that $N \epsilon = T= t_b - t_a$, it is easy to see that the limit operation in Eq. \eqref{eq:Path_integral_free_particle_full} can be readily performed, leading us to the final form for the path integral of a free particle, \emph{i.e.},
\beq 
\label{eq:Path_integral_Free_Particle}
	K(x_b,x_a) = \sqrt{\frac{m}{2 \pi i \hbar (t_b - t_a)}} \exp{\left[\frac{i  m (x_b - x_a)^2}{2 \hbar (t_b - t_a)}\right]}.
\eeq
The functional form of the real and imaginary parts of the kernel above for a constant time interval $t_b-t_a=T$ are shown in Fig. \ref{figKernelFree}. It is worth to comment this result a bit. From it, in fact, we see how the quantum dynamics of a free particle (but, more in general, of an arbitrary quantum system) couples with its classical dynamics described by the action \eqref{eq:Action_Free_Particle}. However, this result also allows to shed light on the essential difference between quantum and classical dynamics. While the classical principle of least action localised the propagation of a particle on a specific trajectory $x(t)$ (\emph{i.e.}, the path of minimal action), the propagator $K(x(t),x_a)$ describing the propagation of a particle from the initial state $x_a$ along the classical trajectory $x(t)$, yields instead a complex wave function $\psi(x,t)$ delocalising in time during propagation (obviously within the constraints of the Heisenberg principle). 

\subsection{Refraction of Photons at an Interface}\label{SnellSection}
\begin{figure}[!t]
\begin{center}
\includegraphics[width=0.5\textwidth]{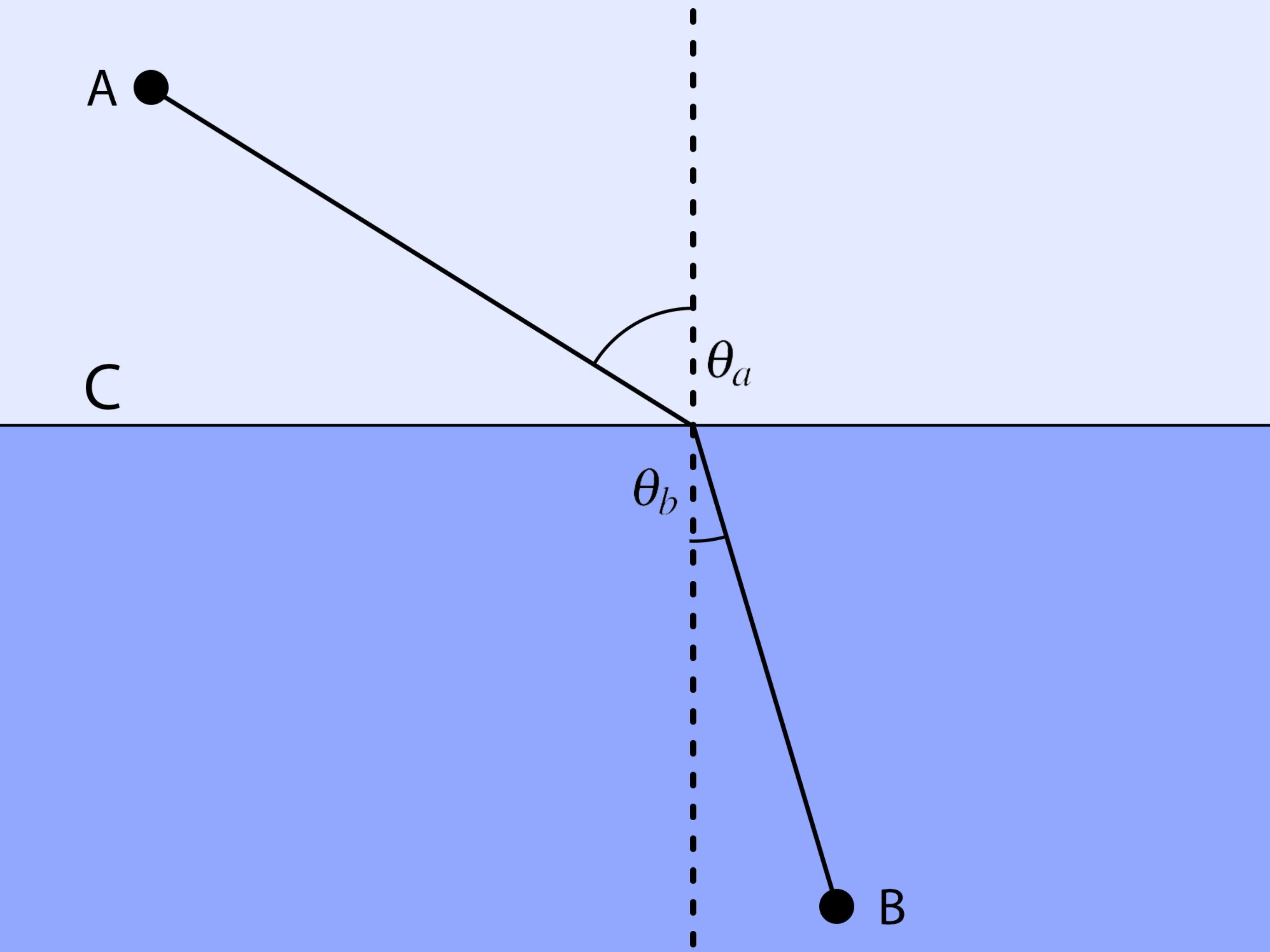}
\caption{Schematic representation of a ray of light being refracted from a planar interface. A light source located at $A$ emits a photon, which gets detected at point $B$ by a suitable detector. Between $A$ and $B$, a planar interface $C$ is placed, separating the two different media in which the photon propagates. In the language of path integrals, refraction can be interpreted as the necessary change in velocity of the photon along the path where the action has the minimal value, with respect to the contact point $x_c$ on the planar interface $C$.}
\label{Figrefraction}
\end{center}
\end{figure}
\begin{figure*}[!t]
\begin{center}
\includegraphics[width=0.8\textwidth]{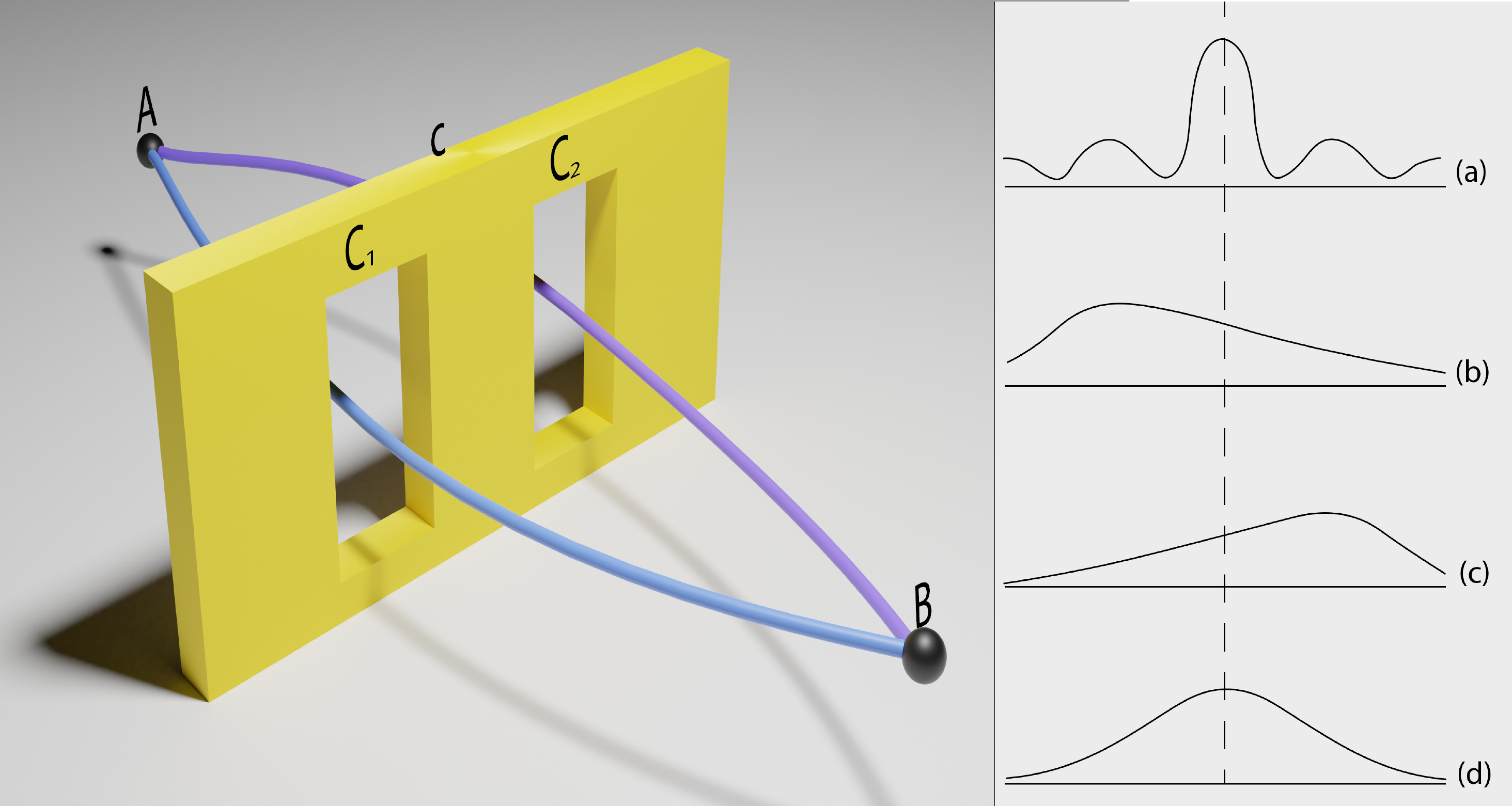}
\caption{Left panel: pictorial representation of the double-slit experiment. A quantum particle is emitted from a source located in $A$ and reaches a detector (or a distribution thereof) located at point $B$ by passing through a screen $C$, on which two slits, $C_{1,2}$ have been carved. Right panel: sketch of the various possible probability distributions that could be observed at the detector plane $B$. (a) Interference pattern $P$ revealing the wave nature of the quantum particle. (b) Probability distribution $P_2$ obtained by closing the slit $C_1$. (c) Probability distribution $P_1$ obtained by closing the slit $C_2$. (d) Estimation of the total probability distribution $P_1+P_2$, obtained by simply summing the distributions obtained in the cases (b) and (c), as would be valid for a classical particle, where no interference occurs.} 
\label{FigDoubleSlit}
\end{center}
\end{figure*}
With the above notations, consider now the space divided in two parts by a planar interface $C$, and let $A$ and $B$ be two points located at opposite sides of the interface, as it is shown in Fig. \ref{Figrefraction}.  Then, we assume constant but different potentials at opposite sides of the interface.  At both sides we expect the path of the photons to be those of a free particle.  With this assumption, we can project the path connecting the points $A$ and $B$ onto a one dimensional subspace, for simplicity. Notice, however, that despite we will perform the calculations in this one dimensional subspace, we still need to think in terms of three dimensional space when considering deflection of light rays from the interface $C$.

Without loss of generality, we can assume that total energy is conserved in both separate sides of the interface, and while passing through it. However, since the two sides might have different values of the potentials (\emph{i.e.}, different refractive indices), the particle passing through the interface $C$ needs to change its velocity from $\dot{x}_a$ to $\dot{x}_b$, to account for this variation in potential energy. The change of velocity, moreover, occurs at some position $x_c$ on the interface.

As shown in Fig. \ref{Figrefraction}, in case the straight line from $A$ to $B$ is not perpendicular to the interface, the observed path becomes deflected at $x_c$.

Now, we can write the action for the path as
\barr\label{eq:refraction_action}
S_{cl}(x_b,x_a)&=&S_{cl}(x_b,x_c)+S_{cl}(x_c,x_a)\nonumber\\
&=&\frac{m}{2}\left[\dot{x}_b\left(x_b-x_c\right)+\dot{x}_a\left(x_c-x_a\right)\right],
\earr
where the coordinate $x_c$ is a free parameter to optimise following the principle of least action. For the case of a photon refracting from interface $C$, therefore, the "minimum optical path length" from $A$ to $B$ is found by optimising the coordinate $x_c$. Following Eq. \eqref{eq11}, the kernel for the refraction of a quantum particle from a planar interface $C$ is given, up to an inessential constant $F\equiv F(t_b-t_a)$, by
\beq
K(x_b,x_a)=F\exp{\left\{\frac{i\,m}{2\hbar}\left[\dot{x}_b\left(x_b-x_c\right)+\dot{x}_a\left(x_c-x_a\right)\right]\right\}}.
\eeq

We leave as an exercise for the reader to figure out the explicit expression of the normalisation constant $F\equiv F(t_b-t_a)$.

In the equations above, the two different constant velocities $\dot{x}_a$ to $\dot{x}_b$ of the photon follow from two different refractive indices.  In optics, the optimisation of the optical path length is called Fermat's principle. If we consider the problem from a geometrical optics perspective, in fact,  it is easy to see how the optimisation of $x_c$ leads directly to the celebrated Snell's law of refraction \cite{luneburg}, \emph{i.e.},
\beq
n_a\sin\theta_a=n_b\sin\theta_b,
\eeq
where $n_{a,b}=1/\dot{x}_{a,b}$ are the refractive indices of the two media separated by interface $C$ (that can be expressed as the inverse velocity of the particle in each side of $C$), and $\theta_{a,b}$ the angle the optical rays emerging from $A$ and $B$, resepctively, make with it. The angle difference between the two sides of the interface $C$ is reminiscent of the different wavelength that a photon with velocity $\dot{x}_a$ and one with velocity $\dot{x}_b$ experiences.

In this simple example, the classical and quantum approaches give identical explanations for observations.  So, we see that the "quantum corrections", though not absent may give not only small, but even vanishing contribution.
\subsection{Diffraction from a Double Slit}\label{DoubleSlitSection}
This is the well known ``classic experiment" for demonstrating the wave nature of light, or, if conducted with quantum particles like electrons, to expose the wave nature of particle dynamics \cite{feynmanQED}. The results of this section can be then thought as valid for both a photon, or a quantum particle. Experimentally, to prove the wave nature of light one would need to perform this experiment with monochromatic light, while mono-energetic particles are needed to unravel the wave nature of particle dynamics. From a path integral perspective, the interference pattern typical of such experiments naturally arises when the possible paths the particle can take to traverse the double slit are explicitly taken into account.

Our first task is then to construct the kernel for this problem. To this aim, let us consider the situation depicted in Fig. \ref{FigDoubleSlit}, where a particle is emitted from a suitable source located at point $A$ and propagates to a detector (or a collection thereof) placed at $B$, through a screen $C$ with two slits $C_1$ and $C_2$ carved in it. Without loss of generality, we can assume the evolution of the particle is that of a free particle, and that the only potential it encounters is represented by the double-slit structure (which, as a matter of fact, acts as a transfer function for the particle). 

To reach the detector at $B$, the particle takes a time $\tau$ to travel from $A$ to the screen $C$ (and, in particular, to one of the slits), pass through the screen, and then arrive at $B$ after a time $T-\tau$, where $T=t_b-t_a$ is the total time the particle takes to go from $A$ to $B$. However, since we don't know exactly the time at which the particle arrives on the screen, we need to integrate over all possible times. In other terms, since we don't know the exact path the particle will take to reach $C$ from $A$, we need to integrate over all possible paths it might undertake. With this in mind, and by remembering that $a=(x_a,t_a)$, $b=(x_b,t_b)$, and $c_{1,2}=(x_{c_{1,2}},t_{c_{1,2}})$ represent the position and time at points $A$, $B$, and $C_{1,2}$, respectively, the kernel for the propagation of a particle through a two-slit screen is given by
\barr
K(b,a)&=&\int\,d\tau\,\Big[K(c_1,b;\,\tau)K(a,c_1;\,T-\tau)\nonumber\\
&+&K(c_2,b;\,\tau)K(a,c_2;\,T-\tau)\Big],
\earr
where the first term accounts for the particle passing through the slit $C_1$, while the second one for it passing through the slit $C_2$. From the expression above, it is clear that interference must occur, since the probability to detect the particle at point $B$ is then proportional to $|K(b,a)|^2$.

In fact, we can arrive at the same conclusion by considering the case above in terms of the probability amplitudes and observed probabilities in Sect. \ref{sectCLvsQU}. In particular, let us have a look at the probability distribution of the particle observed on the screen at $B$ for different cases, namely the probability distribution $P$ when  interference pattern is observed (both slits are open) [panel (a)], the probability distribution $P_2$ observed when the slit $C_1$ has been closed [panel (b)], the probability $P_1$ observed when the slit $C_2$ has been closed [panel (c)], and finally the probability distribution $P_1+P_2$ obtained by summing the results of observations in (b) and (c). 

As it can be clearly seen, the probabilities do not sum up, as $P \not= P_1 + P_2$. If, on the other hand, we first sum the probability amplitudes, \emph{i.e.},  $\psi  =  \psi_1 + \psi_2$, and then calculate the probability distribution as $P=|\psi|^2$ we get

\beq
P  =  | \psi |^2  =  | \psi_1 + \psi_2 |^2  =  P_1 + P_2 + 2\operatorname{Re}\{ \psi_1 \psi_2^*\}, 
\eeq

where the third term is responsible for adding interference on top of the sum of probabilities $P_1+P_2$ in Fig. \ref{FigDoubleSlit} (d), thus leading to the correct result of Fig. \ref{FigDoubleSlit} (a). The reader familiar with wave theory would immediately recognise that the probability distributions shown in Fig. \ref{FigDoubleSlit} (a) and (d) occur as well for classical waves. Specifically, the distribution (a) occurs for the interference of correlated (i.e., first-order coherent) classical waves, while distribution (d) results from incoherent classical wave mixing. This is yet another indicator of the wave nature of quantum particles \cite{feynmanLectures}. 
\subsection{Path Integral for the Harmonic Oscillator}\label{harmonicOscillator}
\begin{figure}[!t]
\begin{center}
\includegraphics[width=0.48\textwidth]{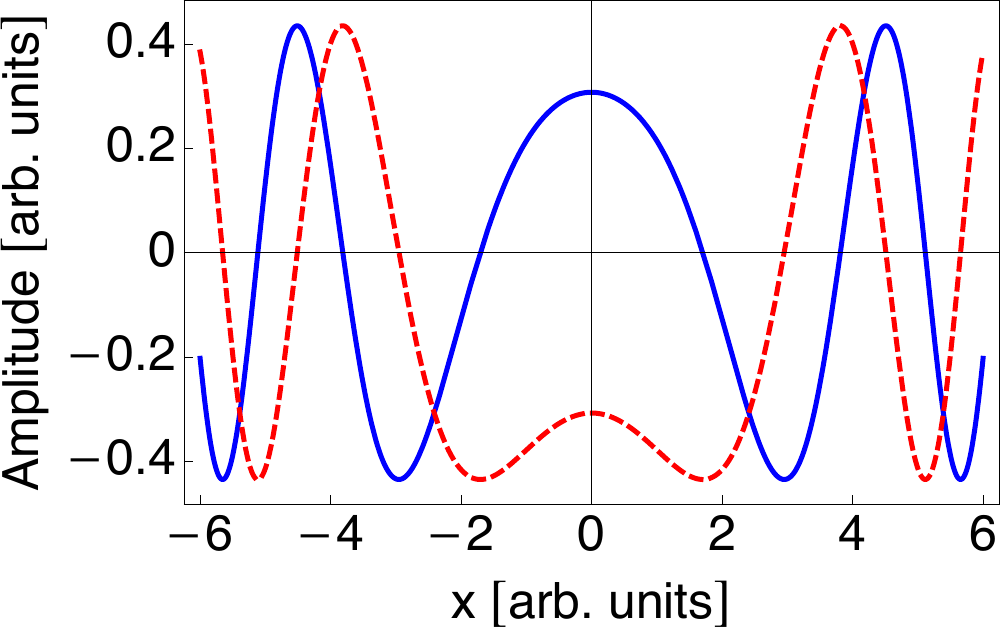}
\caption{Real (blue, solid line) and imaginary (red, dashed line) parts of the kernel $K(x,0)$ of a harmonic oscillator, as defined by Eq. \eqref{pathIntegralHO}, assuming that the initial condition for the oscillator is $x_a=0$, and for a given time interval $T=1$. Analogously to the convention adopted in Fig. \ref{figKernelFree}, $m=1=\hbar$ have been assumed to plot this figure as well.}
\label{figkernelHO}
\end{center}
\end{figure}
As our next example, we consider a simple harmonic oscillator, described by the following Lagrangian
\beq\label{eq:h_o_lagrangian}
	L(x,\dot{x}) = \frac{m \dot{x}^2}{2} - \frac{ m  \omega^2 x^2 }{2},
\eeq
where $m$ is the mass of the oscillator, and $\omega$ its characteristic resonance frequency. Using the Euler---Lagrange  Eq. \eqref{eq:Euler}, the equation of motion reads $\ddot{x}+\omega^2x=0$, and the action $S = \int_{t_a}^{t_b} L(\dot{x},x,t) d t$ is explicitly given by \cite{hibbs}
\beq\label{eq:h_o_Action_}
	S = \frac{m}{2T \text{sinc}(\omega T)} \left[(x_{a}^{2} - x_{b}^{2})\, \cos( \omega T) - 2 x_a x_b)\right],
\eeq
where $T=t_b-t_a$, and $\text{sinc}(x)=\sin(x)/x$.

To calculate the path integral for the harmonic oscillator, we take a slightly different approach, than the one taken above, which allows us to perform calculations in an easier and more intuitive manner. Let us then assume that $x_{cl}(t)$ and $\dot{x}_{cl}(t)$ represent the classical path of the oscillator, and its velocity, respectively. We can then express any other possible path taken by the oscillator, as a deviation from the classical path, \emph{i.e.}, $x(t)=x_{cl}(t)+y(t)$ and $\dot{x}(t)=\dot{x}_{cl}(t)+\dot{y}(t)$, and choose the appropriate boundary conditions on $y(t)$ and $\dot{y}(t)$, \emph{i.e.}, $y(t_a)=0=y(t_b)$ and $\dot{y}(t_a)=0=\dot{y}(t_b)$ as required by the principle of least action.

If we apply this change of variables to the Lagrangian above, we get three different terms, namely
\beq
L(x_{cl},\dot{x}_{cl})+L(y,\dot{y})+m(\dot{x}_{cl}\dot{y}+\omega x_{cl}y),
\eeq
where the first term has the form given by Eq. \eqref{eq:h_o_lagrangian} and it represents the Lagrangian for the classical trajectory, the second term is the Lagrangian \eqref{eq:h_o_lagrangian} for the deviation, and the third term amounts to a vanishing term (for the Euler---Lagrange equations) and can therefore be neglected \cite{hibbs}. Using this result, we can then factor the propagator $K(x_2,x_1)$ for the harmonic oscillators into two terms as follows

\beq
\label{eq:h_o_kernel_only_classical_action}
K(x_b,x_a) = F(T)\exp{\left(\frac{i}{\hbar}S_{cl}\right)},
\eeq

where $S_{cl}=\int_{t_a}^{t_b}\,dt\,L(x_{cl},\dot{x}_{cl})$, and the explicit expression of the function $F(T)$, which depends upon the time interval solely, is
\beq
\label{eq:h_o_factor_expression}
F(T) = \int\,Dy(t)\, \exp{ \left[ \frac{i}{\hbar} \int_{0}^{T}\,dt\, \frac{m}{2} (\dot{y}^2 - \omega^2 y^2)\right]}.
\eeq
Because of our choice of boundary conditions for the deviations $y(t)$ and $\dot{y}(t)$, namely that they must both be zero at the endpoints, we can significantly simplify the calculation of the path integral above if we allow the various paths $y(t)$ to be represented as a Fourier series as
\beq
\label{eq:h_o_y_fourier_series}
y(t) = \sum_{n = 1}^{\infty} a_n \sin\left( \frac{ n \pi t}{T}\right).
\eeq
The requirement that  $y(t_a)=0=y(t_b)$ and $\dot{y}(t_a)=0=\dot{y}(t_b)$, in fact, corresponds to say that both $y(t)$ and $\dot{y}(t)$ are $T$-periodic functions, that can then be represented in a Fourier series.

This representation of $y$ gives possibility to specify a path through the coefficients $a_n$ instead of values of $y$ at any particular time $t$. This can be seen as a linear transformation of coordinates, whose Jacobian $J$ is a dimensionless constant independent of $\omega$, $m$ or $\hbar$. Moreover, we don't really need to evaluate $J$ explicitly, since we can always recover the correct normalisation factor at the end of our calculation, by requiring that
\beq\label{freePartLimit}
\lim_{\omega\rightarrow 0}F(T)=\sqrt{\frac{m}{2\pi i\hbar T}},
\eeq
\emph{i.e.}, that in the limit of $\omega\rightarrow 0$, where the Lagrangian of the harmonic oscillator reduces to that of a free particle, we find the appropriate normalisation coefficient for a free particle. For this reason, we omit $J$ from the following calculations, and we restore the correct normalisation factor only at the very end of them. 

Before substituting Eq. \eqref{eq:h_o_y_fourier_series} into Eq. \eqref{eq:h_o_factor_expression}, a couple more assumptions are needed, in order to easily compute the path integral. First, we truncate the Fourier series to a finite number $N$, so that
\beq
\int\,Dy(t)\hspace{0.5cm}\rightarrow\hspace{0.5cm}\frac{1}{A(\varepsilon)^N}\int\,da_1\,\int\,da_2\,\cdots\,\int\,da_N,
\eeq
and we can calculate it recursively as we did for the case of a free particle. We will then take the limit $N\rightarrow\infty$ at the end of the calculation.

Putting everything together, we obtain the following expression for the term $F(T)$, after performing the trigonometric integrals \cite{byronFuller}, as
\barr
\label{eq:h_o_factor_inserted_fourier}
F(T) &=& \frac{1}{A(\varepsilon)^N} \int\,da_1\,\cdots\,\int\,da_N \exp \Bigg\{ \frac{i m}{2 \hbar} \frac{T}{2} \\
&&\sum_{n = 1}^{N} \left[ \left( \frac{\pi n}{T} \right) ^{2} - \omega^2 \right] a_{n}^{2} \Bigg\}.
\earr
Notice, once more, that the integrals above are all Gaussian in the integration variables $a_n$. As we did for the case of the free particle, therefore, we can perform them individually, and then obtain the final result recursively. The result of a single integration over $a_n$ is then given by
\beq
\int\,\frac{da_n}{A}\,\exp{\left[\frac{imT}{4\hbar}f_n(\omega,T)a_n^2\right]}=\sqrt{\frac{2}{T f_n(\omega,T)}},
\eeq
where 
\beq\label{effeN}
f_n(\omega,T)=\frac{n^2\pi^2}{T^2}-\omega^2.
\eeq

Since there are no linear terms in $a_n$ in any of the integrals above, the final result of the path integration will be proportional to simply the product of $N$ independent terms $f_n(\omega,T)$, one for each value of $n$. This allows us to write
\beq
F(T)\propto\prod_{n=1}^N\frac{1}{\sqrt{f_n(\omega,T)}}.
\eeq
We now need to take the limit of the expression above for $N\rightarrow\infty$. To do that, let us first rewrite the product above in the following way, using Eq. \eqref{effeN} 
\barr
\prod_{n=1}^{N}\frac{1}{\sqrt{f_n(\omega,T)}}&=&\prod_{n=1}^{N} \left( \frac{n^2 \pi^2}{T^2} - \omega^2 \right)^{-1/2} = \prod_{n=1}^{N}\left(\frac{n^2\pi^2}{T^2}\right)^{-1/2}\nonumber\\
&\times& \prod_{n=1}^{N} \left( 1 - \frac{\omega^2 T^2}{ n^2 \pi^2} \right)^{-1/2}.
\earr
The first product does not depend on $\omega$ and, together with the Jacobian and the terms $\sqrt{2/\varepsilon T}$ deriving from the various integrations, can be collected into an overall normalisation factor $C$. The second product, on the other hand, admits the following limit
\beq
 \lim_{N\rightarrow\infty}\prod_{n=1}^{N} \left( 1 - \frac{w^2 T^2}{ n^2 \pi^2} \right)^{-1/2}=\sqrt{\frac{1}{\text{sinc}(\omega T)}}.
\eeq
Putting everything together and evaluating $C=\sqrt{m/2\pi i\hbar T}$ from the free-particle-limit \eqref{freePartLimit} we get the final expression of the term $F(T)$ as
\beq
F(T) =\sqrt{\frac{m}{2\pi i\hbar T\text{sinc}(\omega T)}},
\eeq
and, after substituting this result into Eq. \eqref{eq:h_o_kernel_only_classical_action}, we obtain the final form of the path integral for the harmonic oscillator to be
\beq\label{pathIntegralHO}
 K(x_b,x_a) = \sqrt{\frac{m}{2\pi i\hbar T\text{sinc}(\omega T)}}\exp{\left(\frac{i}{\hbar}S_{cl}\right)},
\eeq
where $S_{cl}$ is given by Eq. \eqref{eq:h_o_Action_}. The real and imaginary pary of the kernel $K(b,a)$ for the harmonic oscillator are shown in Fig. \ref{figkernelHO}.

\section{An Example from Classical Optics: Path Integral Description of Light Dynamics in an Inhomogeneous Medium}\label{sect3}
We now take a look at how path integrals can be used to solve problems outside quantum mechanics, and apply this formalism to describe the propagation of the electromagnetic field inside a weakly inhomogeneous medium, using the case of gradient-index (GRIN) media as explicit reference. This problem has been solved by C. G\~{o}mez-Reino and J. Li\~{n}ares in 1987 \cite{gomez-reino1987}. In their work, G\~{o}mez-Reino and Li\~{n}ares first represent the electromagnetic field in a GRIN medium as a superposition of optical rays, and use this picture to calculate the propagator as a path integral over the rays' trajectories. They then provide an explicit expression for it, parametrised in the so-called paraxial and field rays of an arbitrary (paraxial) optical system \cite{luneburg}. 

Here, we take a different approach, with which we want to show how the free propagation of light in a medium can be seen as, essentially, the evolution of a massive quantum particle in a harmonic oscillator potential with a suitably defined frequency, which, in general, can be $z$-dependent. We will identify such massive particle with a photon propagating inside the medium, and all the possible trajectories the particle can take as the possible optical rays linking the initial ($z=0$) and final ($z=z$) propagation plane in the medium. We will then calculate the diffraction kernel by means of path integrals, essentially following the results of Sect. \ref{harmonicOscillator} and show how our calculations naturally suggest a representation of the diffraction kernel in terms of Hermite---Gaussian functions. 

Without loss of generality, and for the sake of simplicity of exposition, we consider light propagating in a $1+1$-dimensional GRIN medium, characterised by the following refractive index profile
\beq\label{refractiveIndex}
n^2(x,z)=n_{0}^{2}[ 1- g^2(z) x^2 ].
\eeq
where $n_0=n(0,z)$ is the background index and $g(z)$ is a smooth-enough function that describes the evolution of the refractive index along the $z$ axis. 
\subsection{Diffraction Kernel as a Path Integral}
Let us first assume paraxial propagation of light in a medium described by the refractive index given by Eq. \eqref{refractiveIndex}. In general, if we know the field distribution at an initial plane $z=0$ to be $E(x_a,y_a)$, we can calculate the field distribution at a plane $z>0$ by means of the diffraction integral \cite{jackson}
\beq
\label{eq:integral_equation_grin}
E(x_b,z) =  \int\, dx_a\, K(x_b,x_a,z) E(x_a,0),
\eeq
where $K(x_b,x_a,z)$ is the diffraction kernel, \emph{i.e.}, the Green's function of the paraxial equation
\beq\label{paraxialEq}
2\,i\,k\,n_0\frac{\partial K}{\partial z}=-\left[\frac{\partial^2}{\partial x^2}+2k^2\,n_0\,n(x,z)\right]K
\eeq
with the boundary condition that $K(x,x_a,z)\rightarrow\delta(x-x_a)$ for $z\rightarrow 0$. If we imagine the electromagnetic field propagating in the medium described by $n(x,z)$ as a bundle of optical rays, then the diffraction kernel can be interpreted as a path integral over all the possible trajectories of the optical rays contained in the field as
\barr\label{eq:grin_path_integral}
K(x_b,x_a,z) &=&\int_{x_a}^{x_b}\,Dx(z)\, \exp{\left[\frac{i}{\lambdabar} S(x_b,x_a,z)\right]}
\earr
where $\lambdabar = \lambda / 2 \pi=1/k$, and  the action functional $S$ for an optical ray propagating in the medium described by Eq. \eqref{refractiveIndex} is defined as
\beq
S = \int_{0}^{z} L(x,\dot{x},z) dz.
\eeq
The Lagrangian $L$ for an optical ray propagating in a medium with refractive index $n(x,z)$ can be written as follows
\beq
\label{eq:grin_optical_lagrangian}
L(x,\dot{x},z) = n(x,z)\sqrt{1 + \dot{x}^2}.
\eeq
To justify our starting assumption, \emph{i.e.}, that the electromagnetic field in the medium can be seen as a collection of rays, we can use different arguments. One possibility would be to represent the field in its plane wave components and consider each plane wave as an optical ray \cite{jackson}. Another, more inspiring, possibility is to notice that the paraxial equation \eqref{paraxialEq} is formally equivalent to the Schr\"odinger equation for a quantum particle of mass $n_0$ in a potential $n(x,z)$, where the propagation direction $z$ plays the role of time, and $k$ plays the role of $\hbar$. Thanks to this formal analogy, we can identify a single optical ray as a (massive) photon propagating in the medium, and then easily understand as the diffraction kernel can be seen as a path integral over all the possible trajectories that such quantum particle can take when evolving from the initial state to the final one.

In the form above, the Lagrangian is of little use, since it provides no analytical solution for the trajectory of the optical ray and, by extension, doesn't really allow for an easy handling of the path integral in Eq. \eqref{eq:grin_path_integral}. To circumvent this problem we can however assume that the medium is weakly inhomogeneous, so that $\Delta n=n(x,z)-n_0\ll n$ holds, and we only consider rays propagating in a small region around the $z$-axis, which corresponds to assuming $\dot{x}\ll 1$ (this is equivalent to assuming that the paraxial approximation holds). With these assumptions, we can Taylor expand the square root in Eq. \eqref{eq:grin_optical_lagrangian}  and the refractive index profile obtaining, to the leading order in $\dot{x}$ and $x$
\beq
\label{eq:grin_opt_lag_simpl}
L(x,\dot{x},z) \simeq n_0 \left[ 1 + \frac{ \dot{x}^2}{2} - g^2(z)\frac{x^2}{2} \right].
\eeq
Notice, that the optical Lagrangian is quadratic in both $x$ and $\dot{x}$, and can be therefore interpreted as the Lagrangian of a (shifted) harmonic oscillator with mass $n_0$ and $z$-dependent resonance frequency $\Omega(z)=g(z)$. Thanks to this analogy, we can calculate the kernel $K(x_b,x_a,z)$ in the same manner we did for the harmonic oscillator in the previous section, with the difference that now we need to account for the fact that the frequency of the oscillator varies with $z$. In particular, we can employ the same trick of writing the components of the trajectories as $x(z)=x_{cl}(z)+y(z)$, with the boundary conditions that the deviation $y(z)$ is  zero at the endpoints $z=0$ and $z=z$, and we can then write the propagator, in analogy with Eq. \eqref{eq:h_o_kernel_only_classical_action} as 
\beq
\label{eq:grin_propagator}
K(x_b,x_a,z) = F(z)\exp{\left[ \frac{i}{\lambdabar}S_{cl}(x_b,x_a,z)\right]},
\eeq
where
\beq
F(z)=\sqrt{\frac{i}{2\pi\lambdabar}\frac{\partial^2}{\partial x_b\partial x_a}S_{cl}(x_a,x_b,z)}.
\eeq
Following this line of reasoning, and representing the classical action $S_{cl}(x_a,x_b,z)$ in terms of the so-called paraxial ($H_1(z)$) and field ($H_2(z)$) rays of a general optical system, we can obtain a similar result, in our $(1+1)$-dimensional model, than the $(2+1)$-dimensional result obtained by G\~{o}mez-Reino and J. Li\~{n}ares \cite{gomez-reino1987}, namely \footnote{Notice that the pre-factor of Eq. \eqref{eq:propagator_full} differs slightly from Eq. (25) of Ref. \citenum{gomez-reino1987}. This is due to the different dimensionality of the problem at hand. While Ref. \citenum{gomez-reino1987} deals with a $(2+1)$-dimensional problem, in our model we are only considering one transverse dimension, and therefore the pre-factor has the form of $\sqrt{X}$ instead of $X$.}
\barr\label{eq:propagator_full}
K(x_b,x_a,z)&=&\sqrt{\frac{kn_0}{2\pi i H_1(z)}}\exp{(ik\,n_0\,z)}\nonumber\\
&\times&\exp{\Bigg[i\,k\,n_0\,\frac{\dot{H}_1(z)x_b^2+H_2(z)x_a^2-2x_ax_b}{2H_1(z)}\Bigg]},
\earr
where the dot stands for derivative with respect to $z$ and $H_{1,2}(z)$ are two independent solutions of the following equation of motion
\beq
\ddot{H}_{1,2}+g^2(z)\,H_{1,2}=0.
\eeq
This result, however, is not particularly insightful, and the solution presented above (or its two-dimensional counterpart presented in Ref. \citenum{gomez-reino1987}) is quite hard to intuitively link to known results. For this reason, we present below a much clearer and intuitive approach, which we hope will help the reader in appreciating the universality and versatility of path integrals beyond quantum mechanics.

\subsection{Paraxial Propagation as a Harmonic Oscillator}
Let us go back to the analogy to the Lagrangian \eqref{eq:grin_opt_lag_simpl} and calculate the path integral deriving from it. The fact that the frequency of the oscillator is now $z$-dependent doesn't really allow us to repeat one-to-one the calculations in Sect. \ref{harmonicOscillator}. In particular, after we introduce the deviations $y(z)$, we cannot represent $F(z)$ in terms of a Fourier series anymore, since now the oscillator frequency depends on $z$ as well. Instead of doing that, then, we just follow the same line of reasoning that we used to calculate the path integral for a free particle in Sect. \ref{freeParticle}, namely we divide the propagation ``interval" $Z=z_b-z_a$ into $N$ smaller intervals of length $\varepsilon=z_{i+i}-z_i$, such that $Z=N\varepsilon$, so that we can write the action correspondent to the Lagrangian  \eqref{eq:grin_opt_lag_simpl} as
\beq
S_i(x_i,x_{i-1})=\frac{n_0}{2\varepsilon}(x_i-x_{i-1})^2-\frac{n_0\varepsilon}{2}\Omega_i^2 x_i^2,
\eeq
where $x_0\equiv x_a$, $x_N\equiv x_b$, and $\Omega_i^2\equiv g^2(z_i)$. This allows us to approximate the path integral in Eq. \eqref{eq:grin_path_integral} using Eq. \eqref{equation12}, and evaluate it first over a finite set of $N$ trajectories, and then to get back to the actual result by taking the limit $N\rightarrow\infty$. By doing this, moreover, we gain the advantage, that in each infinitesimal interval $\varepsilon$, the frequency $\Omega_i$ of the oscillator is constant, and so we can use the results for the standard oscillator within each interval of length $\varepsilon$. If we then introduce the quantities
\bseq
\begin{align}
\beta &= \frac{n_0}{\varepsilon\lambdabar},\\
\alpha_i &= \beta\left(1-\frac{1}{2}\Omega_i^2\varepsilon^2\right),
\end{align}
\eseq
we can then rewrite the path integral in Eq. \eqref{eq:grin_path_integral} as $N$ nested Gaussian integrals, \emph{i.e.},
\barr\label{equation50}
K_N(x_b,x_a,z)&=&\exp{\left[\frac{i\beta}{2}\left(x_a^2+x_b^2\right)\right]}\nonumber\\
&\times&\int\,dx_1\cdots dx_{N-1}\,\exp{\left[i\sum_{n=1}^{N-1}\alpha_nx_n^2\right]}\nonumber\\
&\times&\exp{\left(-i\beta x_0x_1\right)}\cdots\exp{\left(-i\beta x_{N-1}x_N\right)},
\earr
and then the diffraction kernel can be calculated as 
\beq
K(x_b,x_a,z)=\lim_{N\rightarrow\infty}K_N(x_b,x_a,z).
\eeq
The explicit expression for $K_N$ can be calculated by recursively applying the following Gaussian integral result to Eq. \eqref{equation50}
\barr
I&=&\int\,dx\,\exp{\left\{i\left[\alpha x^2-\left(a+b\right)x\right]\right\}}\nonumber\\
&=&\sqrt{\frac{i\pi}{\alpha}}\exp{\left[-i\frac{a^2+b^2}{4\alpha}\right]}\exp{\left[-i\frac{ab}{2\alpha}\right]}.
\earr
This procedure will lead to the quite compact result
\beq
K_N(x_b,x_a,z)=\sqrt{\frac{a_N}{2\pi}}\exp{\left[i\left(p_Nx_a^2+q_Nx_b^2\right)-a_Nx_ax_b\right]},
\eeq
where $a_N$, $p_N$, and $q_N$ are quantities that depend on $\beta$ and $\alpha_i$, and their explicit expression can be found in Ref. \citenum{harmonicTimeDep}. We can then use the result above and take its limit for $N\rightarrow\infty$ to get the final form of the propagator, which is given explicitly as follows
\barr
K(x_b,x_a,z)&=&\sqrt{\frac{n_0\sqrt{\dot{\gamma}_a\dot{\gamma}_b}}{2\pi i\lambdabar\sin\phi(z)}}\exp{\left[\frac{in_0}{2\lambdabar}\left(\frac{\dot{s}_b}{s_b}x_b^2-\frac{\dot{s}_a}{s_a}x_a^2\right)\right]}\nonumber\\
&\times&\exp{\Bigg\{\frac{in_0}{2\lambdabar\sin\phi(z)}\Bigg[(\dot{\gamma}_bx_b^2+\dot{\gamma}_ax_a^2)\cos\phi(z)}\nonumber\\
&-&2\sqrt{\dot{\gamma}_b\dot{\gamma}_a}\,x_ax_b\Bigg]\Bigg\}
\earr
where the proper limit for the quantities $a_N$, $p_N$, and $q_N$ has been taken as instructed in Ref. \citenum{harmonicTimeDep} (also see the same discussion for the simple harmonic oscillator in Sect. \ref{harmonicOscillator}),  $\phi(z)=\gamma(z)-\gamma(0)$, and the quantities $s(z)$ and $\gamma(z)$ are determined from the solution of the following differential equation
\beq\label{equation55}
\frac{d^2}{dz^2}\xi(z)+\Omega^2(z)\xi(z)=0
\eeq
where $\xi(z)=s(z)\exp{[i\gamma(z)]}$, and $\Omega(z)\in\mathbb{R}$. First of all, notice the similarity between this result and the form of the propagator for a harmonic oscillator, as given by Eq. \eqref{pathIntegralHO}. In fact, the result above reduces to the propagator of a harmonic oscillator with constant frequency $\omega$ by means of the substitution $s(z)=\sqrt{n_0/\omega}$ and $\gamma(z)=\omega\, z$. We can then interpret the propagation of light in a medium described by the refractive index \eqref{refractiveIndex} as basically being given by the propagator of a harmonic oscillator with a suitably chosen $z$-dependent frequency (which depends on the longitudinal properties of the refractive index). The propagator, moreover, is uniquely determined by the amplitude and phase of the solution of a harmonic-oscillator-like equation of motion for $\xi(z)$, where the $z$-dependent profile of the medium determines, again, the characteristic frequency of the oscillator. 

Following Ref. \citenum{harmonicTimeDep} we can make this connection appear more evident, by rewriting the propagator above in terms of the eigenstates of the refractive index potential given by Eq. \eqref{refractiveIndex}. This can be done by first rewriting the $\sin\phi(z)$ and $\cos\phi(z)$ as complex exponentials and then using  Mehler's formula \cite{olver} 
\barr
&&\frac{\exp{\left[-(X^2+Y^2-2X\,Y\,Z)/(1-Z^2)\right]}}{\sqrt{1-Z^2}}\nonumber\\
&=&\exp{\left[-\left(X^2+Y^2\right)\right]}\sum_{n=0}^{\infty}\frac{Z^n}{2^nn!}\text{H}_n(X)\text{H}_n(Y)
\earr
with $Z=\exp{(-i\phi(z))}$, $X=x_b\sqrt{n_0\dot{\gamma}_b/\lambdabar}$, and $Y=x_a\sqrt{n_0\dot{\gamma}_a/\lambdabar}$ to transform the resulting expression in terms of Hermite polynomials and finally obtain (we generalise $z=0$ to a generic $z=z_a$ for convenience)
\beq
K(x_b,x_a,z_a,z_b)=\sum_{n=0}^{\infty}\psi_n^*(x_a,z_a)\psi_n(x_b,z_b),
\eeq
where $\psi_n(x,z)$ are the eigenfunctions of a harmonic oscillator with $z$-dependent frequency $\Omega(z)$, and their explicit expression is given as
\barr
\psi_n(x,z)&=&\sqrt{\frac{1}{2^nn!}\sqrt{\frac{n_0\dot{\gamma}}{\pi\lambdabar}}}\exp{\left[-i\left(n+\frac{1}{2}\right)\gamma(z)\right]}\nonumber\\
&\times&\exp{\left[\frac{in_0}{2\lambdabar}\left(\frac{\dot{s}}{s}+i\dot{\gamma}\right)x^2\right]}\text{H}_n\left(\sqrt{\frac{n_0\dot{\gamma}}{\lambdabar}}x\right),
\earr
where $s(z)$ and $\gamma(z)$ are solutions of Eq. \eqref{equation55}. For light propagating in vacuum, $g(z)=0$ (and, therefore $\Omega(z)=0$) and the above expression of the propagator reduces to the well-known result of the resolution of the diffraction kernel in terms of Hermite-Gaussian eigenstates of the paraxial equation \cite{AJP_harmonic_oscillator_paraxial}. The expression above is moreover fully equivalent to that found by C. G\~{o}mez-Reino and J. Li\~{n}ares \cite{gomez-reino1987}, but it gives more insight on how path integrals can be used to easily solve complicated problems in optics, such as the propagation of light in a GRIN medium, whose analytical solution, even in the simplest cases, in not really available. By establishing the analogy with a time-dependent harmonic oscillator, on the other hand, path integrals give a rather elegant and remarkably simple result, which can be intuitively hinted at using simple arguments, such as the propagation of light in free space.
\section{An Example from Quantum Optics: Path Integral Description of Degenerate Parametric Amplifiers}\label{sect4}
Path integrals, in the form defined by Eq. \eqref{eq:pathIntegral}, have also been used to approach various problems in nonlinear optics. In this case, as we will see throughout this section, the integral over all the possible trajectories of the quantum particle will be replaced, in the formulation originally proposed by Hillery and Zubairy in 1982 \cite{hillery2}, by an integral over all possible configurations in the complex plane spanned by coherent states. 

\begin{figure}[!t]
    \centering
    \includegraphics[width=0.45\textwidth]{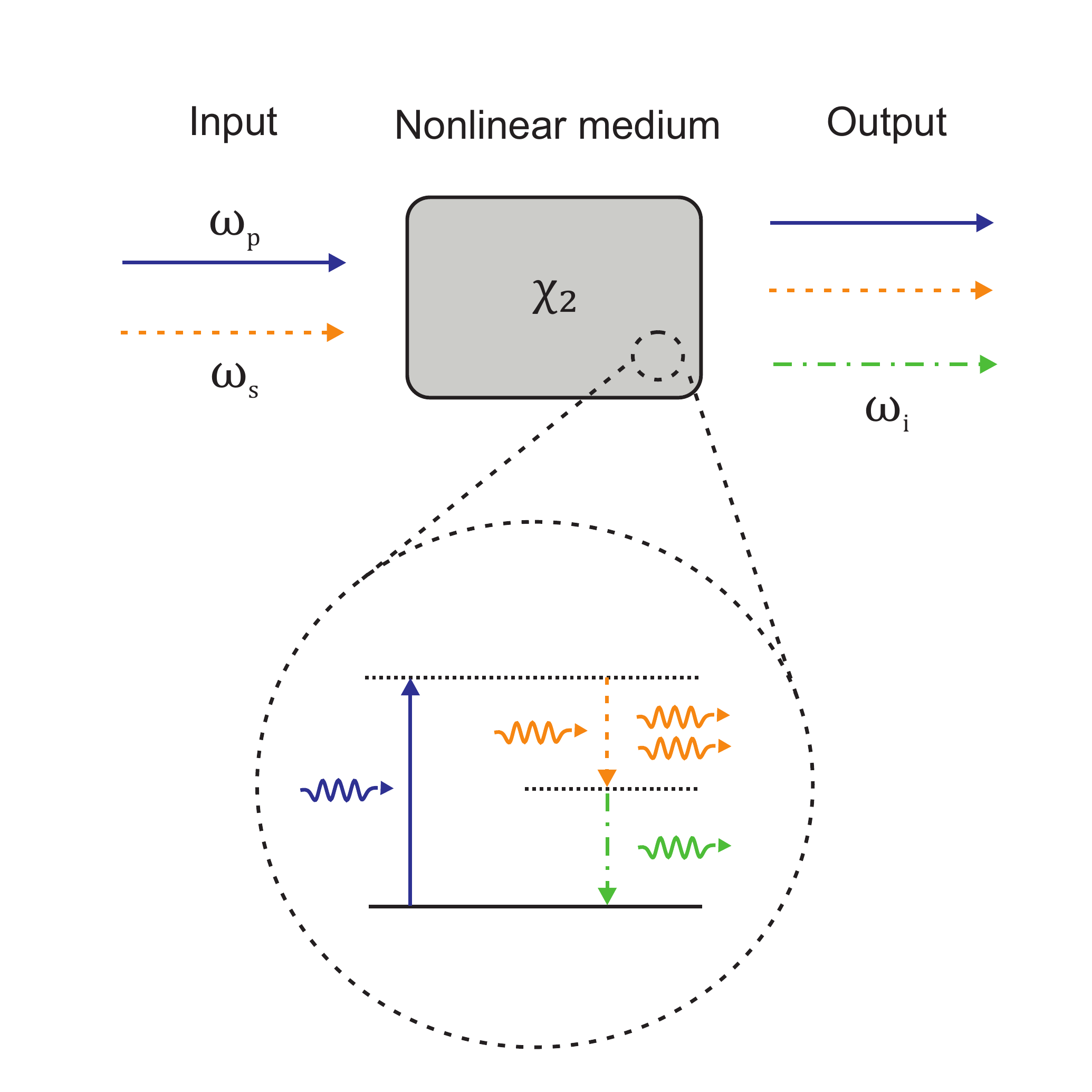}
    \caption{Pictorial representation of optical parametric amplification. A nonlinear material with second-order susceptibility ($\chi_{2}$) is impinged upon by pump photons at angular frequency $\omega_{\rm{p}}$ and signal photons at $\omega_{\rm{s}}$. Due to the excitation of the medium to a virtual energy level by the pump (shown at the bottom of the figure), the signal photon stimulates the emission of photons at $\omega_{\rm{s}}$ and, due to energy conservation, produces an idler field at $\omega_{\rm{i}}$. For the case of a degenerate parametric amplifier, the signal and idler fields have the same frequency, \emph{i.e.}, $\omega_{\rm{s}}=\omega_{\rm{i}}$.}
    \label{fig:DPA_1}
\end{figure}

As an explicit example, we consider here the case of parametric amplification, whereby a signal incident on an optically nonlinear material is amplified \cite{fox,boyd}. A pump field impinging on a material characterised by a $\chi_{2}$ nonlinearity (see Fig. \ref{fig:DPA_1}) can combine with a signal field, leading to a depleted pump and an amplified signal (and an idler field, in order to conserve energy). 

In this section, the contents of which based on Ref. \citenum{hillery2}, we will then apply the tools presented so far in this tutorial to the case of the degenerate parametric amplifier, thus providing an example of how path integrals can prove useful to solve problems in quantum optics. 

For the sake of simplicity, let us consider a single mode of the radiation field (generalisations of this formalism to multimode fields is covered explicitly in Ref. \citenum{hillery2}), and we represent the field using coherent states, a natural choice here as the Hamiltonians under consideration will be expressed in terms of creation and annihilation operators ($a^{\dagger}$ and $a$, respectively). This assumption, in particular, will allow us to define the path integral in terms of coherent states, \emph{i.e.}, as a path integral in the field's phase space.
\subsection{Propagator as Path Integral Over Coherent States}
We begin by defining the time-evolution operator as $\hat{U}(t_{b},t_{a})$, which evolves the system's state at time $t_{a}$ to the state at time $t_{b}$ by $|\psi (t_b)\rangle = \hat{U}(t_b,t_a) |\psi (t_a)\rangle $ \cite{sakurai}. Denoting the time-dependent Hamiltonian of the system by $H(t)$, the time-evolution operator is then $\hat{U}(t_b,t_a) = \hat{T}\mathrm{exp}\left[ -i \int_{t_a}^{t_b}H(t')dt' \right]$, where $\hat{T}$ is the time-ordering operator \cite{gifted}.

%Coherent states $|\alpha_{i} \rangle$ are the eigenstates of the annihilation operator $a$, with eigenvalues $\alpha_{i}$ at time $t=0$ \cite{knight}.

If the electromagnetic field is represented in terms of coherent states, we can readily give a definition of the propagator using the time-evolution operator defined above as follows
\beq
K(\alpha_{b},t_b;\alpha_{a},t_a) = \langle \alpha_{b} |\hat{U}(t_b,t_a)|\alpha_{a}\rangle,
\eeq
where $\ket{\alpha}$ represent the set of coherent states, defined as the eigenstates of the annihilation operator, \emph{i.e.}, $\hat{a}\ket{\alpha}=\alpha\ket{\alpha}$ \cite{knight}. If we introduce the notation $\ket{\alpha,t}\equiv\hat{U}(t,0)\ket{\alpha}$, we can rewrite the expression above in the following form, which will prove useful in the remainder of this section
\beq\label{propagatorCoherent}
K(\alpha_{b},t_b;\alpha_{a},t_a)=\braket{\alpha_b,t_b}{\alpha_a,t_a}. 
\eeq
To understand why the form above is useful to our means, let us show how the quantity above naturally appears when calculating expectation values of operators and, more generally, correlation functions,  in the so-called $P$-representation. We assume that at time $t=0$ the density matrix of the field can be written as
\beq
\rho = \int d^{2}\alpha P(\alpha)|\alpha\rangle \langle \alpha |,
\eeq
where $P(\alpha)$ is the Glauber---Sudarshan $P$-function \cite{sudarshan,glauber}. In this representation, the expectation value of any operator $\hat{O}(t)$ in the Heisenberg picture  can then be written as
\begin{equation}
\langle \hat{O}(t)\rangle = \mathrm{Tr}[\rho \hat{O}(t)] = \int d^{2} \alpha P(\alpha)\langle \alpha | \hat{O}(t) | \alpha \rangle .
\end{equation}
Using the expression above in combination with the completeness relation of coherent states, \emph{i.e.}, 
\beq\label{completeness}
\frac{1}{\pi}\,\int d^{2}\alpha\, |\alpha, t \rangle \langle \alpha , t| = 1,
\eeq 
where $d^2\alpha=d\operatorname{Re}\alpha\,d\operatorname{Im}\alpha$ spans the complex-plane defined by coherent states \cite{knight}, any normal-ordered correlation function of the form $\langle \hat{a}^{\dagger}(t_1)\hat{a}^{\dagger}(t_2)\cdots\hat{a}(t_{N-1})\hat{a}(t_N)\rangle$ can be then expressed in terms of the propagator \eqref{propagatorCoherent}. For the simple case of $\langle\hat{a}(t)\rangle$, this can be readily shown as follows
\barr\label{eq:expec_1}
\langle \hat{a}(t) \rangle &=&\int\,d^2\alpha\,P(\alpha)\expectation{\alpha}{\hat{a}(t)}{\alpha}\nonumber\\
&=&\int\,d^2\alpha\,P(\alpha)\expectation{\alpha}{\hat{U}^{-1}(t,0)\,\hat{a}\,\hat{U}(t,0)}{\alpha}\nonumber\\
&=&\frac{1}{\pi}\,\int\,d^2\alpha\,d^2\,\beta\,P(\alpha)\,\beta\,\left|K(\beta,t;\alpha,0)\right|^2,
\earr
where to pass from the second to the third line we have employed the completeness relation for coherent states. For correlation functions containing more creation and annihilation operators, more propagators, calculated at different times, will appear in the expression above \cite{hillery2}. The propagator is then clearly an important quantity, to evaluate such expectation values.

We now show how to write the propagator as a path integral. To start with, let us assume the Hamiltonian of our system is normal-ordered, \emph{i.e.}, $H=H(\hat{a}^{\dagger},\hat{a},t)$, and as we did in Sect. \ref{freeParticle} assume to divide the evolution interval $t_b-t_a$ into $N$ slices of length $\varepsilon=(t_b-t_a)/N$. We can mirror this choice directly into Eq. \eqref{propagatorCoherent} by inserting $N$ times the completeness relation \eqref{completeness} and set $t\rightarrow t_j=t_{j-1}+\varepsilon$ for each entry. If we do so, we then obtain
\barr\label{eq:coh_prop}
K(\alpha_{b},t_{b};\alpha_{a},t_{a}) &=& \left( \frac{1}{\pi} \right)^{N} \int d^{2}\alpha_{1} \ldots d^2\,\alpha_{N} \langle \alpha_{b},t_{b}|\alpha_{N},t_{N}\rangle\nonumber\\
&\times& \langle \alpha_{N},t_{N}|\alpha_{N-1},t_{N-1}\rangle \ldots \langle \alpha_{1},t_1|\alpha_{a},t_{a}\rangle .
\earr
The quantities $\braket{\alpha_j,t_j}{\alpha_{j-1},t_{j-1}}$ can be readily evaluated using the time-evolution operator defined above, and noticing that since $t_j-t_{j-1}=\varepsilon\ll 1$, we can Taylor expand the time-evolution operator to obtain
\beq
\hat{U}(t_j,t_{j-1})\simeq 1-i\,\varepsilon\,\int_{t_{j-1}}^{t_j}\, d\tau\,H(\hat{a}^{\dagger},\hat{a},\tau).
\eeq
With this result at hand, the individual terms  $\braket{\alpha_j,t_j}{\alpha_{j-1},t_{j-1}}$ can be brought, after a simple algebraic manipulation \cite{hillery2}, in the following form
\barr \label{eq:coh_overlap}
\langle \alpha_{j},t_{j} | \alpha_{j-1},t_{j-1} \rangle &\approx& \exp{\left[ (-1/2) \left( |\alpha_{j}|^{2} + |\alpha_{j-1}|^{2} \right)+ \alpha_{j}^{*} \alpha_{j-1}\right]}\nonumber\\
&\times&\exp{\left[  - i\epsilon H (\alpha_{j}^{*} ,\alpha_{j-1} ,t_{j-1}) \right]},
\earr
where we have defined
\begin{equation}
H(\beta^{*} ,\alpha ,t) = \frac{\langle \beta | H(a^{\dagger} ,a,t)|\alpha \rangle }{\langle \beta| \alpha \rangle }.
\end{equation}
Now that we have an expression of the propagator in terms of $N$ discrete ``trajectories" (\emph{i.e.}, $N$ different coherent states $\ket{\alpha_j}$), we can take the limit $N\rightarrow\infty$ and arrive at the definition of the path integral in coherent state representation, following the same line of reasoning used in Sect. \ref{sect2}. The details of this calculation are reported in Ref. \citenum{hillery2}, and we refer the interested reader therein. The final result of this calculation is then given as follows:
\barr\label{pathIntegralCoherent}
K(\alpha_{b} ,t_{b} ;\alpha_{a} ,t_{a} ) &=& \int \mathcal{D}\alpha (\tau) \exp{ \Bigg\{ \int_{t_{a}}^{t_{b}} d\tau \Big[ \frac{\alpha \dot{\alpha}^{*} - \alpha^{*} \dot{\alpha}}{2}}\nonumber\\
&-&iH(\alpha^{*} ,\alpha ;\tau) \Big] \Bigg\} ,
\earr
with the integration measure taken to mean an integral over all coherent states parameterised by $\tau$, with $\alpha (t_{a}) \equiv \alpha_{a}$ and $\alpha (t_{b}) \equiv \alpha_{b}$.
\subsection{Propagator for Quadratic Hamiltonians}
We now apply the above results, focusing on the class of Hamiltonians which are at most quadratic in the creation and annihilation operators. This class of Hamiltonians is of particular interest in quantum optics, since it describes second-order nonlinear phenomena, within the so-called undepleted pump approximation \cite{boyd}, and it can sometimes also be used to describe harmonic generation in third-order nonlinear systems \cite{boyd, drummond}. 

The most general quadratic Hamiltonian can be written as 
\begin{equation} \label{eq:general_H}
H(\hat{a}^{\dagger} ,\hat{a}, t) = \omega (t)\hat{a}^{\dagger}\hat{a} + \left[f(t)\hat{a}^2+ g(t)\hat{a} + \text{h.c.}\right],
\end{equation}
with $f(t)$ and $g(t)$ arbitrary, but well-behaved, time-dependent functions. The reader might notice a similarity between the Hamiltonian above and that of a forced harmonic oscillator \cite{hibbs}. For this class of Hamiltonians, the integrals appearing in Eq. \eqref{pathIntegralCoherent} are all Gaussian, and can be performed using the methods described in Sect. \ref{sect2}. In particular, to calculate the path integral with the Hamiltonian defined above, one first needs to discretise the paths over the coherent states $\alpha$ (in the same manner we discretised the trajectories for the harmonic oscillator in Sect. \ref{sect2}), then express the exponential appearing in Eq. \eqref{pathIntegralCoherent} in terms of $\operatorname{Re}\{\alpha\}$ and $\operatorname{Im}\{\alpha\}$ explicitly, and notice, that the correspondent integrals are Gaussian in both the real and the imaginary parts of $\alpha$. After having calculated a single term, one could then calculate the remaining integrals iteratively, as we have done for the examples in Sect. \ref{sect2}. After a lengthy but straightforward calculation, which is partially covered in Appendix A of Ref. \citenum{hillery2}, the propagator for a quadratic Hamiltonian assumes the following form
\beq \label{eq:prop_quad_Hamilt}
K(\boldsymbol\alpha,\vett{t}) = F(\boldsymbol\alpha,\vett{t})\exp{\left[-i\,\Sigma(\boldsymbol\alpha,\vett{t})\right]},
\eeq
where $\boldsymbol\alpha=\{\alpha_a,\alpha_b\}$, $\vett{t}=\{t_a,t_b\}$ and the functions $F$ and $\Sigma$ are defined as
\barr
F(\boldsymbol\alpha,\vett{t})&=&\exp{\Big[-\frac{1}{2}(|\alpha_b|^2+|\alpha_a|^2)+Y(t_b)\alpha_b^*\alpha_a}\nonumber\\
&+&X(t_b)(\alpha_b^*)^2+Z(t_b)\alpha_b^*\Big],
\earr
\barr
\Sigma(\boldsymbol\alpha,\vett{t})&=&\int_{t_a}^{t_b}\,d\tau\Big\{f(\tau)\Big[2X(\tau)+Z^2(\tau)+\alpha_a^2Y^2(\tau)\nonumber\\
&+&2\alpha_a Y(\tau)Z(\tau)\Big]+g(\tau)\Big[Z(\tau)+\alpha_a\,Y(\tau)\Big]\Big\},
\earr
where the auxiliary function $X(t)$ is constrained by 
\begin{equation} \label{eq:X_diff_eq}
\frac{dX}{dt} = -2i\omega (t) X - 4if(t)X^{2} - if^{*} (t),
\end{equation}
with initial condition $X(t_{a})=0$. The functions $Y(t)$ and $Z(t)$ are instead defined in terms of $X(t)$ as
\begin{equation}
Y(t) = \mathrm{exp}\left( -i\int_{t_{a}}^{t} d\tau \left[ \omega (\tau) + 4f(\tau ) X(\tau ) \right] \right),
\end{equation}
and
\barr
Z(t) &=& -i \int_{t_{a}}^{t} d\tau \left[ g^{*} (\tau) + 2g^*(\tau ) X(\tau ) \right]\nonumber\\
&\times&\exp{\left( -i \int_{\tau}^{t} d\tau' \left[ \omega (\tau') + 4f(\tau' )X(\tau') \right] \right)} .
\earr
\subsection{Propagator for Degenerate Parametric Amplification}
A degenerate parametric amplifier is characterised by the following quadratic Hamiltonian \cite{boyd}
\begin{equation} \label{eq:DPA_Hamilt_approx}
H(t) = \omega a^{\dagger} a + \kappa \left( e^{2i\omega t}a^{2} + e^{-2i\omega t}a^{\dagger 2} \right),
\end{equation}
where $\omega$ is the angular frequency of the mode and $\kappa$ is a coupling constant. The above Hamiltonian can clearly be seen to fall into the class of Hamiltonians given in Eq. (\ref{eq:general_H}) if we identify $\omega (t) = \omega$, $f(t) = \kappa e^{2i\omega t}$, and $g(t)=0$. Although parametric amplification is formally a 3-wave process involving, as depicted in Fig. \ref{fig:DPA_1}, a pump, a signal, and an idler field, it is common practice in nonlinear optics experiments to work within the so-called undepleted pump approximation \cite{boyd, drummond}, which treats the pump mode as a classical (bright) field, whose number of photons does not change significantly (\emph{i.e.}, the pump field remains undepleted) during the nonlinear interaction. Within this approximation, then, the Hamiltonian becomes effectively quadratic in the signal and idler modes, and can be described by Eq. \eqref{eq:DPA_Hamilt_approx} above. The explicit form of the propagator given below is then valid within this approximation. A fully quantised analysis of parametric amplification beyond the undepleted pump approximation in terms of path integrals has been presented by Hillery and Zubairy in Ref. \citenum{hillery}.

In this case, with the form of $\omega(t)$, $f(t)$, and $g(t)$ given above, the functions $X(t)$, $Y(t)$, and $Z(t)$ can be written in an explicit form (in particular, Eq. \eqref{eq:X_diff_eq} cabn be solved analytically), giving $Z(t)=0$ and \cite{hillery2}
\bseq
\begin{align}
X(t) &= \frac{1}{2i}e^{-2i\omega t}\mathrm{tanh} \left[ 2\kappa (t-t_{a}) \right] ,\\
Y(t) &= e^{-i\omega (t-t_{i})}\mathrm{sech} \left[ 2\kappa (t-t_{a}) \right].
\end{align}
\eseq
These time-dependent quantities can then be directly substituted into Eq. \eqref{eq:prop_quad_Hamilt}), finally yielding the propagator for the degenerate parametric amplifier:
\barr \label{eq:DPA_prop}
K(\alpha_{b} ,t_{b}; \alpha_{a} ,t_{a}) &=& \sqrt{\mathrm{sech} \left[ 2\kappa (t_{b} - t_{a}) \right]}\exp{\left(-\frac{|\alpha_b|^2+|\alpha_a|^2}{2}\right)}\nonumber \\
& \times& \exp{\Bigg\{\alpha_{b}^{*} \alpha_{a} e^{-i\omega (t_{b} - t_{a} )} \mathrm{sech} \left[ 2\kappa (t_{b} - t_{a}) \right]}\nonumber \\
&-&i\left(\frac{\alpha_{b}^{*}}{\sqrt{2}}\right)^2  e^{-2i\omega t_{b}} \mathrm{tanh} \left[ 2\kappa (t_{b} - t_{a}) \right]\nonumber\\
&-&i\left(\frac{\alpha_{a}}{\sqrt{2}}\right)^2 e^{2i\omega t_{a}} \mathrm{tanh} \left[ 2\kappa (t_{b} - t_{a}) \right] \Big]\Bigg\} .
\earr

\section{Predicting Optical Properties of Matter Using Path Integrals}\label{sect4bis}
The examples discussed above only deal with a single particle, either freely propagating, or interacting with a suitable potential. In all these cases, the complexity of the problem is low enough, to allow analytical solutions to the path integrals. As soon as many-body effects are taken into account, as it is the case, for example, for atoms and molecules, handling the path integral analytically becomes impossible, and numerical techniques for its evaluation are necessary. A particularly successful computational method for this task has been Path Integral Monte Carlo (PIMC). The interested reader can download David Ceperley's group PIMC++ open-source code for Path Integral Monte Carlo simulations, available, together with its relative documentation, at \nolinkurl{https://pimc.soft112.com}. 

In this section, we then present some of the results obtained using PIMC to predict the \emph{exact} linear and nonlinear optical properties of simple systems as a function of frequency. These results, recently demonstrated by Tiihonen \emph{et al.} \cite{tiihonen1}, constitute one of the first attempts to use PIMC to investigate the optical properties of matter. There, however, only very simple systems were studied, such as the hydrogen (H) and hydrogen-like ($\text{Li}^+$ and $\text{Be}_2^+$) atoms, the helium atoms He and $\text{He}^+$, the hydrogen molecule $\text{H}_2$ and $\text{H}_2^+$, hydrogen-helium ($\text{HeH}^+$) and hydrogen-deuterium ($\text{HD}^+$) molecules, and the positronium atom. These systems have been studied employing different methods, including finite-field simulations for static polarisabilities \cite{tapio5}, polarisability estimators for simulation without the external field \cite{tapio14}, static field-gradient polarisabilities \cite{tapio6}, and finally, dynamic polarisabilities and van der Waals coefficients \cite{tapio13}. The last one, in particular, is of great interest, as the macroscopic electric susceptibility is constructed starting from the dynamic frequency dependent polarisabilities.

The results presented in the aforementioned references, and summarised in this section represent an exceptional benchmark to gauge the capabilities of PIMC in providing accurate and multidimensional information about the optical properties of matter, and hint at the potential impact PIMC could have in photonics, as a new modelling platform to calculate \emph{exactly} the optical response of exotic materials, such as 2D materials, the understanding of which is still in its infancy \cite{nonlinear2D}. A great advantage PIMC would give, for example, is the possibility to investigate the nonlinear properties of 2D materials far from equilibrium, a physical regime that is currently poorly understood. This could be of particular importance for the nonlinear optical properties of 2D materials, since for such materials, non-equilibrium dynamics can be easily reached at relatively small optical powers.
\subsection{Polarisability in PIMC}
The optical response of matter \cite{jackson}, frequently described in terms of the matter polarisation vector $\vett{P}$, is completely determined by electron dynamics solely within the limits of the the Born-Oppenheimer (BO) approximation, which specifies that the nuclear dynamics occurs on a much longer timescale and can be therefore neglected, it is implicitly assumed \cite{fetter}. In general, however, when thermal effects are explicitly taken into account, this approximation might not be valid anymore, as the role of the nuclei becomes more prominent as the temperature of the system is increased. To properly take into account these effects, then, an approach able to go beyond BO is needed to get the correct results. PIMC, then, is the most viable, if not the only possible, approach to systematically study the optical properties of materials in this particular regime (which, for example, includes the calculation of such properties at room temperature).

While optics makes extensive use of the (linear and nonlinear) susceptibility tensor to describe the optical properties of a material \cite{boyd}, from a PIMC perspective, it is better to work with polarisabilities, as they can be defined quite easily in terms of path integrals. For example, the $ij$-component of the linear (\emph{i.e.}, dipole) polarisability tensor, namely $\alpha_{ij}$, can be simply calculated as the (functional) derivative of the expectation value (\emph{i.e.}, first order correlator) of the dipole moment $\boldsymbol\mu$ oriented, say, along the $i$-direction, with respect to an external electric field oriented along the $j$ direction \cite{tiihonen1}, \emph{i.e.},
\beq
\alpha_{ij}=\frac{\partial}{\partial E_j}\langle\mu_i\rangle,
\eeq
where the expectation value of an operator $\hat{O}$ is defined with respect to the density matrix operator of the system in the so-called imaginary time representation \cite{tiihonen1}, \emph{i.e.},
\beq\label{thermalAverage}
\langle\hat{O}\rangle = \mathcal{N}\operatorname{Tr}\left\{\hat{O}\,\exp{\left[-\frac{S(T)}{\hbar}\right]}\right\},
\eeq
where $\mathcal{N}$ is a suitable normalisation constant, $S(T)=\left(\hat{H}_0-\hat{\vett{Q}}\cdot\vett{E}\right)/k_BT$ is the temperature-dependent action of the system (described by the Hamiltonian $\hat{H}_0$) interacting with the electromagnetic field $\vett{E}$, and $\hat{\vett{Q}}$ is the electric moment operator \cite{tiihonen1}. Notice that to represent the expectation value above as a path integral, it is computationally more convenient to transform the complex phase factor $\exp{\left(i\,S/\hbar\right)}$, naturally arising from path integrals, into an exponentially decaying term $\exp{\left(-S/\hbar\right)}$ by means of a Wick rotation \cite{srednicki} ( \emph{i.e.}, a change in reference frame from real to imaginary time, namely $t\rightarrow i\tau$), since computationally it is much easier to deal with real-valued, exponentially decaying terms, rather than complex-valued, spuriously oscillating ones. In fact, the everywhere positive exponential function can be considered as a probability distribution for sampling the imaginary time paths with Metropolis Monte Carlo algorithm in canonical (NVT) ensemble.

PIMC makes use of several different strategies to compute the quantities defined above. The zero-field polarisability estimators, for example, are Hellmann—Feynman type operators \cite{tiihonen1}, whose expectation values give the polarisabilities of various types and order, without the need of including an external electric field in the simulation.

Computation of dynamic multipolar polarisabilities, on the other hand, is more sophisticated and it requires to first determine, through PIMC, the multipole-multipole correlation function in the so-called imaginary time representation, then analytically continue it to obtain the spectral function in the real time domain (a task that is highly non-trivial, due to the fact that it is an ill-posed numerical problem), and finally both the real and imaginary parts of the polarisability. 
\subsection{From Polarisability to Susceptibility}
\begin{figure*}[!t]
\begin{center}
\includegraphics[width=\textwidth]{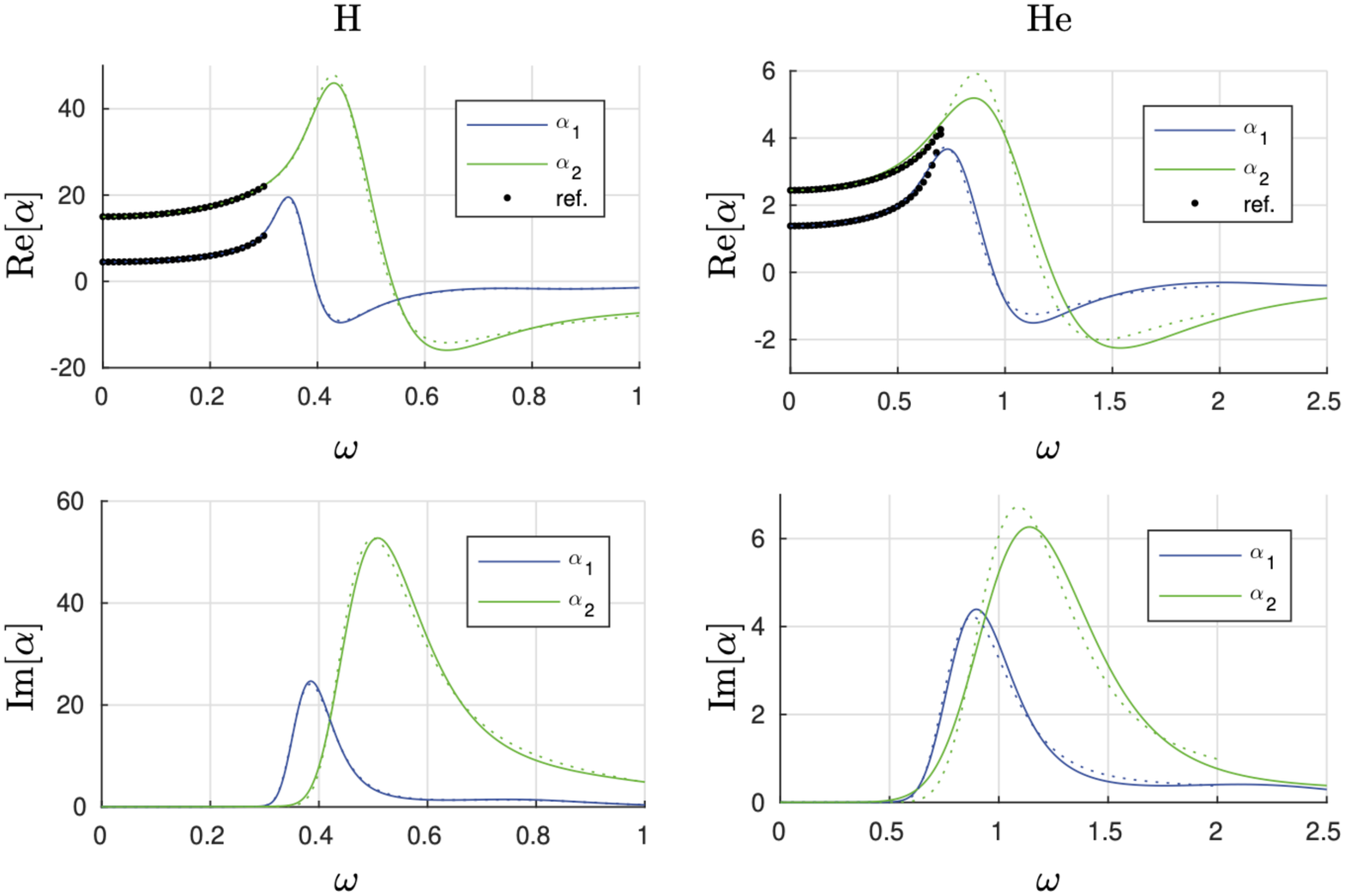}
\caption{Real (top) and imaginary (bottom) part of the dynamic polarisability as a function of frequency for Hydrogen (H, left) and Helium (He, right) atoms at $T=2000$ K, within the BO approximation. In these plots, a blue line represents the dipole polarisability $\alpha_1(\omega)$, while a green line represents the quadrupole polarisability $\alpha_2(\omega)$. The black dots represent the reference simulations of polarisabilities calculated at $T=0$ K. Dotted lines correspond to big time steps, i.e., $\Delta\tau=0.1$ for Hydrogen, and $\Delta\tau=0.025$ for Helium, while solid lines correspond to small time steps, namely $\Delta\tau=0.05$ for Hydrogen, and $\Delta\tau=0.0125$ for Helium. In PIMC, time steps are defined, in imaginary time, as $\Delta\tau=1/(Mk_B T)$, where $k_B$ is the Boltzmann's constant, $T$ is the temperature of the system, and $M$ is the so-called Trotter number, which controls the accuracy of the discretisation of the propagator $\exp{(-\Delta\tau\,\hat{H})}$. In these plots, the frequency is measured in Hartree (1 Ha $=$ 2 Ry $= 27.2$ eV). To get a better grasp of the span of the frequency axis of the plots above, a frequency of $\omega =0.1$ Ha corresponds to a wavelength $\lambda=456$ nm, while a frequency of $\omega=1$ Ha corresponds to a wavelength $\lambda=45$ nm. This figure has been taken from Ref. \citenum{tiihonen1} with permission of the author.}
\label{figure7}
\end{center}
\end{figure*}
Assuming a very low density gas phase of these small atoms or molecules, one can evaluate the corresponding macroscopic susceptibility from the atomic and molecular polarisabilities.  At higher densities, interactions and chemical reactions of these moieties will change the polarisabilities, composition or both, which consequently leads to the density dependent susceptibilities.

With increasing density and, in particular, in case of liquids and solids, where several electrons interact strongly, the Fermion Sign Problem (FSP) emerges.  This is the notorious challenge for Monte Carlo evaluation of essential expectation values of identical fermions, and it discloses partly a still open problem.  The problem emerges from the evaluation of the difference of two large contributions with opposite signs, resulting from the sign of the density matrix (or wave function).  There are practical solutions to FSP in PIMC simulations for many cases \cite{condMatt3,condMatt4,condMatt5,condMatt6,tiihonen1}, but not any general robust approaches, yet.  Most of the solutions are based on finding approximate or iterative nodal surfaces of the density matrix.

The real-time path integral (RTPI) approach \cite{tapio14,tapio15,tapio16,tapio17} may offer a remedy in the future, as it works directly with the wave function with explicit sign, even though only at zero Kelvin.  However, as for most of the cases of interests, electrons at room temperature can be well-described with the zero Kelvin temperature model, this is not an issue. Moreover, including temperature in RTPI should be, in principle, possible.

Overall, the FSP represents a challenging problem, and sits at the forefront of current research in PIMC methods. The interested reader can find in Refs. \citenum{condMatt3,condMatt4,condMatt5,condMatt6}, and references therein, a good starting point to delve deeper in this fascinating, yet challenging, problem. In the examples that we are going to present below, however, we have purposely selected simple systems, where there are no more than two electrons in each moiety, where at low temperatures we can assume the system to be in its singlet ground state, thus enabling us to directly label the light fermions with opposite spins and avoiding FSP entirely.
%Assuming a very low density gas phase of these small atoms or molecules, one can estimate the corresponding macroscopic susceptibility from the atomic and molecular polarisabilities.  At higher densities, interactions or chemical reactions may change the polarisabilities or composition, and consequently, the density dependent susceptibilities.

%Furthermore, with increasing density, and in particular, in case of liquid and solid phases, the Fermion Sign Problem (FSP) appears as a notorious challenge and still represents an open problem.  There are practical solutions to FSP  in PIMC simulations in many cases \cite{condMatt3, condMatt19}, but not any general approaches, yet.  
%
\subsection{Some Examples}
In this section we present some results concerning simple systems, such as one- and two-electron atoms and molecules, obtained using PIMC. In particular, we focus our attention on the electric polarisability, to show what are the capabilities of this computational method, and how it could be used to provide a new interesting platform for photonics, to calculate the optical properties of interesting materials, beyond the standard approaches and approximations frequently used in optics \cite{jackson, boyd}.
\begin{figure*}[!t]
\begin{center}
\includegraphics[width=\textwidth]{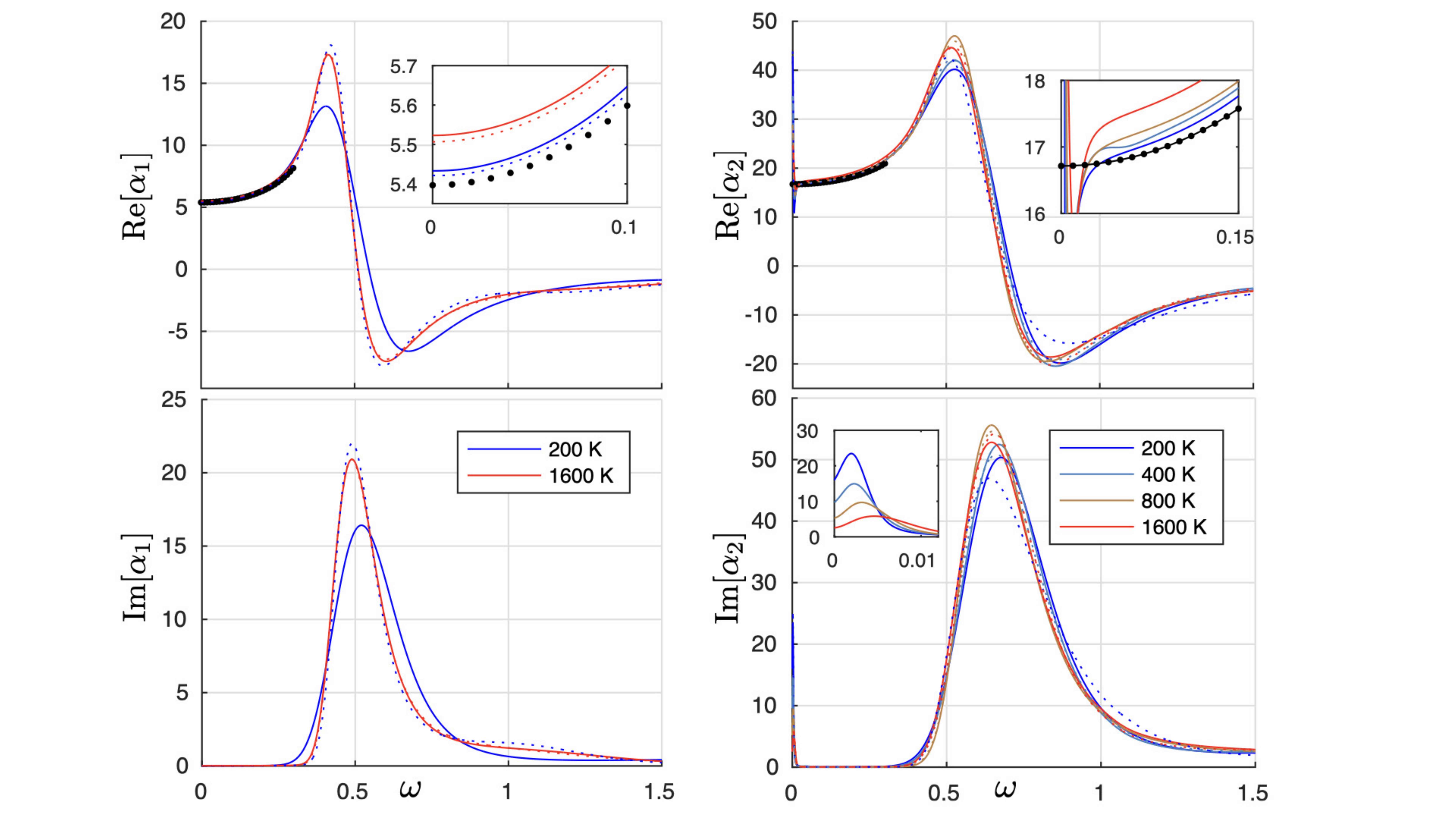}
\caption{Real (top) and imaginary (bottom) part of the dynamic polarisability as a function of frequency for molecular Hydrogen $H_2$ for various temperatures. The left panel presents the results for the dipole polarisability $\alpha_1(\omega)$, while the right panel contains the results for the quadrupole polarisability $\alpha_2(\omega)$. The black dots represent the reference simulations of polarisabilities calculated at $T=0$ K. Dotted lines correspond to big time steps, while continuous lines correspond to smaller time steps. For these simulations, the time step has been adapted according to the chosen temperature, and it varies from $\Delta\tau=0.02$ to $\Delta\tau=0.05$. In these plots, the frequency is measured in Hartree (1 Ha $=$ 2 Ry $= 27.2$ eV). To get a better grasp of the span of the frequency axis of the plots above, a frequency of $\omega =0.1$ Ha corresponds to a wavelength $\lambda=456$ nm, while a frequency of $\omega=1$ Ha corresponds to a wavelength $\lambda=45$ nm. This figure has been taken from Ref. \citenum{tiihonen1} with permission of the author.}
\label{figure8}
\end{center}
\end{figure*}
In both examples below, the polarisability is calculated from the Kubo-like formula for the time-dependent optical susceptibility \cite{fetter}
\beq\label{expectationChi}
\chi(t) = \frac{i}{\hbar}\Theta(t)\,\langle\,[\hat{P}(t),\hat{Q}(0)]\,\rangle = - G_{ret}(t),
\eeq
where the square brackets denote a commutator, the angled brackets denote instead thermal average [see Eq. \eqref{thermalAverage}], $\Theta(t)$ is the Heaviside step function \cite{byronFuller}, $\hat{Q}$ is a multipole operator (which represents dipole or quadrupole interactions, for the cases presented below), $G_{ret}(t)$ is the retarder Green's function of the perturbation $\hat{P}$ and the latter represents the perturbation generating the optical susceptibility. The link between susceptibility and polarisability is better understood in Fourier space, where the following relation is valid \cite{tapio13}
\beq
\chi(\omega)=-\int\,\frac{d\,\Omega}{\pi}\,\frac{A(\Omega)}{\omega-\Omega+i\eta}=\langle\,\alpha(\omega)\,\rangle,
\eeq
where the spectral function $A(\omega)$ is related to the susceptibility via the Kramers-Kronig relation \cite{jackson}, i.e., $A(\omega)=\operatorname{Im}\{\chi(\omega)\}$. PIMC is then used to calculate the value of the expectation value in Eq. \eqref{expectationChi} (or, equivalently, the retarded Green's function), and from it both the real and imaginary part of the polarisability are then calculated. Notice, moreover, that the real part of the polarisability $\operatorname{Re}\{\alpha(\omega)\}$ corresponds to the actual optical response of the material, while its imaginary part $\operatorname{Im}\{\alpha(\omega)\}$ is, basically the spectral function $A(\omega)$.

The first example we present is given in Fig. \ref{figure7}, which depicts the real (top) and imaginary (bottom) part of the dipole (blue line) and quadrupole (green line) polarisabilities $\alpha_1(\omega)$ and $\alpha_2(\omega)$ for the case of atomic Hydrogen (left) and Helium (right) at $T=2000$ K. The simulations have been performed using PIMC with two distinct characteristic time steps $\Delta\tau$, one large (solid green and blue lines in Fig. \ref{figure7}) and one small (dotted green and blue lines in Fig. \ref{figure7}), to rule out possible numerical artefacts \cite{tapio13}. The results have been then compared with polarisability data available in literature, and obtained  using a $T=0$ K approach, with methods other than PIMC \cite{ref0K}. This reference is indicated by black dots in Fig. \ref{figure7}.  For low frequencies, i.e., in the region $0<\omega<0.4$ Ha (corresponding to $0<\lambda<1.14$ $\mu m$), PIMC simulations reproduce very well the standard $T=0$ K result. This, essentially, means, that for the hydrogen atom, the eigenstates at $T=2000$ K are essentially the same as those calculated at $T=0$ K, and, as a consequence of that, the low-frequency polarisability in Fig. \ref{figure7} fits perfectly the results obtained at zero temperature \cite{ref0K}.

As the frequency is increased beyond $\omega=0.5$ Ha (corresponding, for the case of Hydrogen, to its ionisation energy), the zero temperature approach fails to faithfully describe the optical properties of the system. In this high frequency range, i.e., ranging from the near infrared to the deep UV, PIMC produces instead a more accurate and reliable result. This is an important result. Usually, the models used in photonics to describe the optical properties of various materials, are based on a zero temperature approach. Although this is perfectly fine (and enough) for most materials, since the wavelength at which non-equilibrium effects are starting to become non-negligible is either in the far IR or the deep UV, this is not the case for 2D materials, which can be easily be driven out of equilibrium in the wavelength range of interest for different applications (typically ranging from THz to visible). In this case, traditional, zero-temperature-based, methods would fail to correctly describe the influence of out-of-equilibrium effects onto the optical properties of 2D materials, while PIMC, on the other hand, are able to exactly account for these contributions, as they naturally take these into account.

The second example deals with a slightly more complicated system, namely molecular hydrogen $\text{H}_2$. Figure \ref{figure8} depicts the real (top) and imaginary (bottom) part of the dipole (left) and quadrupole (right) polarisabilities $\alpha_1(\omega)$ and $\alpha_2(\omega)$, respectively, for several different values of the ambient temperature $T$. A comparison with reference simulations obtained at $T=0$ K from Ref. \citenum{ref0K} is also shown (black dots in Fig. \ref{figure8}). 

Notice, in particular, the deviation from the zero temperature behaviour in the small frequency regime (insets in the top part of Fig. \ref{figure8}). This discrepancy is due to the fact, that in the case of molecules, the nuclear contribution (mainly nuclear vibrations induced by finite temperature) cannot be neglected, as it significantly changes the behaviour of the polarisability at low frequencies. This change amounts only to a shift of the value of $\operatorname{Re}\{\alpha_1(\omega)\}$ for the dipole polarisability (top let inset in Fig. \ref{figure8}), while it radically changes the form of $\operatorname{Re}\{\alpha_2(\omega)\}$ at low frequencies, for the case of quadrupole polarisability (top right inset in Fig. \ref{figure8}). Notice, moreover, as in this case temperature has also a significant effect in shaping the imaginary part of the quadrupole polarisability at very low frequencies (lower right inset in Fig. \ref{figure8}).

Although they show the optical properties of very simple systems, such as H, He and $\text{H}_2$, the examples presented in this section offer good insight on the capabilities and potential of using PIMC for calculating the optical properties of even more complicated systems. The large frequency range PIMC can span (from THz to the deep UV) represents an enormous advantage to test the current models for susceptibilities and polarisabilities of photonic materials, and could in particular provide tremendous insight on the nonlinear properties of exotic materials, such as epsilon-near-zero materials, for example, as it provides a single, \emph{ab-initio} method to calculate the optical response in an exact manner, without inserting any approximation or without introducing different models for different frequency regimes.

Besides that, the examples provided in this section also help the reader in getting a better grasp of the breadth of path integrals, and how they do not only represent an elegant, alternative formulation of quantum mechanics, but also how their use in atomic and molecular physics can be pivotal for the accurate description of the properties of matter.
\section{Path Integral for Classical and Quantum Fields}\label{sect5}
In the previous sections, we have seen how it is possible to describe the dynamics of a quantum particle (or a collection thereof) in terms of path integrals. In this section, instead, we extend this formalism to the case of quantum fields, where the building blocks for the path integral are not particles and all the trajectories they can possibly take to evolve from the initial to the final state, but rather all the possible configurations of a field, that serve the same purpose.

A quantum field can be seen as a continuous collection of harmonic oscillators, one for each point in space where the field is defined. As such, then the results presented for the harmonic oscillator in Sect. \ref{harmonicOscillator} can serve as guidance to better understand this section, and, more generally, the properties of a QFT. Some of the language and tools used here, however, are slightly different than those introduced above, and it is worthwhile to introduce some key concepts and methods proper of QFT before specialising our attention to the electromagnetic field. To do that in the most beginner-friendly manner, we use the case of a scalar field as a guidance to construct the path integral, and show how interactions can be introduced in such a theory, and what is the consequence of doing that.  Then, we will qualitatively and intuitively extend these results to the case of a vector field, and use those results to look at the linear and nonlinear dynamics of the electromagnetic field in an arbitrary medium.

This section follows closely the notation and line of reasoning of Ref. \citenum{srednicki}. For a more formal discussion of path integrals in field theory, we invite the interested reader to consult Refs. \citenum{srednicki, brown, das, rivers}.  For the sake of simplicity, natural units $c=\hbar=m_e=1$ are implicitly assumed. If needed, these constants can be easily reinserted in the final results of calculations, by means of simple dimensional analysis. Moreover, summation over repeated indices (Einstein summation convention) is also understood.
\subsection{Path Integral for a Scalar Field}
To start with, let us then consider a free scalar field $\phi(x)$ (where $x$ is a shorthand for $\{\vett{r},t\}$), whose Lagrangian density is given by
\beq\label{LagrangianFree}
\mathcal{L}_0(\phi(x))=\frac{1}{2}\eta^{\mu\nu}\partial_{\mu}\phi\partial_{\nu}\phi(x)-\frac{1}{2}m^2\phi(x)^2,
\eeq
where $\eta^{\mu\nu}=\text{diag}(-1,1,1,1)$ is the Minkowski flat space metric tensor \cite{srednicki}, and the subscript $0$ indicated that the Lagrangian is that of the free field, \emph{i.e.}, $\phi(x)$ is not interacting with either itself, or its environment. 

Before going any further, it is instructive to discuss the physical meaning of the parameter $m$ appearing in the above equation. In QFT, it is customary to give $m$ the meaning of mass of the scalar field $\phi$, as it could represent a scalar massive boson such as the Higgs boson \cite{kaku}, or even a scalar fermion in more complicated theories involving dark matter \cite{darkMatter}. In optics, on the other hand, we frequently refer to the case $m=0$, as the electromagnetic field is composed by photons, which are massless bosons \cite{kaku}. However, when propagating inside a medium, the electromagnetic field acquires a mass, proportional to the refractive index of the medium \cite{luneburg}. This can be easily proven by recalling that the dispersion relation for a monochromatic electromagnetic field propagating inside a homogeneous medium characterised by a refractive index $n$ is given by \cite{jackson} $k^2=\omega^2n^2$. The dispersion relation for a massive scalar field $\phi$, on the other hand, can be readily calculated from its equation of motion (i.e., the Klein-Gordon equation \cite{srednicki}), which can be readily derived from Eq. \eqref{LagrangianFree} using the Euler-Lagrange equations for fields, i.e.,
\beq
\partial_{\mu}\left[\frac{\partial\mathcal{L}_0}{\partial\left(\partial_{\mu}\phi\right)}\right]-\frac{\partial\mathcal{L}_0}{\partial\phi},
\eeq
which leads to $(\nabla^2-\partial_t^2-m^2)\phi=0$. If we now assume plane wave solutions for $\phi$, the dispersion relation for a massive scalar field can be written as $k^2=\omega^2-m^2$. Notice, now, that upon defining $m^2=\omega^2(1-n^2)$, the dispersion relation for a massive scalar field $\phi$ matches that of an electromagnetic field inside a homogeneous medium. This allows us to identify the mass (squared) of the scalar field $\phi$ with $m^2=1-n^2=1-\varepsilon=\chi$, i.e., the medium's susceptibility \cite{jackson}. In this sense, therefore, while a massless scalar field describes a (scalar \footnote{A scalar electromagnetic field is defined as an electromagnetic wave whose polarisation is the same in any point of its wavefront, and always remains the same during propagation. If the interaction of a scalar electromagnetic field with a system causes its state of polarisation to vary, this variation will happen in the same manner in every point of its wavefront. A good example of a scalar electromagnetic field is an optical beam, which is typically written as $\vett{E}(\vett{r},t)=\psi(\vett{r},t)\,\uvett{u}$, where $\uvett{u}$ is a constant (unitary) transverse vector (with respect to the rpopagation direction of the field), which describes the field polarisation. The dynamics of an optical beams are then fully determined by $\psi(\vett{r},t)$, which, in general,  is a solution of the scalar wave equation.}) electromagnetic field in vacuum, a massive one describes instead the (scalar) electromagnetic field in a homogeneous medium.

We now proceed to the definition of path integral for the scalar field $\phi$. To do so, let us notice that the Lagrangian density in Eq. \eqref{LagrangianFree} has a similar structure to that of a harmonic oscillator, as given in Eq. \eqref{eq:h_o_lagrangian}. The first term in Eq. \eqref{LagrangianFree}, in fact, is quadratic in the derivative of the field, and can be associated with its ``kinetic energy", while the second term, quadratic in the field, is its ``potential energy". Analogously to what we have done in the previous sections, we can introduce the action of the field as
\beq\label{actionQFTscalar}
S_0(\phi(x),J)=\int\,d^4x\,\left[\mathcal{L}_0(\phi(x))+J(x)\phi(x)\right],
\eeq
where $d^4x=dt\,d^3r$ and we have introduced the source term $J(x)$ (analogous to the external force acting on a harmonic oscillator \cite{hibbs}), which, for the sake of this tutorial, can be imagined to simply be a convenient mathematical tool, introduced to make calculations easier. The physical meaning of $J(x)$, as its name suggests, is to represent a source of the field. The $x$-dependence of the field $\phi(x)$ is from here henceforth, and if not specified otherwise, implicitly assumed, i.e., we define $\phi\equiv\phi(x)$, to make our notation less cumbersome.

We can then define the path integral for a free scalar field in the same manner as we did in Eq. \eqref{eq:pathIntegral}, thus obtaining
\beq\label{eq1}
Z_0(J)=\int\,\mathcal{D}\phi\,e^{iS_0[\phi,J]}
\eeq
where the functional measure $\mathcal{D}\phi(x)$ plays the same role as the sum over all possible trajectories $Dx(t)$ in Eq. \eqref{eq:pathIntegral}, with the conceptual difference, however, that in this case the sum (integral) is extended to all possible configurations of the function $\phi(x)$. Practically, this means that instead of slicing time into small intervals (as we did in the previous sections for the case of quantum particles), we divide the volume of spacetime $\mathcal{V}$  into $N$ elementary cubes, i.e., $\mathcal{V}=\bigcup_{k=1}^NV_k$, with  $V_k=\delta^4$, and assume that in each $V_k$ the field $\phi$ is constant, i.e., $\phi\simeq\phi(t_i,x_j,y_n,z_m)=\phi(x_k)\equiv\phi_k$ (this implies that since each field $\phi$ is a function of 4 variables, each taking $N$ possible values, this discretisation of spacetime results in having a total of $N^4$ fields). This then means, that the integration measure can be written as
\beq
\mathcal{D}\phi=\lim_{N\rightarrow\infty}\prod_{i,j,n,m=1}^Nd\phi(t_i,x_j,y_n,z_m)\equiv\lim_{N\rightarrow\infty}\prod_{k=1}^{N^4}d\phi_k.
\eeq
In doing so, each field configuration $\phi_k$ becomes an independent integration variable, and the integral $\int\,\mathcal{D}\phi$ reduces to a usual Riemann/Lebesgue integral in $N^4$-dimensions.

The above definition of path integral is valid both for a classical, as well as a quantum field. For the latter, the field $\phi(x)$ needs to be equipped with a suitable set of commutation relations, that define the quantum nature and structure of the algebra behind it. To do that, we can revert back to the analogy with the harmonic oscillator, and use it to find a straightforward definition of such quantities. First, we discretise the field $\phi(x)$ and associate a harmonic oscillator to each point in space. Then, for each oscillator, we introduce the usual commutation relations between position and momentum, \emph{i.e.}, $[q_a(t),p_b(t)]=i\delta_{ab}$, where $\delta_{ab}$ simply indicates that the oscillators are mutually decoupled. If we then take the continuum limit (to revert back to the field $\phi(x)$) the commutation relations become those of the field $\phi$ itself with its canonically conjugated momentum $\Pi=\partial\mathcal{L}/\partial(\partial_{t}\phi)$, namely
\bseq
\begin{align}
\left[\phi(\vett{r},t),\Pi(\vett{r}',t)\right]&=i\,\delta(\vett{r}-\vett{r}'),\\
\left[\phi(\vett{r},t),\phi(\vett{r}',t)\right]&=\left[\Pi(\vett{r},t),\Pi(\vett{r}',t)\right]=0.
\end{align}
\eseq

Equation \eqref{eq1}, moreover, has implicitly been defined such that the time interval used for the time integration in the action spans the whole real axis, \emph{i.e.}, $-\infty<t<\infty$.  Moreover, we require that $Z_0(J)$ is normalised in such a way that $Z_0(0)=1$. 

For a classical field, $Z_0(J)$ represents its partition function $Z_0(J)$, from which, in analogy with statistical mechanics, it is possible to derive all the relevant quantities of the classical field, such as energy, momentum, entropy, and Green's function. 

For a quantum field, it represents the vacuum-to-vacuum correlation function, which allows to calculate transition probabilities and correlation functions of field operators with respect to the ground state (or vacuum), which is typically a stable and well-defined state of the QFT at-hand. In other terms, 
$Z_0(J)$ for a quantum field only describes a QFT at equilibrium (\emph{i.e.}, at zero temperature). To deal with non-equilibrium systems, a further step in the definition of the path integral is needed, which is beyond the scope of this tutorial. The interested reader is then referred to Ref. \citenum{calzetta} for further information about non-equilibrium QFTs.
\subsection{The Propagator for a Free Scalar Field}
To get familiar with the formalism of QFT, in this section we calculate the propagator for the free scalar field, which constitutes, to a certain extent, the QFT analogue of the free quantum particle described in Sect. \ref{freeParticle}. To this aim,  let us first introduce the Fourier transform of the field as
\bseq
\begin{align}
\varphi(k)&=\int\,d^4x\,\phi(x)\,e^{-ikx},\\
\phi(x)&=\int\,\frac{d^4k}{(2\pi)^4}\,\varphi(k)\,e^{ikx},
\end{align}
\eseq
where $kx=\vett{k}\cdot\vett{r}-\omega t$. If we substitute the expression of $\phi$ above into Eq. \eqref{actionQFTscalar} we arrive, after some straightforward calculations, at the following expression for the action in Fourier space
\barr
S_0&=&\int\,\frac{d^4k}{(2\pi)^4}\,\Big[-\frac{1}{2}\varphi(k)\left(k^2+m^2\right)\varphi(-k)\nonumber\\
&+&\tilde{J}(k)\varphi(-k)+\tilde{J}(-k)\varphi(k)\Big],
\earr
where $k^2=|\vett{k}|^2-\omega^2$, and $\tilde{J}(k)$ is the Fourier transform of $J(x)$. The trick to use here for calculating the path integral \eqref{eq1} with the action above is the same as the one we employed in Sect. \ref{freeParticle}, namely we need to discretise the integral appearing in the action $S_0$, and consider a discrete set of field configurations, such that $\int\,\mathcal{D}\phi$ can be written as a product of integrals. If we do so, we can then use the fact that the integrand above is Gaussian in $\varphi(k)$ and repeatedly apply Gaussian integration to it to get the final result.

Alternatively, we can operate a change of path integration, by performing the change of variables $\chi(k)=\varphi(k)-\tilde{J}(k)/(k^2+m^2)$, such that $\mathcal{D}\chi=\mathcal{D}\phi$, to isolate, in the action above, a term that only depends on the field $\chi$, from a term independent from it, \emph{i.e.},
\barr
S_0&=&\frac{1}{2}\int\,d^4k\Big[-\chi(k)(k^2+m^2-i\varepsilon)\chi(-k)\Big]\nonumber\\
&+&\frac{1}{2}\int\,d^4k\,\frac{\tilde{J}(k)\tilde{J}(-k)}{k^2+m^2}.
\earr
If we now substitute the expression above into Eq. \eqref{eq1}, together with $\mathcal{D}\chi=\mathcal{D}\phi$ from the change of field variables, we obtain the following result
\barr
Z_0(J)&=&\exp{\left[\frac{i}{2}\,\int\,\frac{d^4k}{(2\pi)^4}\,\frac{\tilde{J}(k)\tilde{J}(-k)}{k^2+m^2}\right]}\int\,\mathcal{D}\chi\,\exp{\Big\{}\nonumber\\
&-&i \int \frac{d^4k}{(2\pi)^4}\left[\chi(-k)\frac{k^2+m^2}{2}\chi(k)\right]\Big\}.
\earr
Notice, how the first term has been brought out of the path integral, as it is independent on $\chi$. Notice moreover, that the path integral can now be evaluated, using a generalisation of the Gaussian integration formula (see Appendix \ref{appendixA}), and it essentially amounts to a normalisation factor (which, for convenience, will be henceforth neglected). We then arrive at the final result for the path integral, which can be written in a very similar form to that of a harmonic oscillator, as
\barr\label{eq2}
Z_0(J)&=&\exp{\left[\frac{i}{2}\,\int\,\frac{d^4k}{(2\pi)^4}\,\frac{\tilde{J}(k)\tilde{J}(-k)}{k^2+m^2-i\epsilon}\right]}\nonumber\\
&=&\exp{\left[\frac{i}{2}\,\int\,d^4x\,d^4y\,J(x)G(x-y)J(y)\right]},
\earr
where 
\beq\label{greenScalar}
G(x-y)=\int\,\frac{d^4k}{(2\pi)^4}\,\frac{e^{ik(x-y)}}{k^2+m^2-i\epsilon},
\eeq
and a small imaginary part $\varepsilon\ll 1$ has been introduced to regularise the $\omega$-integral and allow its calculation as a contour integral in the complex plane \cite{byronFuller}. The quantity $G(x-y)$ is the propagator (Green's function) for the free scalar field, which solves the correspondent equation of motion (derived from the Euler-Lagrange equations of the field), \emph{i.e.},
\beq
\left(\nabla^2-\partial_t^2+m^2\right)G(x-y)=\delta(x-y).
\eeq
The Green's function can also be interpreted as the vacuum-to-vacuum two-point correlation function between the field at point $x$ and the field at point $y$ as follows \cite{srednicki}
\beq\label{eq3}
G(x-y)=i\expectation{0}{\hat{T}\phi(y)\phi(x)}{0},
\eeq
where $\hat{T}$ is the usual time-ordering operator. The two-point correlation function is defined in terms of path integrals as follows
\beq
\expectation{0}{\hat{T}\phi(x)\phi(y)}{0}=\int\,\mathcal{D}\phi\,\phi(x)\,\phi(y)\,\exp{\left[iS_0(\phi,J)\right]}.
\eeq
From this definition, by noticing that, in analogy with the derivative of the exponential of a function we can write \cite{kleinert}
\beq\label{trick}
\phi(y)e^{i\,S_0(\phi,J)}=-i\frac{\delta}{\delta J(y)}e^{i\,S_0(\phi,J)},
\eeq 
where the symbol $\delta/\delta J(y)$ stands for the functional derivative, and it is a generalisation of the concept of derivative to functionals \cite{kleinert} defined by the operative relation
\beq\label{functionalDerivative}
\frac{\delta f(x)}{\delta f(y)}=\delta(x-y).
\eeq
With this in mind, then, we can formally introduce in Eq. \eqref{eq3}  the substitution $\phi(x)\rightarrow -i\delta/\delta J(x)$, thus obtaining an alternative definition for the propagator of a free field as
\barr\label{eq4}
\expectation{0}{\hat{T}\phi(y)\phi(x)}{0}&=&\int\,\mathcal{D}\phi\,\phi(y)\,\phi(x)\,e^{iS_0[\phi,J]}\nonumber\\
&=&\left(\frac{1}{i}\right)^2\frac{\delta^2}{\delta J(y)\delta J(x)}\braket{0}{0}_J\nonumber\\
&=&-\left\{\frac{\delta^2}{\delta J(y)\delta J(x)}Z_0(J)\right\}\Bigg|_{J=0}.
\earr
\section{Nonlinear interactions Through Path Integrals}\label{sect6}
The results above are suitable for describing the free propagation of a field, such as, for example, the electromagnetic field in either free space (\emph{i.e.}, vacuum), or inside a homogeneous or inhomogeneous medium. However, neither the Lagrangian \eqref{LagrangianFree} nor the path integral \eqref{eq1} take into account the possible self-interactions of the field, or any other form of interaction (such as the field-environment interaction, for example). In general, field interactions are very important, because they allow us to gain valuable insight on how the field behaves in different environments. The self-interaction of a field with itself, in particular, plays a big role in photonics, where it is simply known as nonlinear optics. If we then want to use path integrals to describe quantum nonlinear optics, we need a way to include the effect of interactions in our theory. In QFT this is typically done by adding to the Lagrangian density a so-called interaction term, \emph{i.e.},
\beq\label{intLag}
\mathcal{L}=\mathcal{L}_0+\mathcal{L}_{int},
\eeq
where  $\mathcal{L}_{int}$ contains information about the nature of all the interactions taking place in the considered system. For the sake of simplicity of exposition, and because they play an essential role in nonlinear optics, let us focus our attention on the case of a self-interacting field, where the interaction part is polynomial in the field $\phi$ and therefore the interaction Lagrangian can be written as \cite{landau}
\beq\label{intLagAlt}
\mathcal{L}_{int}=\frac{\lambda}{n!}\phi^n,
\eeq
where $\lambda$ is a suitable coupling constant describing the strength of the interaction. Notice, that the first nontrivial interaction is given by $n=3$, as any quadratic term in the Lagrangian density can be absorbed in the free Lagrangian as a ``shift in the potential energy". 

Inserting Eq. \eqref{intLag} into Eq. \eqref{eq1} gives the formal expression of the path integral for the interacting theory, \emph{i.e.},
\beq\label{partitionInt}
Z(J)=\int\,\mathcal{D}\phi\,\exp{\left\{i\,\int\,d^4x\,\left[\mathcal{L}_0+\mathcal{L}_{int}+J\phi\right]\right\}}.
\eeq
In general, $Z(J)$ cannot be evaluated analytically, because the integral at the exponent is intrinsically non-Gaussian, due to the presence of the $n>2$ polynomial terms provided by $\mathcal{L}_{int}$. However, we can use the following argument, to write $Z(J)$ in a form that can be treated with perturbation theory. First, notice that the interaction Lagrangian $\mathcal{L}_{int}$ as defined in Eq. \eqref{intLagAlt} is a polynomial of order $n$ in $\phi$. Formally, then, if we employ Eq. \eqref{trick}, we can say that
\beq\label{trick2}
\mathcal{L}_{int}(\phi)\hspace{1mm}\rightarrow\hspace{1mm}\mathcal{L}_{int}\left(\frac{1}{i}\frac{\delta}{\delta J}\right)
\eeq
holds. 

Now, thanks to Eq. \eqref{trick2}, the interaction term $\mathcal{L}_{int}$ does not depend anymore on the path integral variable $\phi$, because we have formally substituted each occurrence of $\phi$ in it with the correspondent derivative with respect to $J$. This allows us to bring this term outside the path integral, and therefore rewrite Eq. \eqref{partitionInt} as
\beq\label{partInt}
Z(J)=\exp{\left[i\,\int\,d^4x\,\mathcal{L}_{int}\left(\frac{1}{i}\frac{\delta}{\delta J}\right)\right]}Z_0(J),
\eeq
where $Z_0(J)$ is given by Eq. \eqref{eq1}. 

If we now assume that the interaction is ``small enough'' (with respect to a certain energy scale, typical of the problem at-hand) that we can treat it perturbatively, \emph{i.e.}, if we set $\lambda\ll 1$ we can expand the exponential in a power series in $\lambda$ (truncated at the desired perturbation order) and write the path integral for the interacting theory in the following manner
\barr\label{Zpert}
Z(J) &=& \sum\limits_{V=0}^{\infty}\frac{1}{V!}\left[\frac{i\lambda}{n!}\int\,d^4x\,\left(\frac{1}{i}\frac{\delta}{\delta J(x)}\right)^n \right]^V\nonumber\\
&\times&\sum\limits_{P=0}^{\infty}\frac{1}{P!}\left[\frac{i}{2}\int\,d^4y\,d^4z\,J(y)G(y-z)J(z) \right]^P,
\earr
where we have also written $Z_0(J)$ in terms of its power series expansion, for a reason that will become clear in a moment. 

 The expression above allow us to treat the path integral for the interacting theory in a perturbative manner, up to an arbitrary perturbation order in $\lambda$. To choose the appropriate one, in fact, one needs only to truncate the summation in $V$ at the desired order, so only terms proportional to $\lambda^V$ (or lower powers) will appear in the perturbative series expansion.
\subsection{Feynman Diagrams}
In general, calculating Eq. \eqref{Zpert} analytically is not possible, and even writing down the terms contributing to $Z(J)$ at a given perturbation order might be cumbersome. In fact, at order $\lambda^V$ in Eq. \eqref{Zpert}, $Z(J)$ contains $nV$ functional derivatives through the term $(\delta/\delta J)^n$, that acts on the $2P$ different source terms $J$ deriving from the $P-$th order expansion of $Z_0(J)$. This, in general, amounts to $(2P)!/(2P-nV)!$ different ways the $nV$ functional derivatives can act on the $2P$ source terms. It is not difficult to see from this example, then, how the number of required integrals to calculate becomes very high and complex very easily, with increasing $V$ and $P$.

\begin{figure}[!t]
\begin{center}
\includegraphics[width=0.5\textwidth]{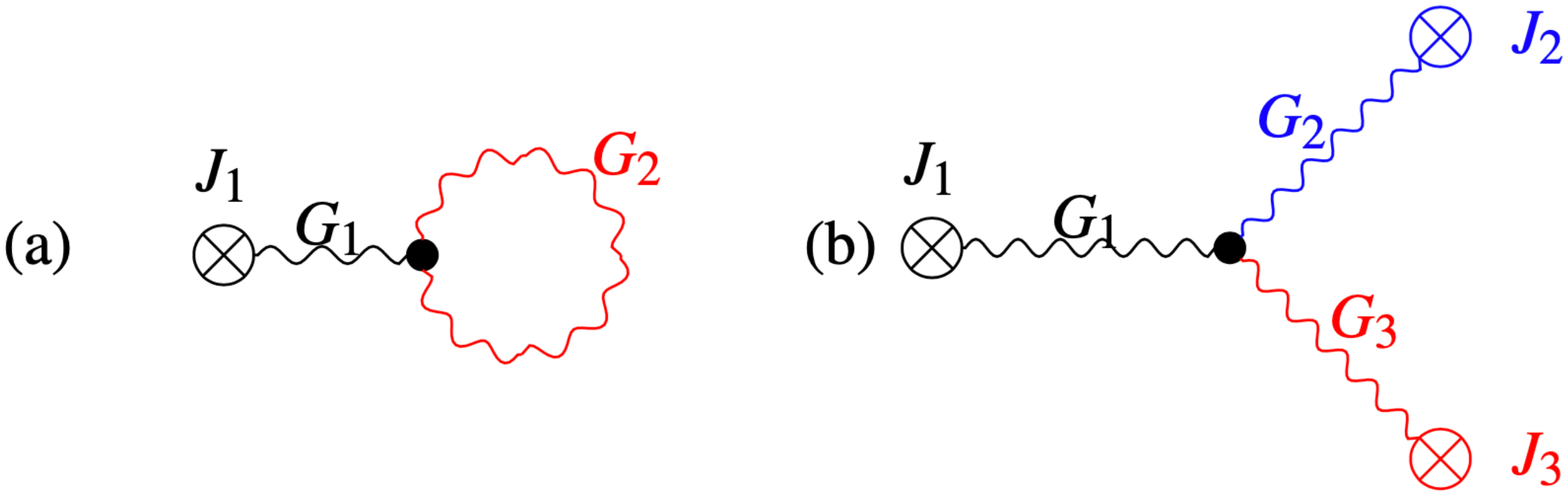}
\caption{(a) Relevant Feynman diagram for a scalar $\phi^3$ interacting QFT for the simple case $V=1$, $P=2$. As it can be seen, the diagram contains one source term corresponding to the current $J(z)$ in Eq. \eqref{firstOrder1}, and two propagators, $G_1$ (black, wiggly line) and $G_2$ (red, wiggly line), which is looping around the vertex. By exchanging the roles of $G_1$ and $G_2$, the diagram doesn't change, \emph{i.e.}, the value of $Z_1[J]$ in Eq. \eqref{firstOrder1} doesn't change. This diagram is known in QFT as a tadpole diagram. (b) Relevant Feynman diagram for the case  of $V=1$ and $P=3$. This diagram can be arranged in $3!=6$ different ways, by cycling through the currents $J_{1,2,3}$ and the propagators $G_1$ (black, wiggly line), $G_2$ (blue, wiggly line), and $G_3$ (red, wiggly line), without changing the value of the correspondent partition function. This diagram will be of particular importance for the case of the electromagnetic field, as its vector counterpart is the basic quantity that describes second-order nonlinear processes in arbitrary media (see Sect. \ref{sect8}). }
\label{QFT_fig1}
\end{center}
\end{figure}

An elegant and quite insightful way to keep track of all these quantities, and to focus on their physical meaning, rather than their complex mathematical expressions is given by the so-called Feynman diagrams. An example of them for the case of $n=3$ is reported in Fig. \ref{QFT_fig1}. 

To understand their meaning, a comprehensive set of rules connecting the various parts of these diagrams with the correspondent integrals and mathematical quantities appearing in Eq. \eqref{Zpert} can be defined, following for example Refs. \citenum{srednicki, das, rivers}, as follows:
\begin{itemize}
\item to represent a propagator $-i G(x-y)$ we use a wiggly line joining the initial ($x$) point  with the final point ($y$) of the free evolution of the field $\phi$;
\item to represent the interaction between fields, we use a so-called vertex, \emph{i.e.}, a filled dot, which joins $n$ lines, thus fulfilling the $\phi^n$ nature of the interaction;
\item to each vertex, moreover, we associate the quantity $i\,\lambda\,\int\,d^4x$, which contains the coupling constant $\lambda$ defining the interaction. In addition, we require that energy and momentum conservation at each vertex holds;
\item finally, to represent the presence of sources $J(x)$ we use a circle, placed at one end of a propagator, which is associated to the quantity $i\,\int\,d^4x\,J(x)$.
\end{itemize}

With these simple rules, the terms at different perturbation orders in Eq. \eqref{Zpert} can be conveniently expressed in terms of diagrams. To illustrate how this works, 
let us consider two examples:
\subsubsection{Feynman Diagram for the Free Propagator}
To obtain the free theory from Eq. \eqref{Zpert}, we simply set $V=0$ (which corresponds to no interaction). Following the rules stated above, the path integral $Z_0(J)$ can then be diagrammatically written as follows
\beq
Z_0(J) =
\begin{tikzpicture}
\begin{feynman}
\vertex [crossed dot] (a) {};
\vertex [above=0.4 cm  of a] (a0) {\(J(x_1)\)};
\vertex [crossed dot, right= of a] (b) {};
\vertex [above=0.4 cm  of b] (b0) {\(J(x_2)\)};
\diagram* {
(a) -- [boson] (b)
};
\end{feynman}
\end{tikzpicture}
+ \text{higher orders}.
\eeq
If we then want to calculate the propagator, we just have to simply take the functional derivative of the diagram above, \emph{i.e.}, remove the source terms $J(x_1)$ and $J(x_2)$ and replace them with simply the labels $x$ and $y$ for the initial and final points of the propagator. This corresponds, in terms of calculations, to the following
\barr
&&<\hat{T}\phi(x) \phi(y)>=\frac{\delta^2Z_0(J)}{\delta J(x)\delta J(y)}\Bigg|_{J=0}\nonumber\\
&=&\frac{\delta^2}{\delta J(x)\delta J(y)}\Bigg\{
\begin{tikzpicture}
\begin{feynman}
\vertex [crossed dot] (a) {};
\vertex [above=0.7 cm  of a] (a0) {\(J(x_1)\)};
\vertex [crossed dot, right= of a] (b) {};
\vertex [above=0.7 cm  of b] (b0) {\(J(x_2)\)};
\diagram* {
(a) -- [boson] (b)
};
\end{feynman}
\end{tikzpicture}
\Bigg\}\Bigg|_{J=0}\nonumber\\
&=&
\begin{tikzpicture}
\begin{feynman}
\vertex (a) {};
\vertex [above= 0.5 cm of a] (a0) { };
\vertex [right=of a] (b) {};
\vertex [ above= 0.5 cm of b] (b0) {};
\vertex [above left=0.2 cm of a] (at) {\((x)\)};
\vertex[above right=0.2 cm of b] (bt) {\((y)\)};

\diagram* {
(a) -- [boson] (b)
};
\end{feynman}
\end{tikzpicture}
\nonumber\\
&\equiv& G(x-y).
\earr
Notice, how the action functional derivatives $\delta/\delta J(x)$, namely to remove a source term from $Z_0(J)$ and replace $x_1$ ($x_2$) with $x$ ($y$), now appears more clear and intuitive.
\subsubsection{First Order Perturbation Diagram for $n=3$}
As a second example, let us consider a $n=3$ interacting theory with first order perturbation (\emph{i.e.}, $V=1$), for the simple cases of $P=2$, and $P=3$ as reported in Fig.  \ref{QFT_fig1} (a), and (b), respectively. 

The case $P=2$ represents the simplest perturbative calculation that we can do on $Z(J)$ for $n=3$, and corresponds to the well-known \emph{tadpole} diagram in QFT, \emph{i.e.}, the case where two propagators interact at a single vertex $V$ [Fig. \ref{QFT_fig1} (a)]. Notice, that this diagram contains $2P-3V=1$ external sources, and can be arranged in $2$ different ways, by exchanging the role of the two propagators $G_1$ and $G_2$. Both of these configurations, however, are algebraically equivalent, and correspond to the same term in the first order perturbative expansion $Z_1(J)$, which for this simple case can be written, following the rules above, as
\barr\label{firstOrder1}
Z_1(J)&=&-\frac{i}{4}\left(\frac{i\lambda}{3!}\right)\int\,d^4x\,\left(\frac{\delta}{\delta J(x)}\right)^3\nonumber\\
&\times&\left[\int\,d^4y\,d^4z\,J(y)\,G(y-z)\,J(z)\right]^2\nonumber\\
&=&\frac{\lambda}{2}\,\int\,d^4x\,d^4z\,J(z)\,G_1(z-x)\,G_2(x-x),
\earr
where the last term $G_2(x-x)$ accounts for the loop appearing in Fig. \ref{QFT_fig1} (a). The factor $1/2$ in front of the integral compensates for the $2$ equivalent ways diagram (a) can be written. The subscript $1$ and $2$ in the propagators have been added to make it easier to recognise the correspondent terms in Fig. \ref{QFT_fig1}, and have no physical meaning.

For the case of three propagators meeting at a vertex ($P=3$ and $V=1$), the actual computation of $Z_1(J)$ would be rather cumbersome, as it requires to perform three successive functional derivatives on an object that has six current terms, and three propagators. If we however rely on the correspondent Feynman diagram, depicted in  Fig. \ref{QFT_fig1} (b), we can readily write the explicit expression of $Z_1(J)$ as
\barr
Z_1(J)&=&\frac{\lambda}{6}\int\,d^4xd^4yd^4zd^4x_1\,J(x)\,G(x-x_1)\Big[\nonumber\\
&&G(x_1-y)J(y)+G(x-z)J(z)\Big].
\earr
The diagram in Fig. \ref{QFT_fig1} can be arranged in $3!=6$ different ways,  by interchanging the roles of the three source terms $J_k$. All these different arrangements of the diagrams, however, are algebraically equivalent, and correspond to the same form of the first-order perturbation $Z_1(J)$ written above. The factor $1/6$ there, moreover, accounts for this symmetry. 
\section{Generalisation to Vector Fields}\label{sect7}
The results obtained in the previous section for a scalar field are valid, if properly generalised, also for vector fields, and, in particular, for the electromagnetic field. The correct way to generalise those results to vector fields, however, is out of the scope of this tutorial. The interested reader is referred to Refs. \citenum{srednicki, das, rivers, kleinert, hibbs, brown, maggiore} for further and more formal details. In what follows, we just give an intuitive and qualitative generalisation.
\subsection{Path Integral for Free Vector Fields}
For a vector field, in fact, $Z_0(J)$ can be qualitatively written in a similar manner to Eq. \eqref{eq2}. To do that, we assume that the source terms are now vectorial in nature, i.e, $J\rightarrow J_{\mu}$, which makes the Green's function become a two-index object, \emph{i.e.}, we need to formally make the substitution $G(x-y)\rightarrow G_{\mu\nu}(x-y)$. This allows us to immediately extend the result of Eq. \eqref{eq2} to the case of a vector field, by simply introducing the dyadic Green's function 
\beq
G_{\mu\nu}(x-y)=\left(\delta_{\mu\nu}+\frac{1}{k^2}\partial_{\mu}\otimes\partial_{\nu}\right)G_0(\vett{r},\vett{r}';t,t'),
\eeq
where $G_0(\vett{r},\vett{r}';t,t')$ is the Green's function of the wave equation, \emph{i.e.}, 
\beq\label{eq86}
\left(\nabla^2-n^2\partial_t^2\right)G_0(\vett{r},\vett{r}';t,t')=-\delta(\vett{r}-\vett{r}')\delta(t-t'),
\eeq
where $n^2$ is a possibly space-dependent parameter describing the properties of the vector field. For the electromagnetic field, for example, $n^2$ is the refractive index profile of the medium the field is propagating through.

We can then write the partition function for the non-interacting vector field simply as
\beq\label{VectorZ0}
Z_0(J)=\exp{\left[\frac{i}{2}\,\int\,d^4x\,d^4y\,J_{\mu}(x)G_{\mu\nu}(x-y)J_{\nu}(y)\right]}.
\eeq
Using Eqs. \eqref{eq3} and \eqref{eq4} we can also write $G_{\mu\nu}(x-y)$ as a two-point correlation function as
\beq\label{greenVector}
G_{\mu\nu}(x-y)=-i\frac{\delta^2}{\delta J_{\mu}(y)J_{\nu}(x)}Z_0(J)\Bigg|_{\vett{J}=0},
\eeq
where the shorthand $\vett{J}=0$ means that all the components of the current need to be set to zero after the calculation. We will make use of this result in the next section, for deriving the propagator of the effective free electromagnetic field in an arbitrary medium.
\subsection{Path Integral for Interacting Vector Fields}
Analogously, Eq. \eqref{Zpert} can be qualitatively extended to the case of interacting vector fields by promoting source terms to vector fields (\emph{i.e.}, $J\rightarrow J_{\mu}$) , and the propagators to tensor fields (\emph{i.e.},  $G(x-y)\rightarrow G_{\mu\nu}(x-y)$), thus obtaining
\barr\label{VectorZ1}
Z(J) &=& \sum\limits_{V=0}^{\infty}\frac{1}{V!}\left[\frac{i\,\lambda_{\{\alpha\}}}{n!}\int\,d^4x\,\left(\frac{1}{i}\frac{\delta}{\delta J_{\alpha}(x)}\right)^n \right]^V\nonumber\\
&\times&\sum\limits_{P=0}^{\infty}\frac{1}{P!}\left[\frac{i}{2}\int\,d^4y\,d^4z\,\vett{J}(y)\overleftrightarrow{\vett{G}}(y-z)\vett{J}(z) \right]^P,
\earr
where $\lambda_{\{\alpha\}}$ represents a rank-$n$ tensor, whose indices are contracting the $n$ indices appearing in the functional derivatives and $\vett{J}(y)\overleftrightarrow{\vett{G}}(y-z)\vett{J}(z)=J_{\mu}(y)G_{\mu\nu}(y-z)J_{\nu}(z)$. In general, moreover, 
 $\lambda_{\{\alpha\}}$ could also be space-dependent. This result will be used in the next section as a starting point to analyse nonlinear quantum effects for photons propagating in arbitrary media.
\section{Path Integral Formalism for Classical and Quantum Electrodynamics in Arbitrary Media}\label{sect8}
We now use the formalism introduced above for describing the propagation and interaction properties of an electromagnetic field in a dispersive medium of arbitrary geometry. The content of this section follows the line of reasoning of Refs. \citenum{bechler, ornigotti2019}, and it is reported here with a higher level of detail, in order to facilitate the reader in the process of adapting the formalism developed in the previous sections to the case of the electromagnetic field.

Let us then start by making a series of assumptions to construct a suitable Lagrangian density, which is the starting point for constructing the path integral. First, we choose to work in the so-called Weyl gauge  \cite{weylGauge} where the scalar potential is zero (which corresponds to assuming that the medium we consider contains no free carriers). This, practically, corresponds to the usual assumption that both the electric and magnetic fields are fully determined by the vector potential, \emph{i.e.}, $\vett{E}=-\partial\vett{A}/\partial t$ and $\vett{B}=\nabla\times\vett{A}$. Then, we assume that light-matter interaction can be fully described within the framework of the dipole approximation \cite{vogelWelsch}. 

We can then describe the evolution of an electromagnetic field in a dispersive medium with arbitrary shape in terms of the Huttner-Barnett Lagrangian density \cite{huttnerBarnett}
\barr\label{LagHB}
\mathcal{L}_{HB}(\vett{r},t)&=&\frac{\varepsilon_0}{2}\vett{E}^2-\frac{1}{2\mu_0}\vett{B}^2+\frac{g(x)}{2\varepsilon_0\omega_0^2\beta(x)}\left(\dot{\vett{P}}^2-\omega_0^2\vett{P}^2\right)\nonumber\\
&+&g(x)\int\,d\omega\,\left[\frac{\rho}{2}\dot{\vett{F}}^2(\omega)-\frac{\rho\omega^2}{2}\vett{F}^2(\omega)\right]\nonumber\\
&-&g(x)\left[\vett{E}\cdot\vett{P}+\int\,d\omega\,f(\omega)\,\vett{P}\cdot\dot{\vett{F}}(\omega)\right].
\earr
Let us first understand what is the physical meaning of all the elements in the expression. $\mathcal{L}_{HB}$ describes the interaction of three fields: the electromagnetic field (represented by $\vett{E}\equiv\vett{E}(\vett{r},t)$ and $\vett{B}\equiv\vett{B}(\vett{r},t)$), the matter (polarisation) field $\vett{P}\equiv\vett{P}(\vett{r},t)$, and the reservoir field $\vett{F}(\omega)\equiv\vett{F}(\vett{r},t;\omega)$, which essentially models all the possible decay channels present in the material, that could affect its optical properties.

The first two lines of Eq. \eqref{LagHB} are the free Lagrangian densities of these three fields, respectively. 

The third line, instead, contains all the relevant interactions, with the first term being the traditional dipole interaction term, and the second one representing the polarisation-reservoir interaction, that models losses in the system. 

The geometry of the system under investigation is contained in the shape factor $g(x)$, which equals $1$ inside the region of interest, and zero otherwise. 

To complete the picture, the medium is characterised by a resonant frequency $\omega_0$, static polarisability $\beta(x)$, mass density (per unit frequency) $\rho(x)$, and spectral coupling $f(\omega)$, which accounts for the frequency distribution of the various decay channels of the medium. The dot in Eq. \eqref{LagHB}, moreover, represents derivation with respect to time. 

Notice, that although $\mathcal{L}_{HB}(\vett{r},t)$ contains interacting terms, from the point of view of the electromagnetic field, it is still considered as a free Lagrangian density, as no terms proportional to powers of $\vett{E}$ or $\vett{B}$ (or, equivalently, the vector potential $\vett{A}$) higher than two is present. 

We can then substitute Eq. \eqref{LagHB} into Eq. \eqref{eq1} to obtain the full path integral of the Huttner-Barnett model as
\barr\label{partHB}
Z(J,J_P,J_F)&=&\int\,\mathcal{D}\vett{A}\mathcal{D}\vett{P}\mathcal{D}\vett{F}\,\exp{\Big\{i\,S_{HB}}\nonumber\\
&+&i\int\,d^4x\,\Big[\vett{J}\cdot\vett{A}+\vett{J}_P\cdot\vett{P}+\vett{J}_F\cdot\vett{F}\Big]\Big\},
\earr
where $\vett{J}$ is the source term for the electromagnetic field, $\vett{J}_P$ the source term for the matter polarisation field, $\vett{J}_F$ the source term for the reservoir field, and $S_{HB}=\int\,d^4x\,\mathcal{L}_{HB}$ is the Huttner-Barnett action.

Written in this way, $Z(J,J_P,J_F)$ allows one to describe, in a fully quantum manner, not only the dynamics of photons (through the current $J$), but also those of polaritons (through the current $J_P$), \emph{i.e.}, quantum excitations of the matter polarisation field, and to also account for the quantum features of the loss reservoir, through the current $J_F$. However, for the purposes of this tutorial, we shall focus our attention on photons only, and therefore we set $\vett{J}_P=0=\vett{J}_F$, leaving $\vett{J}$ as the only free parameter. In this limit, the path integrals for $\vett{P}$ and 
$\vett{F}(\omega)$ can be performed explicitly, resulting in an effective path integral for the electromagnetic field propagating in a medium macroscopically described by an effective dielectric constant. We call the system electromagnetic field plus effective medium \emph{dressed} electromagnetic field.

The full procedure on how to do that can be found in Refs. \citenum{bechler,ornigotti2019}, and it is summarised in Appendix \ref{appendixA}, for completeness.
\subsection{Dressed Free Electromagnetic Field}\label{dressedE}
After performing the path integration with respect to $\vett{P}$ and $\vett{F}(\omega)$ as illustrated in Appendix \ref{appendixA}, the partition function in Eq. \eqref{partHB} reduces to an effective partition function $Z_{eff}(J)$ given by
\beq\label{partDem}
Z_{eff}(J)=\int\,\mathcal{D}\,\vett{A}\exp{\left[i\,S_{eff}(A)+i\int\,d^4x\,\vett{J}\cdot\vett{A}\right]},
\eeq
where the effective action $S_{eff}(A)$ has the following explicit form
\barr\label{effectiveAction}
S_{eff}(A)&=&\int\,d^4x\,\left[\frac{\varepsilon_0}{2}\dot{\vett{A}}^2-\frac{1}{2\mu_0}\left(\nabla\times\vett{A}\right)^2\right]\nonumber\\
&+&\frac{1}{2}\int\,dt\,d\tau\,d^3r\,g(\vett{r})\dot{\vett{A}}(t)\,\boldsymbol\Gamma(t-\tau)\,\dot{\vett{A}}(\tau),
\earr
where the tensor quantity $\boldsymbol\Gamma(\vett{r},t-\tau)$ is directly connected with the dielectric function of the effective medium the ``dressed" electromagnetic field is propagating into. An exhaustive discussion on the physical meaning and origin of $\Gamma(\vett{r},t-\tau)$ is given in Ref. \citenum{bechler}.

If we substitute in the action above the expressions of the electric and magnetic fields in terms of the vector potential $\vett{A}$, \emph{i.e.}, $\vett{E}=-\partial_t\vett{A}$ and $\vett{B}=\nabla\times\vett{A}$, we can see that $S_{eff}(A)$ is at most quadratic in $\vett{A}$ and $\partial_t\vett{A}$, meaning that, \emph{de-facto}, it describes the free evolution of the dressed field. This means, that the path integral in Eq. \eqref{partDem} can be reduced to a form similar to Eq. \eqref{VectorZ0}. In the present case, however, it is more convenient to write the path integral in Fourier space, as a function of the frequency $\omega$, rather than time. By doing this we then obtain
\beq\label{dressedPart}
\mathcal{Z}_0(J)=\exp{\left[\frac{i}{2}\int\,d\omega\,dR\,\vett{J}(\vett{r},\omega)\overleftrightarrow{\vett{G}}(\vett{R},\omega)\vett{J}(\vett{r}',\omega)\right]},
\eeq
where $dR=d^3rd^3r'$, $\vett{R}=\vett{r}-\vett{r}'$, and $\overleftrightarrow{\vett{G}}(\vett{R},\omega)$ is the dyadic Green's function of the dressed field in Fourier space, and it is a solution of the following equation of motion for the dressed field
\barr\label{dressedGreen}
&&\left[\left(-\delta_{\mu\alpha}\nabla^2+\frac{\partial^2}{\partial x_{\mu}\partial x_{\alpha}}\right)-\omega^2\varepsilon(\vett{r},\omega)\delta_{\mu\alpha}\right]G_{\alpha\nu}(\vett{r}-\vett{r}',\omega)\nonumber\\
&=&\mu_0\delta_{\mu\nu}\delta(\vett{r}-\vett{r}'),
\earr
where $\varepsilon(\vett{r},\omega)=1+g(\vett{r})\tilde{\Gamma}(\vett{r},\omega)/\varepsilon_0$ (with $\tilde{\Gamma}$ being the Fourier transform of $\Gamma$ appearing in Eq. \eqref{effectiveAction}). 

Alternatively, the Green's function can be also derived from Eq. \eqref{dressedPart} by using Eq. \eqref{greenVector} as the correlation function between the electromagnetic field at point $\vett{r}$ and $\vett{r}'$, \emph{i.e.},
\barr
G_{\mu\nu}(\vett{r}-\vett{r}',\omega)&=&\frac{\delta^2\mathcal{Z}_0[J]}{\delta J_{\mu}(\vett{r},\omega)\delta J_{\nu}(\vett{r}',\omega)}\Bigg|_{\vett{J}=0}\nonumber\\
&\equiv&\langle A_{\mu}(\vett{r},\omega)A_{\nu}(\vett{r}',\omega)\rangle.
\earr
\subsection{Nonlinear Quantum Electrodynamics}
In photonics, the description of nonlinear interactions of the electromagnetic field in a medium are often expressed in terms of the induced polarisation, which is, in general, expressed as a power series expansion upon the electric field \cite{boyd}, \emph{i.e.}, $\boldsymbol\Pi=\varepsilon_0(\chi_1\vett{E}+\chi_2\vett{E}^2+\chi_3\vett{E}^3+\cdots)$, where $\chi_n$ is the $n$-th order susceptibility tensor, containing the linear ($n=1$) and nonlinear ($n\geq 2$) response of the medium to the electric field. 

This form of self-interaction can be generated with an interaction Lagrangian similar to that of Eq. \eqref{intLagAlt}, \emph{i.e.},
\beq\label{intHB}
\mathcal{L}_{int}(A)=\sum_{n=3}^{\infty}\frac{1}{n!}\boldsymbol\chi_{n-1}(\vett{r},\omega)\cdot\vett{A}^n.
\eeq
The lowest order nonlinear effect described by the interaction Lagrangian above is that of the interaction of three electromagnetic fields (photons), \emph{i.e.}, a second-order nonlinear process \cite{boyd}. 

We can then write the path integral for the nonlinear interaction in analogy with Sect. \ref{sect6} and expand it perturbatively in powers of the various $\chi_n$. However, for the particular case of nonlinear optics, the magnitude of the various nonlinear effects gets progressively smaller, as $n$ increases, \emph{i.e.}, that $|\boldsymbol\chi_{n+1}|\ll|\boldsymbol\chi_n|$ holds for any $n$. 

If we now expand the path integral in powers of $\chi$ using Eq. \eqref{VectorZ1} as guidance, we can decompose the path integral in the following sum of three contributions
\beq
\mathcal{Z}(J)=\mathcal{Z}_0(J)+\sum_{k=2}^{\infty}\mathcal{Z}^{(k)}(J)+\mathcal{Z}_{cross}(J),
\eeq
where $\mathcal{Z}_0(J)$ is the free theory given by Eq. \eqref{dressedPart}, $\mathcal{Z}^{(k)}(J)$ accounts for the $k-$th order nonlinearity, \emph{i.e.}, it contains terms proportional to $\chi_k$ solely, and $\mathcal{Z}_{cross}(J)$ describes the combined interaction of several orders of nonlinearities (for example, it contains terms of the form $\chi_k\chi_n$, with $k\neq n$ etc.). Since $|\boldsymbol\chi_{n+1}|\ll|\boldsymbol\chi_n|$ holds, however, we can safely neglect this last term \cite{ornigotti2019}, as it contains terms that are $\mathcal{O}(|\boldsymbol\chi|^k)$. The explicit expression of the term $\mathcal{Z}^{(k)}(J)$ closely resembles that of the perturbative expansion for an interacting vector field theory, defined in Eq. \eqref{VectorZ1}, \emph{i.e.},
\barr\label{ZpertHB}
\mathcal{Z}^{(k)}(J)=\sum_{n=1}^{\infty}\left[\int d\omega\,d^3r\,\frac{\boldsymbol\chi_{k-1}}{(n!)^{1/n}}\cdot\left(\frac{\delta}{\delta\vett{J}}\right)^k\right]^n\mathcal{Z}_0(J).
\earr
This result can be used to calculate the cross section $\sigma^{(N)}(\vett{r}_1,\cdots,\vett{r}_N)$ of a given $N-$th order nonlinear event to take place, as simply the vacuum $N$-point correlation function of the electromagnetic field as
\beq
\sigma^{(N)}(\vett{r}_1,\cdots,\vett{r}_N)=\sigma_0\,\mathcal{A}(\vett{r}_1,\cdots,\vett{r}_N),
\eeq
where $\sigma_0$ is a suitable dimensional constant and 
\barr\label{crossN}
\mathcal{A}(\vett{r}_1,\cdots,\vett{r}_N)&=&\langle A_{\mu_1}(\vett{r}_1,\omega)\cdots\,A_{\mu_N}(\vett{r}_N,\omega)\rangle\nonumber\\
&=&\frac{\delta^N\mathcal{Z}(J)}{\delta J_{\mu_1}(\vett{r}_1,\omega)\cdots\delta J_{\mu_N}(\vett{r}_N,\omega)}\Big|_{\vett{J}=0}.
\earr
The cross section above can also be interpreted as the probability to generate $N$ photons through a $N-$th order nonlinear process. In this case, the probability is given simply by the modulus square of the quantity defined above, \emph{i.e.},
\beq\label{probN}
P_N(\vett{r}_1,\cdots,\vett{r}_N)=P_0\left|\mathcal{A}(\vett{r}_1,\cdots,\vett{r}_N)\right|^2,
\eeq
where, again, $P_0$ is a suitable dimensional constant that transforms $|\mathcal{A}|^2$ into a probability density.
\section{Applications of Path Integrals in Quantum Optics}\label{sect9}
We now present two examples on how this formalism can be useful to describe linear and nonlinear quantum optical phenomena in arbitrary media, hoping that this will serve the reader to get a better insight on the potential applications of path integrals in photonics. 

In this section we then present the solution, by means of path integrals, to two well-known problems: (1) the generation of spontaneous parametric down conversion (SPDC) in a lossy medium, and (2) the calculation of the spontaneous emission rate of an atom. Both examples make use of the dressed electromagnetic field introduced in Sect. \ref{dressedE}, for both the interacting and free case, respectively.
\subsection{Spontaneous Parametric Down Conversion in Lossy Media}\label{sectSPDC}
Let us consider only the term $n=3$ in Eq. \eqref{intHB}. The interaction Lagrangian then becomes
\beq\label{intSHG}
\mathcal{L}_{int}(A)=\frac{1}{3!}\boldsymbol\chi_{2}(\vett{r},\omega)\cdot\vett{A}^3.
\eeq
This interaction term, as it is well-known within the nonlinear optics community, is responsible for second-order nonlinear phenomena, such as second-harmonic generation (SHG), sum- and difference-frequency generation (SFG/DFG), and spontaneous parametric down conversion (SPDC) \cite{boyd}. As it can be seen, $\mathcal{L}_{int}$ involves the interaction of three fields, which are conventionally called \emph{pump}, \emph{signal} and \emph{idler}. A very common assumption (and experimental working condition) is to stimulate the aforementioned phenomena by means of a very bright pump beam, whose quantum properties are typically neglected, and whose intensity (\emph{i.e.}, number of photons) does not change significantly during the interaction \cite{drummond}. This approximation is called undepleted pump approximation. 

If we accept this approximation, we can rewrite the interaction Lagrangian above in a form that allows us to easily calculate the path integral. In doing so, essentially, we make use of the undepleted pump approximation to eliminate the pump field from the dynamics (\emph{i.e.}, we consider only the dynamics of the signal and idler photons), by defining an effective nonlinear susceptibility $\tilde{\chi}_{\mu\nu}=\chi_{\mu\nu\tau}A_{\tau}^{(p)}$, so that Eq. \eqref{intSHG} becomes \cite{ornigotti2019}
\beq
\mathcal{L}_{int}[A]=\tilde{\chi}_{\mu\nu}(\vett{r},\omega)A_{\mu}^{(s)}A_{\nu}^{(i)},
\eeq
where the superscripts $\{p,s,i\}$ stand for pump, signal and idler field, respectively (mimicking the usual notation of nonlinear optics).

Substituting the interaction Lagrangian above into Eq. \eqref{ZpertHB} with $k=3$ (the order of the nonlinearity in the Lagrangian) and $n=1$ (first order perturbation term), we can write down the explicit expression of the partition function for second-order nonlinear optical processes. In terms of Feynman diagrams, the path integral reads
\beq\label{ZSPDC}
\mathcal{Z}_1(J)=
\begin{gathered}
\begin{tikzpicture}
\begin{feynman}
\vertex[] (a0){};
%\vertex [crossed dot] (a0){};
%\vertex [above = 0.4 cm of a0] (c1) {\( J_{\mu}\)};
\vertex [dot, right= of a0] (a){};
\vertex [crossed dot, above right= of a] (w1){};
\vertex [crossed dot, below right= of a] (w2){};
\vertex [right = 0.5 cm of w1] (c2) {\( J_{\mu}\)};
\vertex [right = 0.5 cm of w2] (c3) {\( J_{\nu}\)};
\diagram* {
(a0) --  [scalar, ]  (a) --[boson, edge label= \(G^{(s)}_{\sigma\mu}\)]  (w1),
(a) --  [boson, edge label=\(G^{(i)}_{\sigma\nu}\)]  (w2),
};
\end{feynman}
\end{tikzpicture}
\end{gathered}
+\hspace{1mm} \text{permutations},
\eeq
where the permutations have to be understood as all the possible ways to order the three lines (and the correspondent labels) appearing in the Feynman diagram. The dashed line in the diagram above represents the dressed vacuum state $\ket{0}\equiv\ket{\alpha_{pump};0_{signal};0_{idler}}$, \emph{i.e.}, the effective quantum vacuum seen by the signal and idler photons.

From Eq. \eqref{ZSPDC}, we can calculate the probability for an SPDC event to occur using Eqs. \eqref{crossN} and \eqref{probN} with $N=2$, namely
\barr\label{pSPDC}
P_{SPDC}&=&\Big|\langle A^{(s)}_{\mu}(\vett{x},\omega)\,A^{(i)}_{\nu}(\vett{y},\omega)\rangle\Big|^2\nonumber\\
&=&\Big|\frac{\delta^2\mathcal{Z}(J)}{\delta J_{\mu}(\vett{x},\omega)\delta J_{\nu}(\vett{y},\omega)}\Big|^2\nonumber\\
&=&\Bigg|
\begin{gathered}
\begin{tikzpicture}
\begin{feynman}
\vertex[] (a0){};
%\vertex [crossed dot] (a0){};
%\vertex [above = 0.4 cm of a0] (c1) {\( J_{\mu}\)};
\vertex [dot, right= of a0] (a){};
\vertex [above right= of a] (w1){};
\vertex [below right= of a] (w2){};
%\vertex [right = 0.5 cm of w1] (c2) {\( J_{\mu}\)};
%\vertex [right = 0.5 cm of w2] (c3) {\( J_{\nu}\)};
\diagram* {
(a0) --  [scalar]  (a) --[boson, edge label= \(G_{signal}\)]  (w1),
(a) --  [boson, edge label=\(G_{idler}\)]  (w2),
};
\end{feynman}
\end{tikzpicture}
\end{gathered}
\Bigg|^2\nonumber\\
&=&\left|\mathcal{X}^{(2)}_{\mu\nu}(\vett{x}-\vett{y})\right|^2,
\earr
where $\mathcal{X}^{(2)}_{\mu\nu}(\vett{x}-\vett{y})$ is the biphoton propagator, whose explicit expression is given as
\beq
\mathcal{X}^{(2)}_{\mu\nu}(\vett{x}-\vett{y})=\int\,d^4z\,\tilde{\chi}_{\alpha\beta}(z)G_{\mu\alpha}(x-z,\omega_s)G_{\beta\nu}(z-y,\omega_i),
\eeq
where $\omega_{s,i}$ are the frequencies of the signal and idler fields, respectively. The probability of detecting an SPDC event is then completely determined by the propagators for the signal and idler field. As an explicit example of this, let us consider a lossy isotropic  1D medium of length $L$. Since the medium is isotropic, $\chi_2(\vett{r},\omega)\equiv\chi$. Moreover, let us assume that the pump beam can be modelled as a plane wave of frequency $\omega_p$ propagating along the $x$-direction. The effective nonlinear susceptibility then has the following explicit form 
\beq\label{chi1D}
\tilde{\chi}_{\alpha\beta}(x)=\chi A_p\exp{\left[i\left(k_px-\omega_p t\right)\right]}.
\eeq
To calculate the biphoton propagator, we need the explicit expression for the Green's function. For a lossy, isotropic, 1D material the dyadic Green's function can be calculated explicitly using Eq. \eqref{greenScalar} (with $m=0$, $k^2=\kappa^2-\omega^2$ and by using the residue theorem to perform the $\omega-$integral)  and has the following form \cite{byronFuller}
\barr\label{green1D}
G(x-y,\omega)&=&\frac{1}{2i\kappa(\omega)}\Big\{\Theta(x-y)\exp{\left[i\left(\kappa(\omega)(x-y)-\omega t\right)\right]}\nonumber\\
&+&\Theta(y-x)\exp{\left[i\left(\kappa(\omega)(x-y)-\omega t\right)\right]}\Big\},
\earr
where $\kappa(\omega)=k(\omega)+i\gamma(\omega)$ is the complex wave vector of the field in the medium. 

\begin{figure}[!t]
\begin{center}
\includegraphics[width=0.5\textwidth]{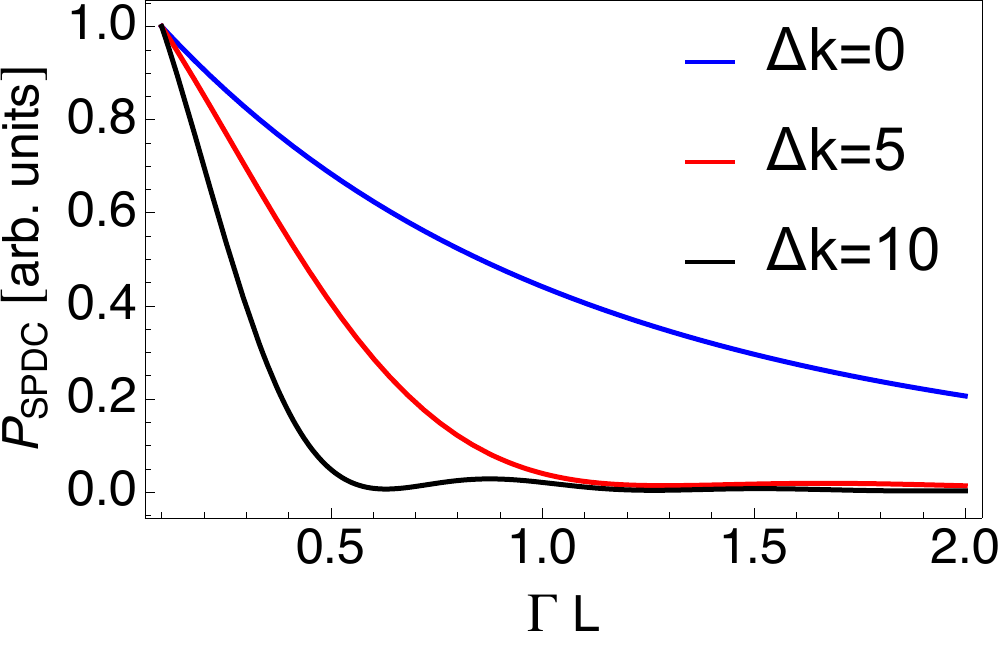}
\caption{Plot of the probability of observing a SPDC event in a 1D homogeneous lossy medium as given by Eq. \eqref{toFig}, as a function of the dimensionless parameter $x=\Gamma L$, and for different values of the phase mismatch parameter, namely $\Delta k=0$ (blue, solid line), corresponding to the case of perfect phase matching, $\Delta k=5$ (red, solid line), and $\Delta k=10$ (black, solid line).}
\label{figurePspdc}
\end{center}
\end{figure}

Substituting Eqs. \eqref{chi1D} and \eqref{green1D} into Eq. \eqref{pSPDC} we obtain
\barr
P_{SPDC}&=&\Bigg|\exp{\left(-\frac{\Gamma L}{2}\right)\text{sinc}\left[\frac{\left(\Delta k-i\Gamma\right)L}{2}\right]}\Bigg|^2,
\earr
where $\Gamma=\gamma_s+\gamma_i$ are the losses seen by the signal ($\gamma_s\equiv\gamma(\omega_s)$) and the idler ($\gamma_i\equiv\gamma(\omega_i)$) and $\Delta k=k(\omega_p)-k(\omega_s)-k(\omega_i)$ is the mismatch parameter. One can readily see that in the limit $\Gamma=0$, the equation above reproduces the well known result from nonlinear optics \cite{boyd}, that $P_{SPDC}\propto\text{sinc}^2(\Delta k L/2)$. For the general case where $\Gamma\neq 0$, instead, the equation above reduces to
\beq\label{toFig}
P_{SPDC}=-\frac{2\exp{\left(-\Gamma L\right)}\left[\cos(\Delta k L)-\cosh(\Gamma L)\right]}{L^2(\Delta k^2+\Gamma^2)},
\eeq
which is depicted in Fig. \ref{figurePspdc} as a function of the dimensionless quantity $x=\Gamma L$, for different values of the phase mismatch parameter $\Delta k$.
\subsection{Spontaneous Emission Decay Rate of a Quantum Emitter in an Arbitrary Medium}
The spontaneous emission decay rate of a quantum emitter embedded in a homogeneous environment can be written as follows \cite{loudon}
\beq\label{spontaneousDecay}
\Gamma_{sp}=\frac{2\pi c\Gamma_0}{\omega_0}\operatorname{Im}\Bigg\{\operatorname{Tr}\left\{\overleftrightarrow{\vett{G}}(\vett{r},\vett{r};\omega_0)\right\}\Bigg\},
\eeq
where $\Gamma_0$ is the free-space spontaneous emission decay rate (the Einstein $A$ coefficient), $\omega_0$ is the characteristic frequency of the quantum emitter, here modelled as a two-level system with a dipole moment $\boldsymbol\mu$, and $\overleftrightarrow{\vett{G}}(\vett{r},\vett{r};\omega)$ is the dyadic Green function of the electromagnetic field in which the emitter is embedded,  calculated at the position of the emitter itself, and $\operatorname{Tr\left\{\vett{A}\right\}=\sum\limits_{i}A_{ii}}$ is the trace operation \cite{byronFuller}. Equation \eqref{spontaneousDecay} is a general result, that holds for different systems, such as atoms in free space \cite{loudon}, cavity QED \cite{cavityQED}, and even metamaterials \cite{atomMeta}. 

If we assume to consider a homogeneous surrounding medium characterised by the complex-valued dielectric function $\varepsilon=\varepsilon_1+i\varepsilon_2$, and assume as well that the dipole moment of the emitter is aligned along the $z$ axis, Eq. \eqref{spontaneousDecay} simplifies to \cite{barnettGamma, greenSing}
\beq\label{spontaneousDecay2}
\Gamma=\Gamma_0\operatorname{Re}\left\{\sqrt{\varepsilon_1+i\varepsilon_2}\right\}
\eeq

We can arrive at the same result of both Eq. \eqref{spontaneousDecay} and Eq. \eqref{spontaneousDecay2} using path integrals. To show that, we can use the interacting theory of the dressed electromagnetic field described in Sect. \ref{sectSPDC}, but rather than calculating the actual nonlinear interaction of the field with itself (\emph{i.e.}, the second-order nonlinear processes) we turn our attention to the nonlinear corrections to the free propagator $\mathcal{Z}_0(J)$. These corrections are known in QFT as loop corrections, or, sometimes, vacuum bubbles \cite{srednicki, das, maggiore, brown} and are represented by Feynman diagrams like the following
\beq
\mathcal{Z}_{corr}=
\begin{gathered}
\begin{tikzpicture}
  \def\vertices{
    \vertex (a)  at (0, 0);
    \vertex (up) at (1, 1);
    \vertex [dot] (md) at (1, 0){};
    \vertex (b)  at (2, 0);
    \vertex (lf) at (1/2,1/2);
    \vertex (rg) at (3/2,1/2);
  }

  \begin{feynman}
    \vertices
    \diagram* {
      (a) -- [scalar] (md) --[scalar]  (b),
      (md) -- [quarter left, photon] (lf) -- [quarter left ,photon] (up),
      (md) -- [quarter right, photon] (rg) -- [quarter right, photon] (up)
    };
  \end{feynman}\end{tikzpicture}
\end{gathered}
+ \text{higher order terms}.
\eeq
This term can be obtained by Eq. \eqref{ZSPDC} by simply identifying the two source terms $J_{\mu}$ and $J_{\nu}$ with each other, and corresponds to the following integral
\beq
\mathcal{Z}_{corr}=-\frac{1}{2}\int\,d^4x\,\lambda_{\alpha\alpha}(x,\omega)G_{\alpha\alpha}(0,\omega),
\eeq
where $\omega_0$ is the transition frequency of the quantum emitter. Notice, that in the integral above, we have used $\lambda_{\alpha\alpha}(x)$, rather than $\tilde{\chi}_{\alpha\alpha}(x)$, since the interaction vertex in this case doesn't necessarily represent an actual nonlinear optical interaction, but rather the quantum emitter which, initially its excited state, emits a photon, which gets absorbed at a later time (from here the loop term appearing in the Feynman diagram above). Moreover, the dipole moment $\boldsymbol\mu$ of the quantum emitter is aligned along the $\alpha$ direction. If, without any loss of generality, we assume that $\alpha\rightarrow z$, and that the quantum emitter is localised at the vertex, and that it is characterised by a transition frequency $\omega_0$, the interaction vertex $\lambda_{\alpha\alpha}(x,\omega)=|\boldsymbol\mu|^2\delta(\omega-\omega_0)$
and therefore
\beq
\mathcal{Z}_{corr}=|\boldsymbol\mu|^2G_{zz}(0,\omega_0).
\eeq
In QFT, these loop corrections need to be properly renormalised, as they amount to a sort of self-energy of the vacuum (in our case self-energy of the dressed electromagnetic field), which typically diverge. However, while the real part of the Green's function above actually diverges, its imaginary part remains bounded, and it is actually the spontaneous emission rate of the quantum emitter, \emph{i.e.},
\beq
\Gamma_{sp}=\operatorname{Im}\left\{\mathcal{Z}_{corr}\right\}\propto\operatorname{Im}\left\{G_{zz}(0,\omega_0)\right\}
\eeq

To prove that, let us first write down the explicit expression of the loop propagator $G_{zz}(0,\omega_0)$, which is now a scalar quantity, using Eq. \eqref{greenScalar} as
\barr
G_{zz}(0,\omega_0)&=&\int\,\frac{d^3k}{(2\pi)^3}\,\frac{1}{k^2+m^2},
\earr
where $m^2=\omega_0^2(\varepsilon_1+i\varepsilon_2)$, \emph{i.e.}, is the ``mass of the dressed photon in the medium described by the complex dielectric function $(\varepsilon_1+i\varepsilon_2)$. Since the environment surrounding the quantum emitter is homogeneous and isotropic, we can perform the integral above in spherical coordinates, readily integrating away the angular degrees of freedom, which amount to an overall $4\pi$ multiplicative term. Separating the real and imaginary part of the fraction in the integral above leads then to the following result
\barr
G_{zz}(0,\omega_0)&=&\frac{1}{2\pi^2}\int_0^{\infty}\,dk\,\frac{k^2(k^2+\omega_0^2\varepsilon_1)}{[k^2-\omega_0^2(\varepsilon_1+i\varepsilon_2)]^2}\nonumber\\
&-&\frac{i\omega_0^2\varepsilon_2}{2\pi^2}\int_0^{\infty}\,dk\,\frac{k^2}{[k^2-\omega_0^2(\varepsilon_1+i\varepsilon_2)]^2}.
\earr
Both integrals above can be readily evaluated, giving $\operatorname{Re}\{G_{zz}(0,\omega_0)\}\rightarrow\infty$, and
\beq
\operatorname{Im}\{G_{zz}(0,\omega_0)\}=\frac{\omega_0}{4\pi}\operatorname{Re}\{\sqrt{\varepsilon_1+i\varepsilon_2}\},
\eeq
which, apart from an inessential multiplicative factor, looks very similar to the result in Eq. \eqref{spontaneousDecay2}.
\section{Conclusions and Future Perspectives}\label{sect10}
In this tutorial, we have introduced the reader to the concept and framework of path integrals. We have discussed their origin, historical development and their interpretation as \emph{sums over all possible trajectories}, and, in Sect. \ref{sect1}, we have provided the foundations upon which the very concept of path integral is rooted. To help the reader familiarise with the formalism, and the kind of mathematics needed to deal with path integrals, in Sect. \ref{sect2} we have presented some introductory standard textbook examples, namely the free particle (Sect. \ref{freeParticle}), the refraction of a light beam from an interface (Sect. \ref{SnellSection}) , the double slit experiment (Sect. \ref{DoubleSlitSection}), and the quantum harmonic oscillator (Sect. \ref{harmonicOscillator}). All these examples have been discussed extensively, pointing out the procedures and techniques to get to the analytical form of the kernel for those special cases.

Starting from Sect. \ref{sect3}, then, we have built upon this knowledge, and presented a series of less trivial applications of path integrals in both classical and quantum optics, with the twofold aim of presenting an alternative derivation, based on path integrals, of results known in optics, and helping the reader contextualise this framework in an accustomed environment. In particular, we have discussed the propagation of light through an inhomogeneous medium in Sect. \ref{sect3}, while in Sect. \ref{sect4}, we have shown how this framework can also be employed to solve problems of quantum nonlinear optics, and we have discussed explicitly the case of parametric amplification.

The first part of our tutorial concludes with Sect. \ref{sect4bis}, where we give the reader an idea of the computational capabilities of PIMC, and briefly discuss how this could be used in photonics. Within the current photonics ecosystem, and considering the present challenges faced by this discipline, in fact, we believe that having at disposal a computational framework that allows calculations of complex light-matter interaction scenarios, and the derivation of optical properties of complex materials in an exact way, without the need to resort to any particular regime of approximation, could represent a very viable resource for photonics. This, somehow, is the very aim of our tutorial, \emph{i.e.}, to introduce the photonics community to this elegant, but also very powerful, formalism, and stimulate cross-contamination between these two seemingly different disciplines.

The second part of our tutorial, starting with Sect. \ref{sect5} gives the reader a very basic introduction on the use of path integrals in QFT. Although at a first glance this might seem too advanced for the spirit of this tutorial, we nevertheless included this topic, for essentially two reasons. First, since many of the modern problems in theoretical physics, such as quantum gravity, string theory, topological field theories, and so on, are formulated in terms of path integrals, introducing the reader to the very basic formalism of path integrals in QFT would greatly help them, were they interested in reading more about these topics. Secondly, sections \ref{sect5}-\ref{sect7} set the stage for Sect. \ref{sect8}, where we present a unified way to describe quantum nonlinear dynamics of the electromagnetic field in arbitrarily shaped media in terms of path integrals, as an example of the potential that this framework has to offer to photonics, not only from a purely computational perspective, like the prediction of optical properties of materials granted by PIMC, but also as a viable and simple analytical tool for complex problems in photonics.

%Since their introduction in the late 1940s, path integrals have always been a fascinating subject in physics. After more than 60 years, they have contaminated many different areas of physics, and not only. The extent of their usefulness in so many different disciplines makes them surely one of the most intriguing framework ever developed, in particular concerning their ability to make very complicated problems treatable in very simple terms.

With this tutorial, we hope to have sparked the curiosity and interest of the reader on the topic, and conveyed the message that path integrals do not only represent an elegant formulation of quantum mechanics , but they also provide the language of modern theoretical physics, and can be applied in many different areas of science. For the specific case of photonics, as extensively discussed in this tutorial, path integrals could represent a valuable analytical and computational resource for investigating complex and multi-physics photonics devices, which would require the assembly of model systems from different disciplines, a task that is not always simple to carry out. The universality of path integrals, the simplicity of adaption of the formalism to very different situations, and their remarkable ability to treat very complex problems in an easy and accessible way, could represent a significant step forward towards this goal.

\section*{Acknowledgements}
The authors acknowledge the financial support of the Academy of Finland Flagship Programme (PREIN - decisions 320165, 320166)
\section*{Data Availability}
The data that support the findings of this study are available from the corresponding author upon reasonable request.
\section*{References}
\bibliography{path_integrals}

\appendix
\section{How to Deal with Path Integrals in QFT: Reducing Eq. \eqref{partHB} to Eq. \eqref{partDem} Using Path Integration}\label{appendixA}
In this appendix, we show how one can perform the path integration over the matter $\vett{P}$ and reservoir $\vett{F}(\omega)$ fields in Eq. (\ref{partHB}), converting the full path integral partition function to the effective partition function $Z_{eff} (J)$ shown in Eq. (\ref{partDem}). This will introduce to the reader not accustomed to this field how one can perform explicit calculations relating to path integration for fields \cite{ornigotti2019,bechler}.

Before we begin our calculations, it is useful to recall the explicit formula for calculating a Gaussian integral in $n$ dimensions \cite{fai,gifted}, which will be useful for the calculations presented in this appendix. To do so, we first need to define two vectors $\vett{x}=(x_{1},x_{2},...,x_{n})\in \mathbb{R}^{n}$, and $\vett{b} \in \mathbb{R}^{n}$, a non-singular $n \times n$ matrix $A\in \mathbb{C}$, such that the scalar product $(\vett{x}, A\vett{x})=x_i^TA_{ij}x_j$ is a quadratic form, the following result holds:
\begin{equation}
\int d^{n}\,x\, \exp{\left[ -\frac{1}{2}(\vett{x},A\vett{x})-(\vett{b},\vett{x}) \right]} = \sqrt{\frac{\pi^{n}}{\mathrm{det}A}}\exp{\left[ \frac{1}{2}(\vett{b},A^{-1}\vett{b}) \right]},\nonumber
\end{equation}
where $A^{-1}$ is the inverse of $A$.

This result can be used to compute Gaussian integrals over field configurations. In analogy with what has been done in Sect. \ref{sect2} for quantum particles, one can in fact discretise the path integral over the field configuration into a sum over a finite set of configurations, \emph{i.e.},  $d\vett{P} \propto \prod_{k} d\vett{P} (x_{k})$, perform Gaussian integration with respect to $x_k$, and then take the usual continuum limit.

We begin by restating the full path integral of the Huttner-Barnett model \cite{huttnerBarnett} as given in Sec. \ref{sect8}, setting $\vett{J}_{P}=\vett{J}_{F}=0$:
\begin{equation} \label{eq:full_path_int}
Z(J) = \int \mathcal{D} \vett{A}\, \mathcal{D} \vett{P}\, \mathcal{D} \vett{F}\, \exp{\left( i\,S_{HB} + i\int d^{4} x\, \vett{J} \cdot \vett{A} \right)} .
\end{equation}
As we have set $\vett{J}_{P}=\vett{J}_{F}=0$ (because, as said in the main text, we are only interested in photon dynamics, and not in the dynamics of polaritons or other quasi-particles), the only free parameter is the source term $\vett{J}$ for the electromagnetic field. Moreover, since we want to reduce the path integral above to Eq. \eqref{partDem}, we can first factor out the parts of the Huttner-Barnett Lagrangian that depend only on the electromagnetic field, including the $\vett{J}\cdot\vett{A}$-term above. To do so, we group them into the term $S_{em}(\vett{A})$, in order to rewrite the integral above in the following, nested form
\begin{equation}
Z(J)=\int\,\mathcal{D}\vett{A}\,\exp{\left[ iS_{em} \left( \vett{A} \right) \right]\,I_{P} \left( \vett{A} \right)},
\end{equation}
with
\begin{equation} \label{eq:IPA}
I_{P} \left( \vett{A} \right) = \int \mathcal{D} \vett{P}\,\exp{\left[ i\left( S_{mat} \left( \vett{P} \right) + S_{mf} \left( \vett{A} ,\vett{P} \right) \right) \right]} I_{F} \left( \vett{P} \right),
\end{equation}
and
\begin{equation} \label{eq:IF}
I_{F} \left( \vett{P} \right) = \int \mathcal{D} \vett{F}\,\exp{\left[ i\left( S_{res} \left( \vett{F} \right) + S_{mr} \left( \vett{F} ,\vett{P} \right) \right) \right]},
\end{equation}
where the various actions $S_k$ appearing above sum to the total Huttner-Barnett action $S_{HB}=\int\,d^4x\,\mathcal{L}_{HB}$, where the Huttner-Barnett Lagrangian is given in Eq. \eqref{LagHB}, from which the following individual pieces can be easily defined:
\begin{equation}\label{lagrangianEM}
\mathcal{L}_{em} \left( \vett{A} \right) = \frac{\epsilon_{0}}{2}\dot{\vett{A}}^{2}
 - \frac{1}{2\mu_{0}}\left( \nabla \times \vett{A} \right)^{2} ,
\end{equation}
\begin{equation}
\mathcal{L}_{mat} \left( \vett{P} \right) = \frac{g(\vett{x})}{2\,\epsilon_{0}\,\omega_{0}^{2}\,\beta (\vett{x})} \left[ \dot{\vett{P}}^{2} - \omega_{0}^{2}\,\vett{P}^{2} \right],
\end{equation}
\begin{equation}
\mathcal{L}_{res} \left( \vett{F} \right) = g(\vett{x}) \int_{0}^{\infty} d\omega \left[ \frac{\rho (\vett{x})}{2}\dot{\vett{F}}^{2} - \frac{\rho (\vett{x})\,\omega^{2}}{2}\vett{F}^{2} \right],
\end{equation}
\begin{equation}
\mathcal{L}_{fm} \left( \vett{A},\vett{P} \right) =- g(\vett{x})\,\dot{\vett{A}}\cdot\vett{P},
\end{equation}
and
\begin{equation}
\mathcal{L}_{mr} \left( \vett{P},\vett{F} \right) = -g(\vett{x}) \int_{0}^{\infty}d\omega f(\omega ,\vett{x})\,\vett{P}\cdot \dot{\vett{F}} .
\end{equation}
In the above equations, a dot indicates derivative with respect to time.

Notice, how Eq. \eqref{eq:IF} only contains quantities depending on the reservoir field $\vett{F}(\omega)$ (with the matter field $\vett{P}$ acting as a parameter, with respect to $\mathcal{D}\vett{F}(\omega)$), and no term of order higher than $\vett{F}^2(\omega)$ appears, meaning that we can compute this integral by means of Gaussian integration.
\subsection{Integration over the reservoir fields}
Let us start by rewriting the exponent in Eq. \eqref{eq:IF} in a form that will explicitly contain a quadratic form $(\vett{F}(\omega),\hat{A}\vett{F}(\omega))$ in the reservoir field. We then first integrate by parts (with respect to time), use the identity
\begin{equation}\label{identity}
\left( \frac{\partial \phi}{\partial t} \right)^{2} = \frac{\partial}{\partial t} \left( \phi \frac{\partial \phi}{\partial t} \right) - \phi \frac{\partial^{2} \phi}{\partial t^{2}} ,
\end{equation}
and integrate again by parts, to get the following result:
\barr \label{eq:res_mr_actions}
S_{res} \left( \vett{F} \right) &+& S_{mr} \left( \vett{F} ,\vett{P} \right) = \int dt\, dt'\, d^{3} x\, \int_{0}^{\infty} \,d\omega\, g(x)\Big\{\nonumber\\
 &-& \frac{1}{2} \left( \vett{F} (t'), \,\hat{A}(t,t')\vett{F} (t) \right) - \left( \vett{b} (t'),\vett{F} (t) \right) \Big\} ,
\earr
where $\vett{x}$ is relabelled as $x$ from now on in this appendix, for simplicity, and we have defined
\begin{equation}
\vett{b}(t') = -i\, g(x)\, f(\omega )\, \delta (t-t')\, \dot{\vett{P}},
\end{equation}
and the operator $\hat{A}\equiv\hat{A}(t,t')$ as
\begin{equation}\label{eq:A_oper}
\hat{A}(t,t')=\frac{i\,\rho g(x)}{2} \left( \frac{\partial^{2}}{\partial t^{2}} + \omega^{2} \right) \delta (t-t').
\end{equation}

Notice, how we have added a second time integration (and consequently a Dirac delta function $\delta(t-t')$) in Eq. \eqref{eq:res_mr_actions} to facilitate the definition of the quantities presented above. Notice, moreover, how the operator $\hat{A}(t-t')$ is very similar to the usual wave equation operator \cite{srednicki, loudon}.

It is instructive to repeat the salient steps that led to Eq. \eqref{eq:res_mr_actions}. First, integration by parts with respect to time allows us to rewrite the reservoir-matter field interaction term as
\begin{equation}
\int \,dt\, \left[ -g(x)\, f(\omega )\,\vett{P} \cdot \dot{\vett{F}} \right] = g(x)\, f(\omega )\, \int dt \dot{\vett{P}} \cdot \vett{F} .
\end{equation}
Then, the identity presented in Eq. \eqref{identity} can be used to rewrite the term proportional to $\dot{\vett{F}}^2$ as
\beq
\frac{\rho\, g(x)}{2}\int dt\,dt'\,\delta(t-t') \frac{\partial \vett{F}}{\partial t} \frac{\partial \vett{F}}{\partial t'},
\eeq
where the Dirac delta function $\delta(t-t')$ has been inserted to differentiate the two time derivative terms, thus making the effect of part integration appear more obvious. At this point, if we integrate by parts, we can shift one of the time derivatives in one term onto the other one, thus obtaining
\begin{equation}
- \frac{\rho g(x)}{2} \int\, dt\,dt'\,\delta(t-t') \vett{F}\, \frac{\partial^{2} \vett{F}}{\partial t{^2}}.
\end{equation}
Putting everything together, we can easily reconstruct both the linear and the quadratic form in Eq. \eqref{eq:res_mr_actions}.  
\subsection{How to calculate the operators $\hat{A}$ and $\hat{A}^{-1}$}
The Green's function for the operator $\hat{A}$ in Eq. \eqref{eq:A_oper} can be found in the following manner. The operator contains no spatial derivatives and so the Green's function $K(t-t',x-x')$ associated with it will have a localized spatial dependence of the form $\delta (x-x')$. Hence,
\begin{equation} \label{eq:misc_eq_1}
i\rho g(x) \left( \frac{\partial^{2}}{\partial t^{2}} + \omega^{2} \right) K (t-t',x) = \delta (t-t') .
\end{equation}
Taking the Fourier transform of (\ref{eq:misc_eq_1}) with respect to time produces
\begin{equation}
\left( -\Omega^{2} + \omega^{2} \right) K (\Omega , x) = \frac{\exp{ \left( i\Omega\, t' \right)}}{i\,\rho\, g(x)} .
\end{equation}
From these results, one finds the expression for the Green's function $K(t-t',x)$ by simply integrating the result above with respect to the frequency $\Omega$, obtaining
\barr
K(t-t',x) &=& \frac{1}{i\,\rho g(x)}\, \int\, \frac{d\Omega}{2\pi}\,\frac{\exp{\left[ i\Omega (t-t') \right]}}{(\omega^{2} - \Omega^{2})}\nonumber\\
& \equiv& \frac{D_{F} (t-t',\omega)}{i\,\rho g(x)} ,
\earr
with $D_{F} (t-t',\omega)$ the Feynman propagator of the reservoir field, explicitly given by
\begin{equation}
D_{F} (t-t',\omega) = \int\, \frac{d\Omega}{2\pi}\,\frac{\exp\,\left[ i\Omega (t-t') \right]}{(\omega^{2} - \Omega^{2})} .
\end{equation}
Hence, by recalling that the Green's function of an operator $\hat{A}$ can be interpreted as the inverse of that same operator \cite{byronFuller}, we can make the following formal identification
\begin{equation}
\hat{A}^{-1} \rightarrow K(t-t',x) = \frac{D_{F} (t-t',\omega)}{i\rho g(x)}.
\end{equation}
\subsection{Calculation of the integral $I_F$}
If we now discretise the integration measure, \emph{i.e.}, $\mathcal{D}\vett{F}\rightarrow\prod_k\,d\vett{F}(x_k,\omega)$ we can perform the integral appearing in Eq. \eqref{eq:IF} using first Gaussian integration and then taking the continuum limit. By doing so, we obtain the following result
\begin{equation} \label{eq:IF2}
I_{F} \left( \vett{P} \right) = \mathcal{N}_{F}\, \exp{ \left[ \frac{1}{2} \left( \vett{b} , \,\hat{A}^{-1} \vett{b} \right) \right]},
\end{equation}
with $\mathcal{N}_{F}$ a suitable normalisation constant and
\barr\label{bAb}
\left( \vett{b} , \,\hat{A}^{-1} \vett{b} \right) &=& - i \Big[ \int dt\, d^{3} x \int_{0}^{\infty}\, d\omega\, \frac{|f(\omega )|^{2} \,g(x)}{\rho}\, \vett{P}^{2} (t)\nonumber\\
&+& \int dt \,dt' \,d^{3} x \,\frac{g(x)}{\rho} \vett{P} (t) \,G (t-t',x)\, \vett{P} (t') \Big] ,
\earr
where
\begin{equation}
G (t-t',x) = \int_{0}^{\infty}\, d\omega\, \omega^{2}\, |f(\omega )|^{2}\, D_{F} (t-t',\omega )
\end{equation}
is defined to be the time-domain Green's function of the reservoir field.
\subsection{Integration over the matter fields}

Eq. (\ref{eq:IF2}) is the solution to the integral over the reservoir degrees of freedom. As it can be seen, this result contains a dependence on the matter polarisation field $\vett{P}$. If we wish to solve the integral $I_P$ with the same strategy adopted above for $I_F$, \emph{i.e.}, by employing Gaussian integration, we first need to rewrite it in a way containing a quadratic form of the kind $(\vett{P}, \,\hat{B}\,\vett{P})$. To do that, we can first integrate by parts with respect to time $t$ and to time $t'$ in Eq. \eqref{bAb}, which allows us to rewrite the term containing  $\dot{\vett{P}}\,D_F(t-t',x)\,\dot{\vett{P}}$ as
\begin{equation} \label{eq:misc_eq_2}
\int dt'\, dt''\, \vett{P}(t)\, \frac{\partial^{2} D_{F} (t-t',\omega )}{\partial t \partial t'}\, \vett{P}(t') .
\end{equation}
As the Feynman propagator depends on the difference $t-t'$, one can change variables to $\tau = t-t'$, yielding $\partial^{2}/\partial t \partial t' = \partial^{2} / \partial \tau^{2}$. The fact that the Feynman propagator obeys
\begin{equation}
\left( \frac{\partial^{2}}{\partial \tau^{2}} + \omega^{2} \right) D_{F} (\tau ,\omega ) = \delta (\tau),
\end{equation}
allows us to write
\begin{equation}
\frac{\partial^{2} D_{F}(\tau,\omega )}{\partial \tau^{2}} = \delta (\tau) - \omega^{2} D_{F} (\tau ,\omega ) .
\end{equation}
Converting back to the separate time variables, $t$ and $t'$, and inserting our result into Eq. (\ref{eq:misc_eq_2}), we get
\begin{equation}
\int dt \vett{P}^{2} (t) - \omega^{2}\int dt dt' \vett{P}(t) D_{F} (t-t',\omega )\vett{P} (t') .
\end{equation}
Before proceeding with our calculations, it is useful to introduce the following quantity
\begin{equation}
G(t-t',x) = \int_{0}^{\infty}d\omega \omega^{2} |f(\omega ,x)|^{2} D_{F} (t-t',\omega ) ,
\end{equation}
which can be interpreted as the reservoir-weighted Green's function. With this at hand, the exponent in Eq. \eqref{bAb} becomes
\barr
&-&\frac{i}{2}\Big[ \int dt\, d^{3}x\, \int_{0}^{\infty}\, d\omega\, \frac{|f(\omega)|^{2} g(x)}{\rho}\, \vett{P}^{2} (t)\nonumber\\
&+& \int\,dt \,dt' \,d^{3}x\, \frac{g(x)}{\rho}\, \vett{P}(t)\, G(t-t',x) \,\vett{P}(t') \Big] .
\earr
\subsection{Derivation of the quadratic form for $I_P$}
In the form above, the exponent of $I_{F} \left( \vett{P} \right)$ contains a term proportional to $\vett{P}^{2}$. This term can be summed with the corresponding quadratic term appearing in the free part of the matter action $S_{mat} \left( \vett{P} \right)$ in the expression of $I_P(\vett{A})$ given by Eq. \eqref{eq:IPA}, \emph{i.e.}, 
\barr
\int dt\, d^{3}x\, \Bigg\{ -\frac{g(x)}{2\epsilon_{0}\, \beta} - \int_{0}^{\infty} d\omega \frac{|f(\omega)|^{2} g(x)}{2\rho} \Bigg\} \vett{P}^{2} (t) .
\earr
If we now introduce the quantity $v (\omega ) = f(\omega ) \sqrt{\epsilon_{0} \omega_{0}^{2} \beta \rho}$ and the scaled resonance frequency
\begin{equation} \label{eq:scaled_resonance_freq}
\tilde{\omega_{0}}^{2} = \omega_{0}^{2} + \int_{0}^{\infty} d\omega \frac{|v (\omega )|^{2}}{\rho^{2}},
\end{equation}
we can rewrite the term in the curly brackets above in the following compact form
\begin{equation}
 -\frac{g(x)}{2\epsilon_{0}\, \beta} - \int_{0}^{\infty}\, d\omega\, \frac{|f(\omega)|^{2}\, g(x)}{2\rho} = -\frac{g(x)\,\tilde{\omega}_0^2}{2\epsilon_{0}\, \omega_{0}^{2} \beta} .
\end{equation}
Notice, how the primary effect of the reservoir field is to introduce a frequency-dependent shift in the resonance frequency $\omega_0$ of the material. 
	
The $I_P$-integral now reads
\beq
I_P=\int\,\mathcal{D}\vett{P}\,\exp{\left[i\,\Phi(\vett{P})\right]},
\eeq
where now
\barr
\Phi (\vett{P}) &=& \int \,dt\, d^{3}x\,\left[ \frac{g(x)}{2\epsilon_{0}\,\omega_{0}^{2}\,\beta} \left( \dot{\vett{P}}^{2} - \tilde{\omega}_{0}^{2}\, \vett{P}^{2} \right)+ g(x)\, \dot{\vett{A}} \cdot \vett{P}\right]\nonumber\\
&+& \frac{1}{2}\int \,dt\, dt'\, d^{3}x\, \frac{g(x)}{\rho}\, \vett{P} (t)\,G(t-t',x)\,\vett{P}(t').
\earr
As we did for $I_F$, one can then integrate by parts the term proportional to $\dot{\vett{P}}^{2}$ and introduce an extra integral with respect to time, to be able to write the above quantity as a quadratic form, \emph{i.e.},  $(\vett{P} ,\hat{B}\, \vett{P}) + (\vett{c},\vett{P})$, where
\begin{equation}
\vett{c}(t) = - i g(x)\, \dot{\vett{A}},
\end{equation}
and the expression for the operator $\hat{B}\equiv\hat{B}(t,t')$ is obtained by looking at the terms sandwiched between the terms $\vett{P}(t')$ and $\vett{P}(t)$, \emph{i.e.}, 
\barr\label{operatorB}
\hat{B}(t,t')&=&\frac{i\,g(x)}{\epsilon_{0}\, \omega_{0}^{2}\, \beta} \left( \frac{\partial^{2}}{\partial t^{2}} + \tilde{\omega_{0}}^{2} \right) \delta (t-t')\nonumber\\
&-& \frac{i\,g(x)}{\rho}\, G (t-t',x).
\earr
Notice how the form of the operator $\hat{B}$ is very similar to that of the operator $\hat{A}$ introduced when calculating $I_F$. The extra term appearing above, \emph{i.e.}, $-i\,g(x)\,G(t-t',x)/\rho$, however, makes $\hat{B}$ intrinsically integro-differential, which, in turns, makes the determination of its correspondent Green's function quite complicated (we discuss it briefly in the next subsection, but the interested reader is referred to Ref. \citenum{bechler} for more details).

Once we have transformed the exponent $\Phi(\vett{P})$ into a quadratic form, we can again discretise the path integral, solve using Gaussian integration, and then take the limit of it, to obtain the following result
\begin{equation}
I_{P} \left( \vett{A} \right) = \exp{ \left[ \frac{i}{2} \int \,dt\, dt'\, d^{3}x\, g(x)\, \dot{\vett{A}}(t)\, \Gamma (t-t',x)\, \dot{\vett{A}}(t') \right]},\nonumber
\end{equation}
where $\Gamma(t-t',x)$ is the Green's function of the operator $\hat{B}$.
\subsection{Expression and physical meaning of $\Gamma(t-t',x)$}

The Green's function for the operator $\hat{B}$ just defined can be found in the following way: first, notice that the prefactor $i\,g(x)$ in the definition of $\hat{B}$, \emph{i.e.}, Eq. \eqref{operatorB} can be treated, for the purpose of finding the Green's function, as a ``numerical" constant, and therefore it can be neglected during calculation (we can put it back at the end of it). Using the definition of Green's function \cite{byronFuller} we then know that $\Gamma(t-t',x)$ is a solution of the equation $\hat{B}\,\Gamma(t-t',x)=\delta(t-t')$.

It should be noted that in this case the operator $\hat{B}$ is not simply a differential operator, as was the case earlier in (\ref{eq:A_oper}), but it contains an extra source term. The Green's function for $\hat{B}$ is then found as the solution to the following integro-differential equation
\barr \label{eq:misc_eq_4}
&&\frac{1}{\epsilon_{0}\, \omega_{0}^{2} \beta} \,\left( \frac{\partial^{2}}{\partial t^{2}} +\tilde{\omega_{0}}^{2} \right) \Gamma (t-t',x)\nonumber\\
&-& \frac{1}{\rho}\, \int\, d\tau\, G (t-\tau ,x)\,\Gamma(\tau -t,x) = \delta (t-t').
\earr
We can solve the equation above by means of Fourier transformation with respect to $t-t'$, thus obtaining
\begin{equation}
\frac{1}{\epsilon_{0} \omega_{0}^{2} \beta} \left( \tilde{\omega_{0}}^{2} - \Omega^{2} \right) \tilde{\Gamma}(\Omega ,x) - \frac{1}{\rho }\tilde{G} (\Omega ,x) \tilde{\Gamma}(\Omega ,x) = 1 ,
\end{equation}
where $\tilde{G} (\Omega ,x)$ is the Fourier transform of $G(t-t',x)$. After a lengthy but straightforward calculation \cite{bechler}, one can arrive at the following form for the Fourier transform of $\Gamma(t-t',x)$
\begin{equation}
\tilde{\Gamma} (\Omega ,x) = \frac{\epsilon_{0} \omega_{0}^{2} \beta}{\omega_{0}^{2} - \Omega^{2}\left[ 1+ \lambda_{F} (\Omega) \right]},
\end{equation}
where $\lambda_F(\Omega)$ is the Fourier transform of the Feynman propagator $D_F(t-t',\omega)$. 

The above result can be interpreted as an effective dielectric constant for the ``dressed" electromagnetic field. To see this, let us consider the full expression of the effective Lagrangian coming from $\mathcal{L}_{em}$ as given in Eq. \eqref{lagrangianEM} and the result given by $I_P(\vett{A})$. Combining these two terms and remembering that $\dot{\vett{A}}=-\vett{E}$ and $\nabla\times\vett{A}=\vett{B}$, we get, in Fourier space
\barr
\mathcal{L}_{eff} \left( \vett{E},\vett{B} \right) &=& \epsilon_{0} |\vett{E} (\Omega ,x)|^{2} + \frac{1}{\mu_{0}}|\vett{B} (\Omega ,x)|^{2}\nonumber\\
&+& g(x)\, \Omega^{2}\, \tilde{\Gamma} (\Omega ,x)\, |\vett{E} (\Omega ,x)|^{2},
\earr
where the effective Lagrangian has been written in terms of the electric and magnetic field, since this form will provide an easier understanding of the interpretation of $\Gamma(t-t',x)$. From $\mathcal{L}_{eff}$ we can calculate the displacement vector $\vett{D}(\Omega,x)$, as follows \cite{landau}
\begin{equation}
\vett{D} (\Omega ,x) = \frac{\partial \mathcal{L}_{eff}}{\partial \vett{E}^{*}} = \epsilon_{0} \vett{E} (\Omega ,x) + g(x) \tilde{\Gamma} (\Omega ,x) \vett{E} (\Omega ,x) .
\end{equation}
If we then recall the usual constitutive relation for the displacement vector \cite{jackson}, \emph{i.e.}, $\vett{D}=\varepsilon_0\,\boldsymbol\varepsilon\,\vett{E}$, we can identify the term $\tilde{\Gamma}(\Omega,x)$ with the dielectric constant of a medium, \emph{i.e.},
\begin{equation}\label{epsilonDEF}
\boldsymbol\varepsilon (\Omega ,x) = 1 + \frac{g(x)}{\epsilon_{0}}\tilde{\Gamma} (\Omega ,x).
\end{equation}
It should now appear clear, that $\Gamma(t-t',x)$ represents the dielectric function of an effective medium the electromagnetic field is propagating through. In particular, it contains information on both the loss channels present in the medium (through the reservoir field), and the possible interaction channels that the medium has with the electromagnetic field (through the matter porlarisation). After the integration with respect to $\vett{F}$ and $\vett{P}$ the partition function has the following form
\barr\label{partFinal}
Z(J)&=&\int\,\mathcal{D}\vett{A}\,\exp{\Bigg\{i\, S_{em}+i\,\int\,d^4x\,\vett{J}\cdot\vett{A}}\nonumber\\
&+&\frac{i}{2}\,\int\,dt\,dt'\,d^3x\,g(x)\,\dot{\vett{A}}\cdot\overleftrightarrow{\boldsymbol{\Gamma}}\cdot\dot{\vett{A}}\Bigg\},
\earr
where $\dot{\vett{A}}\cdot\overleftrightarrow{\boldsymbol\Gamma}\cdot\dot{\vett{A}}=\dot{A}_{\mu}(t)\,\Gamma_{\mu\nu}(t-t;,x)\,\dot{A}_{\nu}(t')$, and now we have restored the current-dependent term, because now we are going to perform the path integration with respect to the vector potential $\vett{A}$. The result above then matches the form of the effective partition function introduced in Eq. \eqref{partDem}.

\end{document}